\documentclass[12pt]{article}
\pdfoutput=1
\usepackage{putex}
\usepackage{mathtools}
\usepackage{graphicx}
\usepackage{caption}
\usepackage{subcaption}
\usepackage{epstopdf}
\usepackage{enumerate}
\usepackage{cite}
\usepackage{youngtab}
\usepackage{tensor}
\usepackage{slashed}
\usepackage[aligntableaux=center]{ytableau}

\numberwithin{equation}{section}

\newcommand{\HH}{\mathbb{H}}

\newcommand{\abs}[1]{\left\lvert #1 \right\rvert}

\newcommand {\be} {\begin {equation}}
\newcommand {\ee} {\end {equation}}

\newcommand {\bes} {\begin {equation*}}
\newcommand {\ees} {\end {equation*}}

\newcommand{\es}[2] {\begin{equation} \label{#1} \begin{split} #2 \end{split} \end{equation}}

\newcommand{\Z}{\mathbb{Z}}

\newcommand{\R}{\mathbb{R}}
\newcommand{\C}{\mathbb{C}}

\newcommand{\beq}{\begin{equation}}
\newcommand{\eeq}{\end{equation}}

\def\SSP#1{{}}
\def\RY#1{}
\def\MD#1{}

\def\<{\langle}
\def\>{\rangle}

\newcommand{\bC}{\ensuremath{\mathbb{C}}}

\newcommand{\bR}{\ensuremath{\mathbb{R}}}

\newcommand{\bZ}{\ensuremath{\mathbb{Z}}}

\newcommand{\cA}{\ensuremath{\mathcal{A}}}
\newcommand{\cB}{\ensuremath{\mathcal{B}}}

\newcommand{\cD}{\ensuremath{\mathcal{D}}}

\newcommand{\cG}{\ensuremath{\mathcal{G}}}
\newcommand{\cH}{\ensuremath{\mathcal{H}}}
\newcommand{\cI}{\ensuremath{\mathcal{I}}}

\newcommand{\cK}{\ensuremath{\mathcal{K}}}
\newcommand{\cL}{\ensuremath{\mathcal{L}}}
\newcommand{\cM}{\ensuremath{\mathcal{M}}}
\newcommand{\cN}{\ensuremath{\mathcal{N}}}
\newcommand{\cO}{\ensuremath{\mathcal{O}}}
\newcommand{\cP}{\ensuremath{\mathcal{P}}}
\newcommand{\cQ}{\ensuremath{\mathcal{Q}}}
\newcommand{\cR}{\ensuremath{\mathcal{R}}}

\newcommand{\cV}{\ensuremath{\mathcal{V}}}
\newcommand{\cW}{\ensuremath{\mathcal{W}}}

\newcommand{\cZ}{\ensuremath{\mathcal{Z}}}
\newcommand{\dd}{\mathrm{d}}

\newcommand{\ed}{\,.}
\newcommand{\ec}{\,,}
\newcommand{\ecq}{\ec\quad}

\newcommand{\tpsi}{\widetilde{\psi}}
\newcommand{\tq}{\widetilde{q}}
\newcommand{\tQ}{\widetilde{Q}}
\newcommand{\tc}{\widetilde{c}}
\newcommand{\tX}{\widetilde{X}}
\newcommand{\vphi}{\varphi}
\DeclareMathOperator{\trace}{Tr}

\begin{document}
	
\preprint{CALT-TH 2016-029\\ PUPT-2509}

\institution{PU}{Joseph Henry Laboratories, Princeton University, Princeton, NJ 08544, USA}
\institution{Caltech}{Walter Burke Institute for Theoretical Physics, California Institute of Technology, \cr Pasadena, CA 91125, USA}
\institution{Weizmann}{Weizmann Institute of Science, Rehovot 76100, Israel}

\title{A one-dimensional theory for Higgs branch operators}

\authors{Mykola Dedushenko,\worksat{\PU,\Caltech} Silviu S.~Pufu,\worksat{\PU} and Ran Yacoby\worksat{\PU, \Weizmann}}

\abstract {
	We use supersymmetric localization to calculate correlation functions of half-BPS local operators in 3d ${\cal N} = 4$ superconformal field theories whose Lagrangian descriptions consist of vectormultiplets coupled to hypermultiplets.  The operators we primarily study are certain twisted linear combinations of Higgs branch operators that can be inserted anywhere along a given line.  These operators are constructed from the hypermultiplet scalars. They form a one-dimensional non-commutative operator algebra with topological correlation functions. The 2- and 3-point functions of Higgs branch operators in the full 3d $\cN=4$ theory can be simply inferred from the 1d topological algebra. After conformally mapping the 3d superconformal field theory from flat space to a round three-sphere, we preform supersymmetric localization using a supercharge that does not belong to any 3d ${\cal N} = 2$ subalgebra of the ${\cal N}=4$ algebra. The result is a simple model that can be used to calculate correlation functions in the 1d topological algebra mentioned above.  This model is a 1d Gaussian theory coupled to a matrix model, and it can be viewed as a gauge-fixed version of a topological gauged quantum mechanics. Our results generalize to non-conformal theories on $S^3$ that contain real mass and Fayet-Iliopolous parameters.  We also provide partial results in the 1d topological algebra associated with the Coulomb branch, where we calculate correlation functions of local operators built from the vectormultiplet scalars.
}

\date{}

\maketitle
	
\tableofcontents
\setlength{\unitlength}{1mm}

\newpage
\section{Introduction}
\label{INTRO}

Correlation functions of local operators are fundamental observables in quantum field theory.  In more than two spacetime dimensions, there are relatively few examples in the literature where such correlation functions have been calculated non-perturbatively. In some examples, such calculations can be preformed by using non-renormalization theorems, such as in \cite{Lee:1998bxa,DHoker:1999ea} who showed that the non-perturbative answer for 3-point functions of $\frac{1}{2}$-BPS operators in 4d $\cN=4$ Yang-Mills theory are captured entirely by the tree-level result.\footnote{See also \cite{Baggio:2012rr} for a proof, and \cite{Beem:2013sza} for a generalization to 4d $\cN=2$ theories.} The conformal bootstrap approach also allows one to calculate certain correlators of BPS operators in some particular supersymmetric conformal field theories (SCFTs) with $8$ Poincar\'e supersymmetries in various dimensions \cite{Beem:2013sza,Beem:2014kka,Beem:2014rza,Chester:2014mea,Beem:2016cbd}. Other examples use the technique of supersymmetric localization (for recent reviews, see \cite{Tachikawa:2016kfc,Dumitrescu:2016ltq,Morrison:2016bps,Pasquetti:2016dyl,Kim:2016usy,Pestun:2016jze,Zarembo:2016bbk,Marino:2016new,Willett:2016adv,Minahan:2016xwk,Hosomichi:2016flq,Dimofte:2016pua,Qiu:2016dyj,Pestun:2016qko,Benini:2016qnm,Pufu:2016zxm,Rastelli:2016tbz} and references therein) allowing for the calculation of two-point functions of conserved flavor or R-symmetry currents in 3d ${\cal N} = 2$ superconformal field theories (SCFTs) \cite{Closset:2012vg, Closset:2012ru}, and of Coulomb branch operators in 4d ${\cal N} =2$ SCFTs \cite{Gerchkovitz:2016gxx}. (See also \cite{Papadodimas:2009eu,Baggio:2014sna,Baggio:2014ioa,Baggio:2015vxa}.)
	
Our goal here is to provide more instances of such exact computations of correlation functions of local operators.   We focus on 3d quantum field theories with ${\cal N} = 4$ supersymmetry defined by general Lagrangians constructed from vectormultiplets coupled to hypermultiplets.\footnote{Such theories have been extensively studied in the literature starting with Refs.~\cite{Seiberg:1996nz, Intriligator:1996ex}. For string theory constructions of such theories, see, for instance, \cite{deBoer:1996mp, Hanany:1996ie, deBoer:1996ck, Gaiotto:2008ak}.}  In these theories, we provide new formulas for calculating correlation functions of certain $\frac{1}{2}$-BPS operators.  The derivation of these formulas also relies on supersymmetric localization, albeit using a different supercharge from the one used in previous supersymmetric localization studies of such theories.

The $\frac{1}{2}$-BPS operators whose correlation functions we compute fall within two classes of more general operators. The first class, referred to as Higgs branch operators, consists of gauge invariant operators constructed from the scalar fields in the hypermultiplet, while the second class, referred to as Coulomb branch operators, contains operators constructed from the scalars in the vectormultiplet as well as from $\frac 12$-BPS scalar monopole operators.  The naming of the two classes reflects that, when the QFT is defined on $\R^3$, these operators acquire non-zero expectation values on either the Higgs or Coulomb branch of the moduli space of supersymmetric vacua.

We restrict our attention to the origin of the moduli space, where the ${\cal N} = 4$ theories we consider flow in the IR to SCFTs.  Let us denote the gauge group of such a theory by $G$ and assume, without loss of generality, that there is only one hypermultiplet transforming in a unitary representation ${\cal R}$ of $G$, where ${\cal R}$ may be reducible.\footnote{\label{HalfHyperFootnote}Let $N_H$ be the total number of hypermultiplets in the absence of gauging.  The $2N_H$ complex scalars and $2N_H$ complex two component fermions transform in the pseudoreal fundamental representation of the flavor symmetry $USp(2N_H)$.  We consider the $U(N_H)$ subgroup of $USp(2N_H)$ under which the fundamental of $USp(2 N_H)$ decomposes as $\mathbf{N}_H \oplus \overline{\mathbf{N}}_H$, and we take the gauge group $G$ to be a subgroup of $U(N_H)$.  At the level of Lie algebras, we have that a subalgebra $\mathfrak{g}$ of $\mathfrak{u}(N_H)$ is gauged, and we define the representation map ${\cal R} : \mathfrak{g} \to \mathfrak{u}(N_H)$ from the gauge algebra into $N_H \times N_H$ hermitian matrices.  

Our results can be easily extended to include half-hypermultiplets, namely to the case where the gauge group $G$ is a subgroup of $USp(2N_H)$ not contained in $U(N_H)$.  However, we do not discuss this possibility here for simplicity.}   In a nutshell, we have three main results:
\begin{itemize}
	\item We present a relatively simple matrix model coupled to a 1d Gaussian theory, 
	\es{ZPreviewHiggs}{
		Z_\text{Higgs} = \frac1{|\cW|}\int_\text{Cartan} d \sigma \,\det{}'_\text{adj} \left[ 2 \sinh ( \pi \sigma) \right] \int DQ\, D \tQ \exp \left[ {- \ell \int_{-\pi}^\pi d\vphi \left( \tQ \partial_\vphi Q  + \tQ \sigma Q \right)} \right] \,,
	}
	which can be used to calculate all 2- and 3-point correlators of Higgs branch operators at the IR fixed point.  In addition, \eqref{ZPreviewHiggs} can be used to calculate $n$-point correlators of certain {\em twisted} Higgs branch operators of the SCFT\@. These twisted operators are specific position-dependent linear combinations of Higgs branch operators, to be defined precisely in Section~\ref{COHOMOLOGY1}, obtained by contracting the various R-symmetry components of Higgs branch operators with position-dependent polarization vectors.

	Let us describe \eqref{ZPreviewHiggs}. $|\cW|$ is the order of the Weyl group of $G$. The variable $\sigma$ is the matrix degree of freedom and takes values in the Cartan of the Lie algebra $\mathfrak{g} = \text{Lie}(G)$.\footnote{Very roughly, the theory \eqref{ZPreviewHiggs} can be interpreted as a 1d gauged quantum mechanics with gauge group $G$, in the gauge where $A_\vphi = \sigma$.  The determinant factor in \eqref{ZPreviewHiggs} is precisely the Fadeev-Popov determinant corresponding to the gauge fixing condition $\partial_\vphi A_\vphi = 0$.}  It is coupled to a 1d Gaussian theory defined on a circle parameterized by $\vphi \in [-\pi, \pi)$ whose degrees of freedom are the anti-periodic scalar fields $Q(\vphi)$ and $\tQ(\vphi)$ which transform in the representations ${\cal R}$ and $\overline{\cal R}$ of $\mathfrak{g}$, respectively.   The path integral over $Q$ and $\tQ$ is over a middle-dimensional integration cycle in the space of complex-valued fields $Q$ and $\tQ$.  We will discuss this integration cycle in more detail in Section~\ref{LOCALIZATION}.  Lastly, $\ell$ is a parameter with dimensions of length whose meaning we will explain momentarily.

	The twisted Higgs branch operators of the 3d SCFT whose correlators can be calculated using \eqref{ZPreviewHiggs}, can be inserted anywhere along a line in $\R^3$ or, equivalently, along a great circle on $S^3$ that maps to this line under the stereographic projection.  In \eqref{ZPreviewHiggs}, the angle $\vphi$ parameterizes the great circle on $S^3$, and $\ell \equiv -4 \pi r$ is proportional to the radius of $S^3$.  Moreover, the twisted Higgs branch operators are represented in the 1d model \eqref{ZPreviewHiggs} by gauge-invariant polynomials in $Q(\vphi)$ and $\tQ(\vphi)$.  From the 2- and 3-point functions of the twisted Higgs branch operators, one can extract in a simple way the 2- and 3-point functions of the most general Higgs branch operators. 
		
	The 1d sector consisting of the twisted Higgs branch operators of a general 3d $\cN=4$ SCFT was previously studied in \cite{Beem:2013sza,Chester:2014mea,Beem:2016cbd} abstractly, using only properties of the superconformal algebra.\footnote{At this abstract level, there is no difference between the 1d sector associated with the Higgs branch and that associated with the Coulomb branch.}  These properties imply that the correlation functions of the twisted Higgs branch operators are topological, in the sense that they do not depend on the relative separation between the insertion points, but do depend on the ordering of the insertions.  Moreover, the 1d OPE algebra of this sector is an associative non-commutative algebra obeying certain very special properties. In some cases, Refs.~\cite{Chester:2014mea,Beem:2016cbd} used bootstrap-type arguments to show that these properties determine the 1d OPE algebra uniquely up to a finite number of parameters.  The model \eqref{ZPreviewHiggs} provides a complementary approach to the analysis in~\cite{Chester:2014mea,Beem:2016cbd} whereby \eqref{ZPreviewHiggs} can be used to calculate explicitly the structure constants of the 1d OPE algebra.

	\item We provide a partial result toward a similar computation of correlation functions of Coulomb branch operators.  In particular, we only consider Coulomb branch operators that are not monopole operators. Such non-monopole operators are given by gauge-invariant polynomials in the vectormultiplet scalars. At the SCFT fixed point, the 2- and 3-point correlators of non-monopole Coulomb branch operators, as well as $n$-point functions of their twisted analogs, can be calculated by inserting gauge-invariant polynomials in $\sigma$ into the matrix model 
	\es{ZPreviewCoulomb}{
		Z_\text{Coulomb} = \frac1{|\cW|}\int_\text{Cartan} d \sigma \, \frac{\det{}'_\text{adj} \left[ 2 \sinh ( \pi \sigma) \right]}{\det_{\cal R} \left[ 2 \cosh ( \pi \sigma) \right]} \,.
	}
	The same matrix model was previously obtained by Kapustin, Willet, and Yaakov \cite{Kapustin:2009kz} as a result of a supersymmetric localization computation that uses only ${\cal N} = 2$ supersymmetry, and with the goal of calculating expectation values of BPS Wilson loop operators.   Its relation to the Higgs branch theory \eqref{ZPreviewHiggs} is that one obtains \eqref{ZPreviewCoulomb} after integrating out $Q$ and $\tQ$ in \eqref{ZPreviewHiggs}.
		
	As was the case with Higgs branch operators, the twisted Coulomb branch operators whose correlation functions can be computed from \eqref{ZPreviewCoulomb} are part of the 1d topological Coulomb branch sector studied abstractly in \cite{Beem:2013sza,Chester:2014mea,Beem:2016cbd}.  In terms of the fields of the 3d SCFT, the twisted Coulomb branch operators represented by gauge-invariant polynomials in $\sigma$ correspond to position-dependent linear combinations of polynomials in the vectormultiplet scalars.  A more complete analysis that includes monopole operators is left for future work.

	\item The above results can be generalized to non-conformal ${\cal N} = 4$ QFTs on $S^3$ that are obtained by introducing real mass and Fayet-Iliopolous (FI) parameters.  The real mass parameters are introduced in \eqref{ZPreviewHiggs}--\eqref{ZPreviewCoulomb} by shifting $\sigma \to \sigma + m r$ in the exponent of \eqref{ZPreviewHiggs} and denominator of \eqref{ZPreviewCoulomb}, where $m$ is a real mass matrix taking value in the Cartan of the flavor symmetry algebra of the hypermultiplet.\footnote{\label{FlavorFootnote}As in Footnote~\ref{HalfHyperFootnote}, let $N_H$ be the total number of hypermultiplets in the absence of gauging.  The full flavor symmetry group $\widehat G_F$ of the $N_H$ hypermultiplets is defined as the normalizer of the gauge group inside $USp(2 N_H)$ modulo the gauge group \cite{Bullimore:2015lsa}.  However, we take our gauge group $G$ to be contained in a $U(N_H)$ subgroup of $USp(2N_H)$, and when defining the flavor symmetry we also consider $G_F = \widehat G_F \cap U(N_H)$.  We sometimes refer to $G_F$ as the flavor symmetry of the hypermultiplet.  The embedding of $G_F$ into $U(N_H)$ induces a map ${\cal F}: \mathfrak{g}_F \to \mathfrak{u}( N_H)$, where $\mathfrak{g}_F = \text{Lie}(G_F)$.  
	
	By $\sigma + m r$ in the main text we then mean ${\cal R}(\sigma) + r {\cal F}(m)$, where ${\cal R} : \mathfrak{g} \to \mathfrak{u}(N_H)$ is the representation map from the gauge algebra into $N_H \times N_H$ hermitian matrices.  
	}   For each abelian factor of the gauge group, one can introduce an FI parameter $\zeta_a$ by including in \eqref{ZPreviewHiggs}--\eqref{ZPreviewCoulomb} an additional factor 
	\es{FITermFactor}{
		e^{-8\pi^2i r \tr_\zeta \sigma} \,.
	}
	In \eqref{FITermFactor}, $\tr_\zeta\sigma \equiv \sum_{a} \zeta_a \sigma_a$ with the sum taken over all abelian factors in $\mathfrak{g}$, while $\zeta_a$ are the corresponding FI parameters and $\sigma_a$ are the components of $\sigma$ that take values in those abelian factors.
		
	The correlators of twisted Higgs (Coulomb) branch operators that are not charged under the flavor (topological) symmetries associated with non-zero real mass (FI) parameters are still topological. We will show that when a real mass parameter triggers an RG flow between two SCFTs, the correlation functions in the 1d theory interpolate between topological correlators in the UV and the IR, even though in the intermediate regime these correlation functions may be position-dependent.

\end{itemize}
	
Let us explain in more detail the procedure by which one arrives at the aforementioned results.  While our results hold in SCFTs defined on any conformally flat space, we find it convenient to first consider more general non-conformal 3d ${\cal N} = 4$ QFTs on a round $S^3$ of radius $r$.  Due to the large amount of supersymmetry, placing the $\cN=4$ theories on $S^3$ is unambiguous.  The $S^3$ Lagrangians we consider are curved space generalizations of the usual flat space ones:  they contain kinetic terms for the hypermultiplets, a Yang-Mills term for the vectormultiplet with Yang-Mills coupling $g_\text{YM}$, and, optionally, real mass and FI parameters, all containing certain curvature couplings that vanish in the limit $r \to \infty$. Setting to zero the real mass and FI parameters and taking both $g_\text{YM}, r \to \infty$, the correlation functions computed in an ${\cal N} = 4$ theory on $S^3$ approach those of the deep infrared limit of the same ${\cal N} = 4$ theory defined on $\R^3$. In the examples we consider, such a deep infrared limit is a non-trivial interacting SCFT\@.  Alternatively, we can first take $g_\text{YM}\to\infty$ at fixed $r$ and then conformally map from $S^3$ to $\R^3$.  As we will see, the $S^3$ correlators we study are independent of $g_\text{YM}$, so the limit $g_\text{YM} \to \infty$ is taken trivially.

After placing the theories of interest on $S^3$, we perform supersymmetric localization with an appropriately chosen supercharge.  The choice of supercharge is guided by the cohomological construction of \cite{Beem:2013sza}, which was elaborated upon in the context of 3d $\cN=4$ SCFTs on $\bR^3$ in \cite{Chester:2014mea,Beem:2016cbd}.  In particular, the authors of \cite{Chester:2014mea,Beem:2016cbd} identified a supercharge $\cQ^H$ in the $\cN=4$ superconformal algebra whose cohomology classes are represented by the twisted Higgs branch operators mentioned in the first bullet point above.   A similar construction for Coulomb branch operators involves the cohomology of a different supercharge $\cQ^C$.  
	
We perform supersymmetric localization using precisely the supercharge $\cQ^H$ or $\cQ^C$, mapped to $S^3$ using the stereographic map.   At first, this statement may seem puzzling for the following reason.  In performing supersymmetric localization, one adds to the action a $\cQ^{H, C}$-exact localizing term.  The standard localizing term for the vectormultiplet is usually constructed by acting with two supercharges on fermion bilinears, and it thus has scaling dimension $4$ and breaks conformal invariance.  However, both supercharges $\cQ^H$ and $\cQ^C$ were constructed in \cite{Chester:2014mea,Beem:2016cbd} as specific linear combinations of Poincar\'e and conformal supercharges on $\bR^3$, so conformal symmetry seemed important. Consequently it seems confusing why the standard vectormultiplet localizing term could even be invariant under $\cQ^{H, C}$, let alone $\cQ^{H, C}$-exact, as required by the supersymmetric localization technique.  We will show, however, that the supercharges $\cQ^{H,C}$ belong to an ${\cal N} = 4$ supersymmetry algebra  $\mathfrak{su}(2|1)_\ell \oplus \mathfrak{su}(2|1)_r$, which, upon mapping to $S^3$, can be seen to contain only the isometries of $S^3$ and a $U(1)^2$ R-symmetry, without any conformal generators.\footnote{We stress that upon contraction $r\to\infty$ the supercharges $\cQ^{H,C}\in\mathfrak{su}(2|1)_\ell \oplus \mathfrak{su}(2|1)_r$ we define on $S^3$ reduce to ordinary Poincar\'e supercharges on $\bR^3$, and \textbf{not} to the supercharges constructed in \cite{Chester:2014mea,Beem:2016cbd}. The latter were also denoted by $\cQ^{H,C}$ above, in a slight abuse of notation.}  Thus, $\cQ^{H,C}$-invariant theories on $S^3$ are not necessarily conformal invariant;  they include the more general non-conformal QFTs on $S^3$ mentioned above. 
	
It is worth commenting on the relation between the localization computation using $\cQ^{H,C}$ and that preformed by Kapustin, Willett and Yaakov (KWY) in \cite{Kapustin:2009kz} for $\cN \geq 3$ theories that yielded the matrix model \eqref{ZPreviewCoulomb}.  This computation was later generalized to ${\cal N} = 2$ theories in \cite{Jafferis:2010un,Hama:2010av}.  The supercharge $\cQ^{\text{KWY}}$ used for localization in \cite{Kapustin:2009kz} thus also resides in an $\cN=2$ sub-algebra, namely $\mathfrak{su}(2|1)\oplus\mathfrak{su}(2)$, of the full $\cN=4$ algebra $\mathfrak{su}(2|1)_\ell \oplus \mathfrak{su}(2|1)_r$. The supercharges $\cQ^H$ and $\cQ^C$ do not reside in any such $\cN=2$ sub-algebra, but are instead different linear combinations of supercharges in the two $\mathfrak{su}(2|1)$ factors of the ${\cal N} = 4$ superalgebra.  In spite of these differences, we find that the results of localizing with $\cQ^C$ or $\cQ^H$ are very much related to the KWY matrix model, as was briefly described above.

Concretely, our calculation proceeds as follows. Just as in \cite{Hama:2010av}, we find that the Yang-Mills action is ${\cal Q}$-exact (with respect to $\cQ^H$ or $\cQ^C$) and can be added with a large coefficient, thus localizing the ${\cal N} = 4$ vectormultiplet in precisely the same way as in \cite{Kapustin:2009kz,Hama:2010av}. In other words, the localization of the vectormultiplet is realized by taking the theory to small gauge coupling.  At any point on the vectormultiplet localization locus, the hypermultiplet is thus free, but massive, with its mass matrix depending on the precise location on the localization locus.  In this weakly coupled theory, the correlation functions of the hypermultiplet can be computed by first using Wick's theorem at a fixed point on the vectormultiplet localization locus, and then integrating over it with a measure given by the determinant of fluctuations of all the fields.  If we focus our attention on ${\cal Q}^{H,C}$-closed operators, which as we show are twisted Higgs (or Coulomb) branch operators inserted along a great circle of $S^3$, a standard argument shows that these correlation functions are independent of the Yang-Mills coupling.  
	
We conclude that once the vectormultiplet has been localized, calculating correlators of twisted Higgs branch operators does not require a further localization of the hypermultiplet. Indeed, the hypermultiplet action in the background of the localized vectormultiplet is Gaussian, and thus the remaining path integral is trivially solvable. For localization with $\cQ^H$, however, it is instructive to also localize the hypermultiplet, which leads to the explicit description \eqref{ZPreviewHiggs} of the correlators in terms of the 1d Gaussian theory coupled to the matrix model obtained from localizing the vectormultiplet.

Localization of the hypermultiplet with $\cQ^H$ has several features that are worth mentioning.  Unlike the ${\cal N} = 4$ vectormultiplet, the supersymmetry algebra does not close off-shell on the hypermultiplet.  Since the supersymmetric localization arguments require a supercharge that does close off-shell, the first step is to add a number of auxiliary fields and modify the supersymmetry transformation rules such that the algebra generated by $\cQ^H$ does close off-shell on the hypermultiplet fields.\footnote{A similar construction in the case of ${\cal N} = 4$ supersymmetric Yang-Mills theory in 4d was used in \cite{Pestun:2007rz,Pestun:2009nn} following the work of \cite{Berkovits:1993hx}.}  The next step is to add to the action a ${\cal Q}^H$-exact term whose bosonic part is positive-definite.  To describe the localization locus, let us think of $S^3$ as a circle fibered over a disk with the circle shrinking at the boundary of the disk.  We find that the hypermultiplet localizes on field configurations that are independent of the coordinate parameterizing the circle and that obey additional differential constraints in the disk directions.  These constraints imply that the hypermultiplet action, when evaluated on the localization locus, becomes a total derivative on the disk and reduces to a boundary term.  This boundary term, living on the boundary of the disk, is the Gaussian action for our localized theory.  We will argue that the one-loop determinant of fluctuations around this configuration equals $1$, so there is no additional determinant factor coming from the hypermultiplet.  The situation presented here for the localization of the hypermultiplet is very similar to the one encountered in \cite{Pestun:2009nn} in the case of 4d ${\cal N} = 4$ supersymmetric Yang-Mills theory on $S^4$.
	
We apply \eqref{ZPreviewHiggs}--\eqref{ZPreviewCoulomb} in a few examples where we calculate explicitly several correlation functions of twisted Higgs and Coulomb branch operators, with the main focus on the Higgs branch.  As in \cite{Beem:2016cbd}, we interpret the 1d OPE algebra as a non-commutative star product on the Higgs branch chiral ring,\footnote{The twisted Higgs branch operators are in 1-to-1 correspondence with chiral Higgs branch operators.  While a generic Higgs branch operator corresponds to a function on the Higgs branch, a chiral Higgs branch operator corresponds to a holomorphic function on the Higgs branch, for a given choice of complex structure.  These operators form the Higgs branch chiral ring.  Similar statements hold for Coulomb branch operators.} and we compute explicitly the values of the parameters that the bootstrap analysis of \cite{Beem:2016cbd} left undetermined.  As we will discuss, in calculating correlation functions using \eqref{ZPreviewHiggs}--\eqref{ZPreviewCoulomb}, one should be aware of the possibility of operator mixing on $S^3$.  The mixing can be removed by diagonalizing the matrix of 2-point functions using the Gram-Schmidt procedure.  A similar approach was taken in \cite{Gerchkovitz:2016gxx} for the Coulomb branch operators of 4d ${\cal N} = 2$ theories.
	
The rest of this paper is organized as follows.  In Section~\ref{THEORIES} we introduce the ${\cal N} = 4$ QFTs on $S^3$ we will study.  In Section~\ref{COHOMOLOGY1} we review the cohomological construction of \cite{Beem:2013sza,Chester:2014mea,Beem:2016cbd} in the case of ${\cal N} = 4$ SCFTs in flat space and explain how it is mapped stereographically to $S^3$.  In Section~\ref{COHOMOLOGY2} we generalize this construction to QFTs on $S^3$ that do not necessarily possess conformal symmetry.  Section~\ref{LOCALIZATION} contains a description of the localization computation that leads to the results \eqref{ZPreviewHiggs}--\eqref{ZPreviewCoulomb} summarized above.  In Section~\ref{PROPERTIES} we describe in general terms the various properties of the 1d theory \eqref{ZPreviewHiggs} and its applications.  Sections~\ref{CFTAPPLICATIONS} and~\ref{NONCONFORMALAPPLICATIONS} contain applications of our results to specific theories.  We end with a brief discussion in Section~\ref{DISCUSSION}.

\section{3d $\cN=4$ Theories on $S^3$}
\label{THEORIES}
	
In this section we will review the construction of $\cN=4$ supersymmetric Lagrangians using vectormultiplets and hypermultiplets on $S^3$.  We first provide the supersymmetry transformation rules and the supersymmetric actions.  We then discuss the supersymmetry algebras preserved by these actions.

\subsection{Actions with vectormultiplets and hypermultiplets}

The components of the vectormultiplets and hypermultiplets carry Lorentz indices as well as $\mathfrak{su}(2)_C\oplus\mathfrak{su}(2)_H$ R-symmetry indices.  Explicitly, the components of the vectormultiplet $\cV$ transform in the adjoint representation of the gauge group $G$ and will be denoted by
\begin{align}
\cV = (A_{\mu}, \lambda_{\alpha a\dot a}, \Phi_{\dot{a}\dot b}, D_{ab})\ed \label{Vmul}
\end{align}
The vector $A_{\mu}$ in \eqref{Vmul} is the gauge field, the spinor $\lambda_{\alpha a\dot{a}}$ is the gaugino, and $\Phi_{\dot{a}\dot b }$ and $D_{ab}$ are scalars, transforming in the $(\mathbf{1},\mathbf{1})$, $(\mathbf{2},\mathbf{2})$, $(\mathbf{3},\mathbf{1})$ and $(\mathbf{1},\mathbf{3})$ irreps of $\mathfrak{su}(2)_C\oplus\mathfrak{su}(2)_H$, respectively.\footnote{We label the doublet irrep of $\mathfrak{su}(2)_{\text{rot.}}$ frame rotations by indices $\alpha,\beta,\ldots = 1,2$, of $\mathfrak{su}(2)_H$ by $a,b,\ldots=1,2$, and of $\mathfrak{su}(2)_C$ by $\dot{a},\dot{b},\ldots=1,2$. See Appendix \ref{conventions} for more details on our conventions.} The hypermultiplet $\cH$ transforms in some unitary representation $\cR$ of $G$ and has components
\begin{align}
\cH = (q_a, \tq^a, \psi_{\alpha\dot a}, \tpsi_{\alpha\dot a}) \ec \label{Hmul}
\end{align}
where $q_a$, $\tq^a$ are complex scalars transforming in $(\mathbf{1},\mathbf{2})$ and $(\mathbf{1},\mathbf{\overline 2})$ irreps of the R-symmetry and representations $\cR$ and $\overline\cR$ of $G$, and $\psi_{\alpha\dot a}$, $\tpsi_{\alpha,\dot a}$ are their spinor superpartners, which transform, respectively, in the $(\mathbf{2},\mathbf{1})$ and $(\mathbf{\overline 2},\mathbf{1})$ irreps of the R-symmetry, and in the $\cR$ and $\overline\cR$ representations of $G$. The multiplets $\cV$ and $\cH$ also have ``twisted'' versions, in which the roles of $\mathfrak{su}(2)_C$ and $\mathfrak{su}(2)_H$ are interchanged, though we will not consider them in this paper.
	
On $S^3$, superconformal transformations are generated by spinors $\xi_{a\dot a}$ in the $(\mathbf{2},\mathbf{2})$ irrep  of the R-symmetry, which satisfy the conformal Killing spinor equations
\begin{align}
\nabla_{\mu}\xi_{a\dot a} = \gamma_{\mu}\xi'_{a \dot a} \ecq \nabla_{\mu}\xi'_{a\dot a} = -\frac{1}{4r^2}\gamma_{\mu}\xi_{a\dot a} \ec \label{CKSeq}
\end{align}
where $\gamma_{\mu}$ ($\mu=1,2,3$) are curved space Dirac matrices and $r$ is the radius of $S^3$. In particular, the transformation rules for the vectormultiplet fields \eqref{Vmul} are\footnote{The field strength is $F_{\mu\nu}\equiv \partial_{\mu}A_{\nu}-\partial_{\nu}A_{\mu}-[A_{\mu},A_{\nu}]$, and $\cD_{\mu}=\nabla_{\mu}-iA_{\mu}$ is the space and gauge covariant derivative. Brackets $()$ enclosing indices denote an average over their permutations.}
\begin{align}
\delta_{\xi} A_{\mu} &= \frac{i}{2} \xi^{a\dot{b}}\gamma_{\mu}\lambda_{a\dot{b}} \ec \label{Avar}\\
\delta_{\xi} \lambda_{a\dot{b}} &= - \frac{i}{2}\varepsilon^{\mu\nu\rho}\gamma_{\rho}\xi_{a\dot{b}}F_{\mu\nu} - D_a{}^c\xi_{c\dot{b}} -i\gamma^{\mu}\xi_a{}^{\dot{c}} \cD_{\mu}\Phi_{\dot{c}\dot{b}} + 2i \Phi_{\dot{b}}{}^{\dot{c}}\xi_{a\dot{c}}' \notag\\
&+ \frac{i}{2}\xi_{a\dot{d}} [ \Phi_{\dot{b}}{}^{\dot{c}}, \Phi_{\dot{c}}{}^{\dot{d}}] \ec \label{lamvar}\\
\delta_{\xi}\Phi_{\dot{a}\dot{b}} &= \xi^c{}_{(\dot{a}}\lambda_{|c|\dot{b})} \ec \label{phivar}\\
\delta_{\xi} D_{ab} &= -i\cD_{\mu}(\xi_{(a}{}^{\dot c}\gamma^{\mu}\lambda_{b)\dot c}) - 2i\xi'_{(a}{}^{\dot c}\lambda_{b)\dot c} + i [\xi_{(a}{}^{\dot{c}}\lambda_{b)}{}^{\dot{d}}, \Phi_{\dot{c}\dot{d}}] \ec\label{dvar}
\end{align}
and those of the hypermultiplet \eqref{Hmul} are
\begin{align}
\delta_{\xi} q^a = \xi^{a\dot{b}} \psi_{\dot{b}}\ecq \delta_{\xi} \psi_{\dot{a}} = i\gamma^{\mu}\xi_{a\dot{a}} \cD_{\mu}q^a + i\xi'_{a\dot{a}}q^a - i\xi_{a\dot{c}}\Phi^{\dot{c}}{}_{\dot{a}}q^a \ec \label{qpsivar}\\
\delta_{\xi} \tq^{a} = \xi^{a\dot{b}} \tpsi_{\dot{b}} \ecq \delta_{\xi} \tpsi_{\dot{a}} = i\gamma^{\mu}\xi_{a\dot{a}} \cD_{\mu}\tq^a + i\tq^a\xi'_{a\dot{a}} + i\xi_{a\dot{c}}\tq^a\Phi^{\dot{c}}{}_{\dot{a}} \ed\label{qtpsitvar}
\end{align}
One can check that acting twice with the transformation rules presented above realizes the superconformal algebra $\mathfrak{osp}(4|4)$ up to gauge transformations and fermionic equations of motion.  We will return to this point with more details shortly.
	
With the supersymmetry transformation rules in hand, one can construct supersymmetric actions.  The action of a hypermultiplet coupled to a vectormultiplet is 
\begin{align}
S_{\text{hyper}}[\cH,\cV] &= \int d^3x \sqrt{g} \left[ D^{\mu}\tq^{a} D_{\mu} q_a - i\tpsi^{\dot{a}}\slashed{D}\psi_{\dot{a}} + \frac{3}{4r^2} \tq^{a} q_a + i \tq^{a} D_a{}^b q_b - \frac{1}{2}\tq^a \Phi^{\dot{a}\dot{b}}\Phi_{\dot{a}\dot{b}}q_a\right.  \notag\\
&\left.-i\tpsi^{\dot{a}}\Phi_{\dot{a}}{}^{\dot{b}}\psi_{\dot{b}} +i\left( \tq^ a\lambda_a{}^{\dot{b}}\psi_{\dot{b}} + \tpsi^{\dot{a}}\lambda^b{}_{\dot{a}}q_b\right)\right] \ed \label{Shyper}
\end{align}
This action is invariant under the full $\mathfrak{osp}(4|4)$ algebra.   Indeed, one can check that it is invariant under the transformations \eqref{Avar}--\eqref{qtpsitvar} provided that \eqref{CKSeq} is obeyed.  This action could have been deduced from the analogous flat space action by simply covariantizing all derivatives and introducing the conformal mass term $ \frac{3}{4r^2} \tq^{a} q_a$ for the hypermultiplet scalars.  

Multiplets $\cV$ and $\cH$ as defined above have too many bosonic components and, in the path integral, have to be integrated over the middle-dimensional cycle determined by the following reality conditions on bosons:
\begin{align}
\label{RealqAPhD}
\tq^a &= (q_a)^*\, ,\cr
(A^I_\mu T^I)^*&=A_\mu^I (T^I)^*\, ,\cr
(\Phi_{\dot a\dot b}^I T^I)^* &= -\Phi^{I\dot a\dot b}(T^I)^*\, ,\cr
(D_{ab}^I T^I)^* &= -D^{Iab}(T^I)^*\, ,
\end{align}
where, for the vectormultiplet fields, we made the representation matrices $T^I$ by which they act on $\cR$ explicit. In Lorentzian signature, the fermions would also obey reality constraints, namely  $\tpsi_{\dot{a}}$ would be the hermitian conjugate of $\psi_{\dot a}$ and $\lambda^{\alpha a\dot b}$ would be the hermitian conjugate of $\lambda_{\alpha a\dot b}$, but in the Euclidean signature, $\psi_{\alpha\dot a}$ and $\tpsi_{\alpha\dot a}$ are independent spinors in the representations $\cR$ and $\overline\cR$ of $G$ respectively, and $\lambda_{\alpha a\dot b}$ do not obey any constraints either.

As far as we know, it is not possible to write down other superconformal actions on $S^3$ with just vectormultiplets and hypermultiplets.\footnote{A Chern-Simons action for the vectormultiplet would be conformal, but preserving ${\cal N} = 4$ supersymmetry would require the presence of twisted hypermultiplets which we do not consider here.}  It is possible, however, to write down actions that are invariant under half the supersymmetries in $\mathfrak{osp}(4|4)$, which anti-commute to the isometries of $S^3$ without any conformal transformations. Such actions provide curved space analogs of the Yang-Mills action or of the actions corresponding to real masses and FI terms, all of which are not conformally invariant on $\bR^3$, and therefore cannot be mapped to $S^3$ using the stereographic map.  The projection from 16 supersymmetries in $\mathfrak{osp}(4|4)$ to the 8 under which these actions on $S^3$ are invariant is described in terms of two $\mathfrak{su}(2)$ matrices, $h_a{}^b$ and $\bar h^{\dot a}{}_{\dot b}$,
\es{hhbDef}{
	h_a{}^b \in \mathfrak{su}(2)_H \ecq \qquad \bar{h}^{\dot a}{}_{\dot b}\in \mathfrak{su}(2)_C \ec
}
normalized such that $h_a{}^ch_c{}^b=\delta_a{}^b$ and $\bar{h}^{\dot a}{}_{\dot c}\bar{h}^{\dot c}{}_{\dot b} = \delta^{\dot a}{}_{\dot b}$, and obeying the tracelessness condition  $h_a{}^a = \bar{h}^{\dot a}{}_{\dot a} = 0$.  We interpret these matrices as representing the Cartan elements of $\mathfrak{su}(2)_H \oplus \mathfrak{su}(2)_C$.  Then one can restrict the $S^3$ Killing spinors \eqref{CKSeq} to those obeying the further condition
\es{massiveSpinors}{
	\xi_{a\dot a}' = \frac{i}{2r} h_a{}^b\xi_{b\dot b}\bar{h}^{\dot b}{}_{\dot a} \,.
}
This condition reduces the number of independent $\xi_{a \dot a}$ by a factor of two.  We will interpret it shortly in terms of generating a subalgebra of $\mathfrak{osp}(4|4)$, but let us first present the actions on $S^3$ that are invariant under the 8 supersymmetry transformations restricted in this fashion.
	
The non-conformal supersymmetric actions depend explicitly on the matrices \eqref{hhbDef}.   The Yang-Mills action is given by
\begin{align}
S_{\text{YM}}[\cV] &= \frac{1}{g^2_\text{YM}}\int d^3x\sqrt{g}\trace\left(F^{\mu\nu}F_{\mu\nu}  - \cD^{\mu}\Phi^{\dot{c}\dot d}\cD_{\mu}\Phi_{\dot{c}\dot d} +i\lambda^{a\dot a}\slashed{\cD}\lambda_{a\dot a} - D^{cd}D_{cd}  -i \lambda^{a\dot a}[\lambda_a{}^{\dot b},\Phi_{\dot{a}\dot b}]\right.\notag\\
&\left.- \frac{1}{4}[\Phi^{\dot a}{}_{\dot b}, \Phi^{\dot c}{}_{\dot d}][\Phi^{\dot b}{}_{\dot a}, \Phi^{\dot d}{}_{\dot c}]- \frac{1}{2r}h^{ab}\bar{h}^{\dot{a}\dot b}\lambda_{a\dot a}\lambda_{b\dot b} + \frac{1}{r}(h_a{}^bD_b{}^a)(\bar{h}^{\dot a}{}_{\dot b}\Phi^{\dot b}{}_{\dot a}) -\frac{1}{r^2}\Phi^{\dot{c}\dot{d}}\Phi_{\dot{c}\dot d}\right) \ed \label{SYM}
\end{align}
Under a decomposition of $\cN=4$ into an $\cN=2$ sub-algebra, one can show that \eqref{SYM} is nothing but the $S^3$ action of an $\cN=2$ vectormultiplet plus an adjoint chiral of R-charge $1$. For each $U(1)$ factor in $G$ we can introduce an FI-term: 
\es{FIAction}{
	S_{\text{FI}}[\cV] &= i\zeta\int d^3x\sqrt{g}\left(h_a{}^b D_b{}^a - \frac{1}{r} \bar{h}^{\dot a}{}_{\dot b}\Phi^{\dot b}{}_{\dot a} \right) \ed
}
Note that while on $\mathbb{R}^3$ the FI parameters $\zeta_{ab}$ take value in the $(\mathbf{1},\mathbf{3})$ irrep of $\mathfrak{su}(2)_C\oplus\mathfrak{su}(2)_H$, only the single component $\zeta = h^{ab}\zeta_{ab}$ invariant under the Cartan of $\mathfrak{su}(2)_H$ survives on $S^3$.  Finally, one can introduce mass terms for the hypermultiplets by coupling them to background vectormultiplets $\cV_{\text{b.g.}}$ in the Cartan of the flavor symmetry. In order to preserve supersymmetry all the components of $\cV_{\text{b.g.}}$ are set to zero except for
\es{PhiDBack}{
	\frac{1}{2}\bar{h}^{\dot a}{}_{\dot b}(\Phi_{\text{b.g.}})^{\dot b}{}_{\dot a} = - \frac{r}{2}h_a{}^b(D_{\text{b.g.}})_b{}^a \ec
}
and the supersymmetry variations \eqref{qpsivar} and \eqref{qtpsitvar} have to be deformed accordingly to account for the masses. As happened with the FI terms, out of the $\mathfrak{su}(2)_C$-triplet of mass parameters that exist on $\mathbb{R}^3$ for each flavor group Cartan element, only one survives on $S^3$.
	
In the remainder of this paper, in order to conform with the conventions of \cite{Chester:2014mea} we will sometimes choose
\begin{align}
h_a{}^b = -\sigma^2 \ecq \bar{h}^{\dot a}{}_{\dot b} = - \sigma^3 \ed \label{cartanchoice}
\end{align}
	
\subsection{Closure of the supersymmetry transformations}
\label{CLOSURE}

Irrespective of the actions presented above, the superconformal transformations \eqref{Avar}--\eqref{dvar} of $\cV$ close off-shell into
\begin{align}
\{\delta_{\xi}, \delta_{\tilde{\xi}}\} \cV = \left(\hat{\cK}_{\xi,\tilde{\xi}} + \hat{\cG}_{\Lambda}\right)\cdot\cV \ec  \label{closure}
\end{align}
where $\hat{\cG}_{\Lambda}$ is a gauge transformation with parameter $\Lambda$ defined as
\es{GotLambda}{
	\Lambda = (\tilde{\xi}^c{}_{\dot a}\xi_{c\dot b})\Phi^{\dot{a}\dot b} - i(\tilde{\xi}^{a\dot a}\gamma^{\mu}\xi_{a\dot a})A_{\mu} \ec
}
while $\hat{\cK}_{\xi,\tilde{\xi}}$ generates bosonic symmetries in $\mathfrak{osp}(4|4)$ and is written explicitly in terms of $\xi$ and $\tilde{\xi}$ in Appendix \ref{closuredetails}. The transformations of the scalars $q_a$ in $\cH$ also close off-shell as in \eqref{closure}, but those of the fermions $\psi_{\dot a}$ do not. Instead, one finds 
\begin{align}
\{\delta_{\xi}, \delta_{\tilde{\xi}}\} \psi_{\dot a} &= \left(\hat{\cK}_{\xi,\tilde{\xi}}+\hat{\cG}_{\Lambda}\right)\cdot\psi_{\dot a} + \tilde{\xi}^{a\dot b}\left(\xi_{a\dot a}\Psi^{\text{e.o.m.}}_{\dot b}\right) + \xi^{a\dot b}\left(\tilde{\xi}_{a\dot a}\Psi^{\text{e.o.m.}}_{\dot b}\right) \ec \label{psiclosure}\\
\{\delta_{\xi}, \delta_{\tilde{\xi}}\} \tpsi_{\dot a} &= \left(\hat{\cK}_{\xi,\tilde{\xi}}+\hat{\cG}_{\Lambda}\right)\cdot\tpsi_{\dot a} - \tilde{\xi}^{a\dot b}\left(\xi_{a\dot a}\widetilde{\Psi}^{\text{e.o.m.}}_{\dot b}\right) - \xi^{a\dot b}\left(\tilde{\xi}_{a\dot a}\widetilde{\Psi}^{\text{e.o.m.}}_{\dot b}\right) \ec\label{tpsiclosure}
\end{align}
where the equations of motion operators are given by
\begin{align}
\Psi^{\text{eom}}_{\dot{a}} &\equiv -i \left[\slashed{D}\psi_{\dot{a}} + \Phi_{\dot{a}}{}^{\dot{b}}\psi_{\dot{b}} + \lambda_{a\dot{a}}q^a\right] \ec \label{eompsi} \\
\widetilde{\Psi}^{\text{eom}}_{\dot{a}} &\equiv i \left[\slashed{D}\tpsi_{\dot{a}} - \tpsi_{\dot{b}} \Phi^{\dot{b}}{}_{\dot{a}} - \tq^{a}\lambda_{a\dot{a}}\right] \ed \label{eompsit}
\end{align}
These are precisely the equations of motion following from the hypermultiplet action \eqref{Shyper}.  
That the supersymmetry algebra closes only up to the fermion equations of motion will be important in Section~\ref{OFFSHELLCL}, since the supercharge used for localization of the hypermultiplet has to be closed off-shell.
	
\subsection{Non-conformal supersymmetry algebra on $S^3$}
\label{NONCONFORMAL}

Let us now return to the projection condition \eqref{massiveSpinors} and interpret it from the point of view of which supersymmetry algebra it is that the actions \eqref{SYM}--\eqref{FIAction} as well as the mass terms introduced via \eqref{PhiDBack} are invariant under.  
	
Let us assume for now that we have not introduced any mass terms and that we set all possible FI parameters to zero.  The anti-commutator of two supersymmetries restricted by \eqref{massiveSpinors} does not produce all the bosonic generators of the superconformal algebra $\mathfrak{osp}(4|4)$, but only a subset of them.  This was to be expected, because we have argued that the Yang-Mills action \eqref{SYM} is invariant under supersymmetries obeying \eqref{massiveSpinors}, and since the Yang-Mills action is not conformal, it must be that the anti-commutator of supersymmetries \eqref{massiveSpinors} does not generate any conformal transformations.
	
Judiciously working out all possible (anti-)commutators, one can check that \eqref{massiveSpinors} parameterize the 8 supersymmetry transformations of the algebra 
\es{MassiveAlgebra}{
	\mathfrak{su}(2|1)_\ell \oplus \mathfrak{su}(2|1)_r \,.
}
(See also  \cite{Assel:2015oxa}.)  The bosonic generators of this algebra consist of the $\mathfrak{so}(4) = \mathfrak{su}(2)_\ell \oplus \mathfrak{su}(2)_r$ isometries of $S^3$ as well as two $\mathfrak{u}(1)$ R-symmetries that we will denote by $\mathfrak{u}(1)_\ell$ and $\mathfrak{u}(1)_r$, reflecting which $\mathfrak{su}(2|1)$ factor they belong to.  The $\mathfrak{u}(1)_\ell \oplus \mathfrak{u}(1)_r$  is a subalgebra of $\mathfrak{su}(2)_H \oplus \mathfrak{su}(2)_C$.  That \eqref{MassiveAlgebra} contains 8 supersymmetries means it is an ${\cal N} = 4$ supersymmetry algebra.

The algebra \eqref{MassiveAlgebra} will be central in our work, so let us describe it in more detail.    Let us denote the generators by $J^{(\ell)}_{\alpha\beta}$, $R_\ell$, and $\cQ_{\alpha}^{(\ell_{\pm})}$ for $\mathfrak{su}(2|1)_\ell$ and $J^{(r)}_{\alpha\beta}$, $R_r$, and $\cQ_{\alpha}^{(r_{\pm})}$ for $\mathfrak{su}(2|1)_r$.  Abstractly, the algebra obeyed by  $J^{(\ell)}_{\alpha\beta}$, $R_\ell$, and $\cQ_{\alpha}^{(\ell_{\pm})}$ is
\begin{alignat}{3}
[J^{(\ell)}_i, J^{(\ell)}_j]  &= i \epsilon_{ijk} J^{(\ell)}_k \ecq
&[J^{(\ell)}_{\alpha \beta}, \cQ^{(\ell_{\pm})}_{\gamma}] &= \frac{1}{2}\left(\varepsilon_{\alpha\gamma} \cQ^{(\ell_{\pm})}_{\beta} + \varepsilon_{\beta \gamma } \cQ^{(\ell_{\pm})}_{\alpha} \right) \ec\\
[R_{\ell}, \cQ_{\alpha}^{(\ell_{\pm})}] &= \pm \cQ_{\alpha}^{(\ell_{\pm})} \ecq &\{\cQ^{(\ell_+)}_{\alpha}, \cQ^{(\ell_-)}_{\beta}\} &= -\frac{4i}{r}\left(J_{\alpha\beta}^{(\ell)} + \frac{1}{2} \varepsilon_{\alpha\beta}R_\ell\right)\ec \label{Commutators}
\end{alignat}
where
\begin{align}
J_{\alpha \beta}^{(\ell)} \equiv \begin{pmatrix}
-(J_1^{(\ell)} + i J_2^{(\ell)}) & J_3^{(\ell)} \\
J_3^{(\ell)} & J_1^{(\ell)} - i J_2^{(\ell)}
\end{pmatrix} \ed
\end{align}
The generators of $\mathfrak{su}(2|1)_r$ obey the same relations with $\ell \to r$.  
	
To be more concrete, let us explain how these generators act on the various operators in the theory.  We will take this opportunity to set up some of the notation we will use later.
	
\subsubsection{Action of $S^3$ isometries}
\label{ISOMETRIES}
	
The commutators with $S^3$ isometries act on a gauge-invariant operator ${\cal O}$ as the Lie derivative
\es{JO}{
	[J^{(\ell)}_i, {\cal O}] = - {\cal L}_{v^\ell_i} {\cal O} \,, \qquad
	[J^{(r)}_i, {\cal O}] = - {\cal L}_{v^r_i} {\cal O}
}
with respect to the Killing vectors $v^\ell_i$ and $v^r_i$ that obey the $\mathfrak{su}(2)$ algebra, $[v^\ell_i, v^\ell_j] = i \varepsilon_{ijk} v^\ell_k$, and similarly for $v^r_i$.  
	
In an explicit description where the three-sphere of radius $r$ is embedded in $\R^4$ via
\es{S3Embedding}{
	X_1^2 + X_2^2 + X_3^2 + X_4^2 = r^2 \,,
}
we can use the parameterization
\es{Param}{
	X_1 + i X_2 = r \cos \theta e^{i \tau} \,, \qquad X_3 + i X_4 = r \sin \theta e^{i \varphi} 
}
in terms of the coordinates $\theta\in[0,\frac{\pi}{2}]$, $\tau, \varphi\in[-\pi,\pi]$.  In this parameterization, the Killing vectors in \eqref{JO} are
\es{Gotv}{
	v^\ell_1 &= \frac{i}{2}\left(-\cos(\tau+\varphi)\partial_{\theta} - \tan(\theta)\sin(\tau+\varphi)\partial_{\tau} + \cot(\theta)\sin(\tau+\varphi)\partial_{\varphi}\right) \ec \\
	v^\ell_2 &= \frac{i}{2}\left(\sin(\tau+\varphi)\partial_{\theta} - \tan(\theta)\cos(\tau+\varphi)\partial_{\tau} + \cot(\theta)\cos(\tau+\varphi)\partial_{\varphi}\right) \ec \\
	v^\ell_3 &= \frac{i}{2}\left(\partial_{\tau}+\partial_{\varphi}\right) \ec \\
	v^r_1 &= \frac{i}{2}\left(\cos(\tau-\varphi)\partial_{\theta} + \tan(\theta)\sin(\tau-\varphi)\partial_{\tau} + \cot(\theta)\sin(\tau-\varphi)\partial_{\varphi}\right) \ec \\
	v^r_2 &= \frac{i}{2}\left(-\sin(\tau-\varphi)\partial_{\theta} + \tan(\theta)\cos(\tau-\varphi)\partial_{\tau} + \cot(\theta)\cos(\tau-\varphi)\partial_{\varphi}\right) \ec  \\
	v^r_3 &= \frac{i}{2}\left(\partial_{\tau}-\partial_{\varphi}\right) \ed
}
We will make significant use of the parameterization \eqref{Param} in the remainder of this paper.   The metric in these coordinates is
\begin{equation}
ds^2(S^3) = r^2 (d\theta^2 + \cos^2(\theta) d\tau^2 + \sin^2(\theta) d\varphi^2).
\end{equation}
Coordinates $\theta$ and $\varphi$ parametrize a disk with the metric $ds^2(D^2)=r^2(d\theta^2 + \sin^2(\theta)d\varphi^2)$, where $\theta$ is the radial coordinate of the disk. The sphere metric then becomes:
\begin{equation}
ds^2(S^3) = ds^2(D^2) + w^2 d\tau^2,\quad w=r\cos\theta,
\end{equation}
which manifests $S^3$ as a $U(1)$-fibration over $D^2$, with the fibers being ``warped'' by $w$ and shrinking to zero size at the boundary of the disk.

\subsubsection{Action of R-symmetries}
	
The action of $R_\ell$ and $R_r$ on the fields of the previous section depends on the precise embedding of $\mathfrak{u}(1)_\ell$ and $\mathfrak{u}(1)_r$ into $\mathfrak{su}(2)_C\oplus\mathfrak{su}(2)_H$ given in terms of the matrices $h$ and $\bar h$ in \eqref{hhbDef} as follows.  Let us first define the operators 
\es{RHECDef}{
	R_H = \frac12 h_a{}^b R_b{}^a \,, \qquad R_C = \frac12 \bar h^{\dot a}{}_{\dot b} R^{\dot b}{}_{\dot a} \,,
}  
where $R_b{}^a$ and $R^{\dot b}{}_{\dot a}$ are the generators of $\mathfrak{su}(2)_H$ and $\mathfrak{su}(2)_C$ respectively. In our conventions, we then have
\es{Ulr}{
	R_\ell = R_H + R_C \,, \qquad R_r = R_H - R_C \,.
} 
This equation provides an identification of $R_\ell$ and $R_r$ with linear combination of the Cartan elements $R_H$ and $R_C$ of the R-symmetry of the superconformal algebra.  In terms of their action on fields, it is sufficient to describe how they act on $\mathfrak{su}(2)_H$ and $\mathfrak{su}(2)_C$ fundamental operators.  We have\footnote{Note that in our conventions $h_a{}^b = h^b{}_a$, and similarly for $\bar h$.}
\es{RHRCAction}{
	[R_H, {\cal O}_a] = \frac12 h_a{}^b {\cal O}_b \,, \qquad
	[R_C, {\cal O}_{\dot a}] = \frac12 \bar h^{\dot b}{}_{\dot a} {\cal O}_{\dot b} \,,
}
with a straightforward generalization to operators with multiple $\mathfrak{su}(2)_H \oplus \mathfrak{su}(2)_C$ indices.  For instance, $[R_H, {\cal O}_{ab}{}^c] = \frac12 h_a{}^d {\cal O}_{db}{}^c + \frac12 h_b{}^d {\cal O}_{ad}{}^c - \frac12 h_d{}^c {\cal O}_{ab}{}^d$.  The action of $R_\ell$ and $R_r$ on operators can then be inferred from simply combining \eqref{Ulr} and \eqref{RHRCAction}.
	
\subsubsection{Action of supersymmetries}
	
The action of the odd generators of $\mathfrak{su}(2|1)_l\oplus\mathfrak{su}(2|1)_r$ on operators in general multiplets can be quite complicated.  As mentioned above, on the vector and hypermultiplet operators their action is just a particular subset of the transformation rules \eqref{Avar}--\eqref{qtpsitvar}.   The precise correspondence between the various $\xi_{a \dot a}$ obeying \eqref{massiveSpinors} and the supercharges ${\cal Q}_\alpha^{(\ell_\pm)}$ and ${\cal Q}_\alpha^{(r_\pm)}$ is given in \eqref{delta2QS} and \eqref{Qlr}.

\subsection{Central extension of non-conformal supersymmetry algebra}
\label{EXTENSION}
	
The discussion in Section~\ref{NONCONFORMAL} was restricted to the case of vanishing mass and FI parameters.  Introducing these parameters amounts to central extensions of the algebra \eqref{MassiveAlgebra}, as we will now describe.

It is not hard to see, using Jacobi identity, that one cannot introduce central charges in (anti-)commutators between left and right algebras, so one can only separately centrally extend $\mathfrak{su}(2|1)_\ell$ and $\mathfrak{su}(2|1)_r$. Each of these algebras admits only one non-trivial central extension, so in total we have two central charges. We denote the centrally extended algebras with a tilde, so the supersymmetry algebra of our theories is $\widetilde{\mathfrak{su}(2|1)}_\ell\oplus\widetilde{\mathfrak{su}(2|1)}_r$. Denoting central charges of the left and right subalgebras by $Z_\ell$ and $Z_r$ respectively, the only place where they appear are the following anti-commutators:
\es{QQCentral}{
	\{\cQ^{(\ell_+)}_{\alpha}, \cQ^{(\ell_-)}_{\beta}\} &= -\frac{4i}{r}\left(J_{\alpha\beta}^{(\ell)} + \frac{1}{2} \varepsilon_{\alpha\beta}R_\ell + \varepsilon_{\alpha\beta}Z_\ell\right)\,, \\
	\{\cQ^{(r_+)}_{\alpha}, \cQ^{(r_-)}_{\beta}\} &= -\frac{4i}{r}\left(J_{\alpha\beta}^{(r)} + \frac{1}{2} \varepsilon_{\alpha\beta}R_r + \varepsilon_{\alpha\beta}Z_r\right)\,.
}
	
Physically, the central charges $Z_\ell$ and $Z_r$ correspond to turning on real masses and FI parameters. We turn on masses by coupling to background vectormultiplets in the Cartan of the flavor symmetry, as explained in Section 2. The only components of these background multiplets which are non-zero are $\bar{h}^{\dot a}_{\,\,\dot b}(\Phi_{\rm b.g.})^{\dot b}_{\,\, \dot a}=-(\Phi_{\rm b.g.})_{\dot1\dot2}$ and $h_a^{\,\, b}(D_{\rm b.g.})_b^{\,\, a}$, as explained in \eqref{PhiDBack}.
	
The supersymmetry algebra has a gauge transformation on the right, as written in Eq.~\eqref{closure}, with the gauge parameter $\Lambda$ of \eqref{GotLambda}. In gauge theories, dynamical gauge fields force us to consider only operators which are not charged under the corresponding gauge symmetry. For such operators, the gauge transformation in the SUSY algebra vanishes. For background gauge fields, this is not so. We can have operators which are charged under the corresponding global symmetry (which would be gauged if the gauge field were dynamical), and for them, such gauge transformations in the algebra will generate central charges. A simple computation, using the expression \eqref{GotLambda} for $\Lambda$, shows that:
\begin{equation}
\frac{1}{r}(Z_\ell + Z_r)=i(\Phi_{\rm b.g.})_{\dot1\dot2}=i\widehat{m}\,, \qquad \frac{1}{r}(Z_\ell - Z_r)=0 \,.
\end{equation}
Here $\widehat{m}={\rm diag}(m_I)$, where $m_I$ are real masses for hypers $q_a^I$, $I$ being the flavor index.
	
Analogously, FI parameters correspond to background twisted vectormultiplets in the Cartan of the gauge group. They similarly generate central charges with:
\begin{equation}
\frac{1}{r}(Z_\ell + Z_r)=0\, ,\quad \frac{1}{r}(Z_\ell - Z_r)=i\widehat\zeta.
\end{equation}
Here, $\widehat\zeta=\zeta_I t^I$ acts non-trivially only on operators charged under the topological symmetry, where $\zeta_I$ is the FI parameter and $t^I$ is the corresponding topological charge. Examples of such operators are monopole operators.

\section{Cohomology in SCFTs}
\label{COHOMOLOGY1}

Our aim in this section and the next is to describe a procedure that generalizes the cohomological truncation of \cite{Beem:2013sza, Chester:2014mea, Beem:2016cbd} from $\cN=4$ SCFTs to the more general non-conformal $\cN=4$ theories on $S^3$ that were described in Section \ref{THEORIES}.  The construction of \cite{Beem:2013sza, Chester:2014mea, Beem:2016cbd} was based on identifying two supercharges $\cQ^H_1$ and $\cQ^H_2$ in $\cN=4$ SCFTs on $\R^3$, such that the OPE restricted to their cohomology gives a certain quantization of the Higgs branch. It was also possible to find another pair of supercharges, $\cQ^C_1$ and $\cQ^C_2$, whose cohomology similarly leads to a quantization of the Coulomb branch, though this second possibility was not explored in detail.   
We will generalize both cases to non-conformal theories on $S^3$, but, just as in \cite{Beem:2013sza, Chester:2014mea, Beem:2016cbd}, our main focus will also be the Higgs branch.  
	
We will find that local operators in the cohomology, both for $\cQ^H_i$ and for $\cQ^C_i$, can only be inserted along a great circle $S^1\subset S^3$.\footnote{There is some freedom in choosing $\cQ^{H,C}_i$, which corresponds precisely to the choice of great circle on $S^3$.} 
The circle is the fixed point locus of the $U(1)$ isometry that appears in the anti-commutator $\{\cQ^H_1,\cQ^H_2\}$ or $\{\cQ^C_1,\cQ^C_2\}$. In the case of $\cQ^H_i$, the operators that can be inserted on $S^1$ will be referred to as ``twisted Higgs branch operators'', because, as we will see, they are in 1-to-1 correpsondence with Higgs branch chiral ring operators. Similarly, operators in $\cQ^C_i$ cohomology will be referred to as ``twisted Coulomb branch operators''.

In Section~\ref{SCFTFLAT}, we start by reviewing the construction of \cite{Beem:2013sza, Chester:2014mea, Beem:2016cbd} in flat space, and then in Section~\ref{SCFTSPHERE} we translate this construction to $S^3$.   In Section~\ref{COHOMOLOGY2} we describe the generalization of this cohomology directly based on the $\mathfrak{su}(2|1)_\ell \oplus\mathfrak{su}(2|1)_r$ algebra.

\subsection{SCFT in flat space}
\label{SCFTFLAT}
	
Consider theories living on a three-dimensional Euclidean space $\R^3$ with the standard coordinates $\vec{x}=(x_1, x_2, x_3)$. The bosonic subalgebra of the $\mathfrak{osp}(4|4)$ superconformal algebra is $\mathfrak{so}(4)\oplus \mathfrak{sp}(4)$, where the generators of $\mathfrak{sp}(4)$ are rotations $M_{\alpha\beta}$, translations $P_\mu$ and special conformal transformations $K_\mu$, and the generators of the $\mathfrak{so}(4) \cong \mathfrak{su}(2)_H \oplus \mathfrak{su}(2)_C$ R-symmetry are denoted by $R_{ab}$ and $\bar{R}_{\dot a\dot b}$ and act on the Higgs and Coulomb branches, respectively. The fermionic generators are $Q_{\alpha a\dot a}$ and $S_{\alpha a\dot a}$, denoting Poincar\'e and conformal supercharges, respectively. The detailed description of this algebra can be found in Appendix C.1.  
	
Define the two supercharges $\cQ^H_1$ and $\cQ^H_2$ by
\begin{align}
\cQ^H_1 = Q_{11\dot 2} + \frac{1}{2r} S^2{}_{2\dot 2} \, ,\quad \cQ^H_2 = Q_{21\dot 1} + \frac{1}{2r} S^1{}_{2\dot 1}\, . \label{Q12}
\end{align}
In \eqref{Q12}, $r$ is some arbitrary parameter with dimensions of length.\footnote{When we map \eqref{Q12} to $S^3$, we will interpret $r$ as the radius of the sphere.} The supercharges $\cQ^H_{1,2}$ are nilpotent, i.e., $(\cQ^H_1)^2=(\cQ^H_2)^2=0$, and their anticommutator is given by
\begin{align}
\cZ = \frac{ir}{4}\{\cQ^H_1,\cQ^H_2\} = -M_{12} + \bar{R}_{\dot 1}{}^{\dot 1} \,. \label{center}
\end{align}
Whether we consider the cohomology of $\cQ^H_1$ or $\cQ^H_2$, the above equation implies that it can be represented by elements from the $\cZ=0$ subspace. In order to satisfy $\cZ=0$, local operators with zero $\bar{R}_{\dot 1}{}^{\dot 1}$ charge can only be inserted at the fixed point locus of the $M_{12}$ rotation, i.e., at the line $x_1=x_2=0$. 
	
There are $\cQ^H_{1,2}$-exact twisted translation and dilatation given by 
\begin{align}
\widehat{L}_- &= -\frac{1}{4}\{\cQ^H_1,Q_{22\dot 1}\} = \frac{1}{4}\{\cQ^H_2,Q_{12\dot 2}\} = P_3+\frac{i}{2r} R_2{}^1 \, \\
\widehat{L}_0 &= -\frac{1}{8} \{\cQ^H_1,\cQ^{H\dagger}_1\} = \frac{i}{8} \{\cQ^H_1,2r Q_{21\dot 1} - S^1{}_{2\dot 1} \} \notag\\
&=-\frac{1}{8}\{\cQ^H_2,\cQ^{H\dagger}_2\} = \frac{i}{8}\{\cQ^H_2,-2r Q_{11\dot 2}+S^2{}_{2\dot 2}\} = -D + R_1{}^1 \, . \label{L0}
\end{align}
The twisted translation generated by $\widehat{L}_-$ can be used to move cohomology classes along the line $x_1=x_2=0$. In particular, every cohomology class defined at the origin can be twisted-translated to the whole line $x_1=x_2=0$. It is those observables on the line which where referred to before as twisted operators. Because $\widehat{L}_-$ is $\cQ^H_1$- and $\cQ^H_2$-exact, this twisted translation is a trivial operation at the level of the cohomology of $\cQ^H_1$ or $\cQ^H_2$. Therefore, to characterize the local operators in cohomology completely, it is sufficient to consider them inserted at the origin. By the state-operator map, this corresponds to studying the state cohomology. Using \eqref{L0},  $\widehat{L}_0 = -\frac{1}{8} \{\cQ^H_1,\cQ^{H\dagger}_1\} =-\frac{1}{8}\{\cQ^H_2,\cQ^{H\dagger}_2\} $, the standard Hodge theory argument proves that the cohomologies of $\cQ^H_1$ and $\cQ^H_2$ are identical and are represented by the kernel of $\widehat{L}_0$. 
	
As shown in \cite{Chester:2014mea,Beem:2016cbd}, these representatives are given by local operators ${\cal O}^{\R^3}_{a_1\cdots a_n}(\vec{0})$ transforming in the $(\mathbf{n+1},
\mathbf{1})$ irrep of the $\mathfrak{su}(2)_H\oplus\mathfrak{su}(2)_C$ R-symmetry and of conformal dimension $\Delta=n/2$. When translated with $\widehat{L}_-$ they give the twisted operator:
\es{TwistedFlat}{
	{\cal O}(s) = {\cal O}^{\R^3}_{a_1\cdots a_n}\Big|_{\vec{x}=(0,0,s)}u_{\R^3}^{a_1}\cdots u_{\R^3}^{a_n} \ecq\quad u_{\R^3}\equiv(1, \frac{s}{2r})\ec
}
which defines a non-trivial cohomology class on the line $x_1=x_2=0$.
	
Local operators in the cohomology form a certain algebraic structure under the OPE of the full theory. In particular, because $\widehat{L}_-$ is zero in cohomology, the OPE of operators in the cohomology does not depend on their positions on the line, but it can depend on their ordering. By moving operators to one point, we then define a product of cohomology classes. This way we get an algebra in the cohomology, which is associative but not necessary commutative.
	
As explained in \cite{Beem:2016cbd}, the operators ${\cal O}(s)$, when inserted at the origin $s=0$, are just the Higgs branch chiral ring operators. However, as we move away from the origin, they become mixed with anti-chiral operators, because of the twisting factor $u_{\R^3}=(1, \frac{s}{2r})$. This twisting factor can be thought of as an $s$-dependent choice of the Cartan generator of $\mathfrak{su}(2)_H$ given by $\frac{1-s^2/(2r)^2}{1+s^2/(2r)^2}\sigma_3 + \frac{s/r}{1+s^2/(2r)^2}\sigma_1$. A twisted operator ${\cal O}(s)$ is in the $\mathfrak{su}(2)_H$ highest weight state with respect to this $s$-dependent Cartan generator. The fact that the twisted operators are not chiral with respect to a fixed Cartan generator is responsible for the fact that the algebraic structure we get is not a chiral ring, but rather its deformation quantization.\footnote{More precisely, the Higgs (or Coulomb) branch chiral ring has a natural Poisson structure, since it corresponds to the ring of holomorphic functions on the moduli space, which for $\cN=4$ theories is a hyperk\"ahler cone. The algebraic structure we obtain is the deformation quantization of this Poisson algebra.} The deformation parameter is $\frac{1}{2r}$, which was denoted by $\zeta$ in \cite{Beem:2016cbd}.
	
In fact, it turns out to be slightly more convenient to study the cohomology of a linear combination $\cQ^H_1+\beta \cQ^H_2$ with some generic $\beta$. This operator squares to the bosonic transformation $\cZ$, and so it plays a role of the equivariant differential. The cohomology problem for this operator therefore involves two steps: one has to restrict to the $\cZ=0$ subspace first, and then compute the cohomology there.
	
Recall that $\cZ$ is a sum of rotation in the $(x_1, x_2)$ plane and a certain R-symmetry transformation. The condition $\cZ=0$ then implies that geometrically, the configuration of operators should be invariant under this rotation. In particular, local operators, as well as line operators, can only be inserted at the line $x_1=x_2=0$, which is the fixed point locus of this rotation. Surface operators, on the other hand, can only span the orthogonal $(x_1, x_2)$ plane and correspond to some fixed value of $x_3$.
	
Including line operators would of course change the answer, and the cohomology of local operators located at the line defect at $x_1=x_2=0$ would give a different protected algebra. Surface operators, on the other hand, are expected to give some modules for the protected algebra to act on. They would describe point defects on the line $x_1=x_2=0$, acted on by the local operators. This action simply corresponds to merging local operators and the defect together.  Including extended operators gives an interesting direction for further explorations, and it would potentially allow one to extract more dynamical information about the theory. However, in this paper, we do not consider any extended operators and study only the protected algebra of local operators.
	
\subsection{SCFT on the sphere}
\label{SCFTSPHERE}
	
Now let us identify the counterpart of the above construction on the sphere. After describing it in some detail, we will be able to see that it generalizes to non-conformal theories in a straightforward fashion.
	
Using the stereographic map, one can place any conformal theory on $S^3$. Under this map, the line $x_1=x_2=0$ maps to a great circle $S^1\subset S^3$, along which the cohomology classes of local operators described in the previous subsection will be inserted. The rotation in $\cZ$ now becomes a $U(1)$ isometry of the sphere, whose fixed point locus is precisely this $S^1$.

As mentioned in Section~\ref{ISOMETRIES}, it will be useful to represent $S^3$ as a $U(1)$ fibration over the disk $D^2$, with fibers shrinking at its boundary $\partial D^2=S^1$. This boundary $S^1$, parameterized by the angle $\varphi$ at $\theta=\frac{\pi}{2}$, is the great circle mentioned above along which local operators in cohomology can be inserted.  The situation here is similar to that in \cite{Pestun:2009nn}, where an analogous representation of $S^4$ was used in the localization of 4d $\cN=4$ Yang-Mills theory to an $S^2$.

\subsubsection{Twisted operators on $S^3$ by stereographic map}
	
In $\R^3$, we were interested in correlators of twisted operators ${\cal O}_i(s_i)$ inserted at points $(0,0,s_i)$. Let us map them on $S^3$: 
\begin{equation}
\label{stereomap}
\langle {\cal O}_1(s_1)\cdots {\cal O}_k(s_k)  \rangle_{\R^3}=\langle  {\cal O}_1(\varphi_{1})\cdots {\cal O}_k(\varphi_k)  \rangle_{S^3} \,.
\end{equation}
Operators on the right are the sphere counterparts of the flat space twisted operators, and are given by contraction of the $S^3$ operators ${\cal O}_{a_1\cdots a_n}^{S^3}(\varphi)\Big|_{\theta=\frac{\pi}{2}}$, inserted on the great circle at $\theta=\pi/2$, with $u=(1,\frac{x_3}{2r})=(1,\tan\frac{\varphi}{2})$.  For every operator of dimension $\Delta$, we have ${\cal O}^{\R^3} = \Omega^\Delta {\cal O}^{S^3}$, with $\Omega$ being the conformal factor, which evaluates to $\Omega =\cos^2\frac{\varphi}{2}$ at $\theta = \pi/2$.  The definition \eqref{TwistedFlat} then implies
\begin{equation}
\label{tw_on_s3}
{\cal O}(\varphi) =\cos^n\frac{\varphi}{2}  {\cal O}^{S^3}_{a_1\cdots a_n}\Big|_{\theta=\frac{\pi}{2}} u_{\R^3}^{a_1}\cdots u_{\R^3}^{a_n}={\cal O}^{S^3}_{a_1\cdots a_n}\Big|_{\theta=\frac{\pi}{2}} {u}_{S^3}^{a_1}\cdots {u}_{S^3}^{a_n} \,,
\end{equation}
where $u_{S^3}=u_{\R^3} \cos\frac{\varphi}{2} = (\cos\frac{\varphi}{2},\sin\frac{\varphi}{2})$.  Note that the twisted operators ${\cal O}$ do not transform with a conformal factor in going from $\R^3$ to $S^3$, and this is why they do not carry an $\R^3$ or $S^3$ superscript and why there is no conformal factor in \eqref{stereomap}.  We will now interpret this construction in a more intrinsic way using the theory on $S^3$ only.

\subsubsection{Interpretation in terms of $\mathfrak{su}(2|1)_\ell\oplus\mathfrak{su}(2|1)_r$ subalgebra}
In Section 2 and Appendix C, we chose an embedding of the $\mathfrak{su}(2|1)_\ell\oplus\mathfrak{su}(2|1)_r$ superalgebra in $\mathfrak{osp}(4|4)$, such that $\mathfrak{su}(2)_\ell\oplus\mathfrak{su}(2)_r\subset\mathfrak{sp}(4)$ corresponds to isometries of the sphere and $\mathfrak{u}(1)_\ell\oplus\mathfrak{u}(1)_r\subset\mathfrak{so}(4)_R\cong \mathfrak{su}(2)_H\oplus\mathfrak{su}(2)_C$ was a Cartan subalgebra of the R-symmetry algebra.  The choice of Cartan subalgebra was parametrized by the matrices $h$ and $\bar{h}$. To be more precise, $h$ parameterizes the Cartan generator $R_H$ in $\mathfrak{su}(2)_H$ and $\bar{h}$ parameterizes the Cartan generator $R_C$ in $\mathfrak{su}(2)_C$. The generators $R_\ell$ and $R_r$ of $\mathfrak{u}(1)_\ell$ and $\mathfrak{u}(1)_r$ are given by \eqref{Ulr}.   The supercharges of $\mathfrak{su}(2|1)_\ell$ were denoted by $\cQ_\alpha^{(\ell\pm)}$, and the supercharges of $\mathfrak{su}(2|1)_r$ by $\cQ_\alpha^{(r\pm)}$. Their expressions in terms of conformal supercharges $Q_{\alpha a\dot a}$ and $S^\alpha_{\,\, a\dot a}$ can be found in Appendix C.2.  
	
Using this embedding, it is easy to identify our supercharges $\cQ^H_1$ and $\cQ^H_2$ as:
\begin{align}
\cQ^H_1 = \cQ_1^{(\ell +)} + \cQ_1^{(r -)},\quad \cQ^H_2 = \cQ_2^{(\ell -)} + \cQ_2^{(r +)}.
\end{align}
Each of these supercharges is of course nilpotent, and
\es{QH1QH2Anti}{
	\{\cQ^H_1,\cQ^H_2\} = \frac{4i}{r}(P_\tau + R_C) 
}
where $P_\tau \equiv -(J^{(\ell)}_3 + J^{(r)}_3)$ is the $\tau$-translation acting as $P_\tau = i \partial_\tau$ on gauge-invariant operators.
	
Using the $\mathfrak{su}(2|1)_\ell\oplus\mathfrak{su}(2|1)_r$ algebra, we can check that:
\begin{align}
\{\cQ^H_1, \frac{r}{4i}(\cQ_2^{(\ell -)} - \cQ_2^{(r +)})\}=\{\cQ_2, \frac{r}{4i}(\cQ_1^{(\ell +)} - \cQ_1^{(r -)})\}= P_\varphi + R_H \equiv \widehat{P}_\varphi \,.
\end{align}
Here $P_\varphi= J_3^{(r)} - J_3^{(\ell)}$ is simply the $\varphi$ translation isometry of $S^3$ acting on gauge invariant operators as $P_\varphi = i \partial_\varphi$.  The generator $\widehat{P}_\varphi$ defined above is a new twisted-translation, which is defined on the sphere purely in terms of the $\mathfrak{su}(2|1)_\ell\oplus\mathfrak{su}(2|1)_r$ superalgebra.
	
Let ${\cal O}_{a_1\cdots a_n}$ be some local operator in the SCFT on the sphere,\footnote{From now on, we drop the superscript or subscript $S^3$ present in the previous subsection.} in the spin-$n/2$ irrep of $\mathfrak{su}(2)_H$. If ${\cal O}_{11\cdots 1}$, when inserted at the point $\theta=\pi/2$, $\varphi=0$ (which corresponds to the origin of $\R^3$ upon stereographic projection), is in the cohomology of $\cQ^H_1$ and $\cQ^H_2$, (recall from previous discussions that it must have the highest $\mathfrak{su}(2)_H$-weight) we can use the $\widehat{P}_\varphi$ translation to move it along the $\varphi$-circle without changing its cohomology class:
\begin{equation}
{\cal O}(\varphi) = e^{i\varphi \widehat{P}_\varphi} {\cal O}_{11\cdots 1}\Big|_{\theta=\frac{\pi}{2},\varphi=0}e^{-i\varphi \widehat{P}_\varphi}= {\cal O}_{a_1\cdots a_n}\Big|_{\theta=\frac{\pi}{2},\varphi=0}{u}^{a_1}\cdots {u}^{a_n},
\end{equation}
where ${u}=(\cos\frac{\varphi}{2}, \sin\frac{\varphi}{2})$.  This expression precisely matches \eqref{tw_on_s3}, which was obtained from the stereographic map from $\R^3$. We conclude that the stereographic map identifies the twisted operators on $\R^3$ defined in \cite{Chester:2014mea, Beem:2016cbd}, with twisted operators on $S^3$ defined purely in terms of the $\mathfrak{su}(2|1)_\ell\oplus\mathfrak{su}(2|1)_r$ subalgebra of the $\mathfrak{osp}(4|4)$ superconformal algebra. This subalgebra has all the necessary ingredients for the cohomological truncation to work.
	
Note that the $\R^3$ construction in \cite{Chester:2014mea,Beem:2016cbd} utilized a different subalgebra of $\mathfrak{osp}(4|4)$, namely a centrally extended $\mathfrak{su}(2|2)$. That algebra is a 1d $\cN=4$ superconformal algebra acting on the $x_1=x_2=0$ line, suggesting that conformal symmetry is somehow important for the construction to work. Our algebra $\mathfrak{su}(2|1)_\ell\oplus\mathfrak{su}(2|1)_r$, on the other hand, is not related to conformal symmetry anymore. In fact, it is the supersymmetry algebra of a general class of non-conformal $\cN=4$ actions on $S^3$, as explained in Section 2 (of course, at the RG fixed points, it becomes enhanced to $\mathfrak{osp}(4|4)$). In the next subsection, we summarize our construction for the theories based on $\mathfrak{su}(2|1)_\ell\oplus\mathfrak{su}(2|1)_r$ and its central extensions.

\section{Cohomology in non-conformal ${\cal N} = 4$ theories on $S^3$}
\label{COHOMOLOGY2}

As we have seen, the cohomolgical construction of \cite{Beem:2013sza,Chester:2014mea,Beem:2016cbd} for SCFTs on $\R^3$ can be readily translated to $S^3$, since SCFTs can be canonically placed on a sphere. Interestingly, once we pass to the sphere, conformal symmetry is no longer necessary, and the cohomology described in the previous section is also defined away from the RG fixed point.  We will explore this construction in this section.
	
For the cohomological reduction to the 1d sector to work, it is enough to preserve the subalgebra $\mathfrak{su}(2|1)_\ell\oplus \mathfrak{su}(2|1)_r\subset \mathfrak{osp}(4|4)$ of the superconformal algebra. The superalgebra $\mathfrak{su}(2|1)_\ell\oplus \mathfrak{su}(2|1)_r$ (or its centrally extended versions), as explained in Section 2, is a possible $\cN=4$ superalgebra on $S^3$. It describes a class of non-conformal theories on $S^3$. At the RG fixed point, the symmetry is of course enhanced to $\mathfrak{osp}(4|4)$, and our results reduce to those of \cite{Beem:2013sza, Chester:2014mea, Beem:2016cbd} as reviewed in Section~\ref{COHOMOLOGY1}.

In an ${\cal N} = 4$ theory on $S^3$ invariant under the centrally extended algebra $\widetilde{\mathfrak{su}(2|1)}_\ell\oplus\widetilde{\mathfrak{su}(2|1)}_r$ discussed in Section~\ref{EXTENSION}, the construction proceeds as follows.  Consider the following linear combinations of supercharges:
\begin{align}
\cQ^H_1 &\equiv \kappa^\alpha_{1\ell}\cQ^{(\ell_+)}_\alpha + \kappa^\alpha_{1r}\cQ^{(r_-)}_\alpha\, , \quad \cQ^H_2 \equiv \kappa^\alpha_{2\ell}\cQ^{(\ell_-)}_\alpha + \kappa^\alpha_{2r}\cQ^{(r_+)}_\alpha \, , \\
\cQ^C_1 &\equiv \kappa^\alpha_{1\ell}\cQ^{(\ell_+)}_\alpha + \kappa^\alpha_{1r}\cQ^{(r_+)}_\alpha \, ,\quad \cQ^C_2 \equiv \kappa^\alpha_{2\ell}\cQ^{(\ell_-)}_\alpha + \kappa^\alpha_{2r}\cQ^{(r_-)}_\alpha\, .
\end{align}
Each of them is nilpotent, and, based on \eqref{QQCentral}, they satisfy:
\begin{align}
\{\cQ^H_1, \cQ^H_2\} &= -\frac{4i}{r}\left(\kappa^\alpha_{1\ell}\kappa^\beta_{2\ell}J_{\alpha\beta}^{(\ell)} + \kappa^\alpha_{1r}\kappa^\beta_{2r}J_{\alpha\beta}^{(r)} + \kappa_{1\ell}\kappa_{2\ell}(\frac{1}{2}R_\ell+Z_\ell) - \kappa_{1r}\kappa_{2r}(\frac{1}{2}R_r+Z_r)\right) \, , \\
\{\cQ^C_1, \cQ^C_2\} &= -\frac{4i}{r}\left(\kappa^\alpha_{1\ell}\kappa^\beta_{2\ell}J_{\alpha\beta}^{(\ell)} + \kappa^\alpha_{1r}\kappa^\beta_{2r}J_{\alpha\beta}^{(r)} + \kappa_{1\ell}\kappa_{2\ell}(\frac{1}{2}R_\ell+Z_\ell) + \kappa_{1r}\kappa_{2r}(\frac{1}{2}R_r+Z_r)\right) \, ,
\end{align}
where $\kappa_1\kappa_2\equiv \varepsilon_{\alpha\beta}\kappa_1^\alpha\kappa_2^\beta$.
We want the rotation which appears on the right to fix a great circle on $S^3$. Before, this circle was determined as an image of the line $x_1=x_2=0$ under the stereographic projection, and it was the fixed point locus of the $\tau$-rotations on $S^3$. But there are many equivalent choices of the large circle on $S^3$, and that is why we have free parameters denoted by $\kappa$ above. 
	
Without loss of generality, we will consider the same $S^1$ as before that is parametrized by $\varphi$ and is a fixed point set for the $\tau$ rotations. To pick such a circle, we use:
\begin{align}
\kappa_{1\ell}=\kappa_{1r}=\left(\begin{matrix}1\cr 0\end{matrix} \right)\, ,\quad \kappa_{2\ell}=\kappa_{2r}=\left(\begin{matrix}0\cr 1\end{matrix} \right).
\end{align}

Then the supercharges become:
\begin{align}
\cQ^H_1 &= \cQ^{(\ell_+)}_1 + \cQ^{(r_-)}_1\, , \quad \cQ^H_2 = \cQ^{(\ell_-)}_2 + \cQ^{(r_+)}_2 \, , \label{QH12} \\
\cQ^C_1 &= \cQ^{(\ell_+)}_1 + \cQ^{(r_+)}_1 \, ,\quad \cQ^C_2 = \cQ^{(\ell_-)}_2 + \cQ^{(r_-)}_2\, , \label{QC12}
\end{align}
and their algebra is:
\begin{align}
\{\cQ^H_1, \cQ^H_2\} &= \frac{4i}{r}\left( P_\tau + R_C + Z_\ell - Z_r \right)=\frac{4i}{r}\left( P_\tau + R_C + ir\widehat\zeta \right) \, , \\
\{\cQ^C_1, \cQ^C_2\} &= \frac{4i}{r}\left( P_\tau + R_H + Z_\ell + Z_r \right)=\frac{4i}{r}\left( P_\tau + R_H + ir\widehat{m} \right) \, ,
\end{align}
where $P_\tau  = -(J_3^{(\ell)} + J_3^{(r)}) = i \partial_\tau$ is the $\tau$-rotation isometry, just as we need, and $R_C=\frac{1}{2}(R_\ell-R_r)$, $R_H=\frac{1}{2}(R_\ell+R_r)$. 
	
Next, we find the $Q_{1,2}^{H/C}$-exact generators given by
\begin{align}
\{\cQ^H_1, \frac{r}{4i}(\cQ_2^{(\ell -)} - \cQ_2^{(r +)})\}=\{\cQ^H_2, \frac{r}{4i}(\cQ_1^{(\ell +)} - \cQ_1^{(r -)})\}= P_\varphi + R_H + Z_\ell + Z_r \,,\\
\{\cQ^C_1, \frac{r}{4i}(\cQ_2^{(\ell -)} - \cQ_2^{(r -)})\}=\{\cQ^C_2, \frac{r}{4i}(\cQ_1^{(\ell +)} - \cQ_1^{(r +)})\}= P_\varphi + R_C + Z_\ell - Z_r \,,
\end{align}
with $P_\varphi = J_3^{(r)} - J_3^{(\ell)} = i \partial_\varphi$ as before. We define two twisted rotations:
\begin{align}
\widehat{P}^H_\varphi = P_\varphi + R_H\, ,\\
\widehat{P}^C_\varphi = P_\varphi + R_C\, ,
\end{align}
which are closed with respect to the corresponding supercharges, i.e., $[\cQ^H_i,\widehat{P}^H_\varphi]=0$ and $[\cQ^C_i,\widehat{P}^C_\varphi]=0$. Therefore, these twisted rotations can still be used to translate cohomology classes along the $\varphi$-circle. Now, however, $\widehat{P}^{H/C}_\varphi$ are not necessarily exact. Rather, $\widehat{P}_\varphi^H$ is cohomologous to $-Z_\ell - Z_r=-ir\widehat{m}$ and $\widehat{P}_\varphi^C$ is cohomologous to $-Z_\ell+Z_r=-ir\widehat\zeta$. Thus, on operators commuting with $\widehat{m}$, twisted translations act in a $\cQ^H_i$-exact way, which is similar to what we had before. For such operators, there is no $\varphi$-dependence of cohomology classes, meaning their correlation functions are position-independent. 
	
For operators that have a non-zero eigenvalue of $\widehat{m}$ (such operators are charged under the Cartan of the flavor symmetry), the corresponding cohomology classes become position dependent. Nevertheless, this position dependence is very simple: it appears in correlators as the factor $e^{r m\varphi}$, where $m$ is the eigenvalue of $\widehat{m}$. We could have included $Z_\ell+Z_r$ into the definition of the twisted translation and removed this position dependence, but we find it more convenient not to do so. 
	
Analogously, operators in $\cQ^C_i$ cohomology, which are not charged under the topological symmetry, have $\widehat\zeta=0$, and correlation functions of twisted-translated operators  are position-independent. For operators charged under the topological symmetry, there is a $\varphi$-dependence given by $e^{r\zeta\varphi}$, where $\zeta$ is the eigenvalue of $\widehat\zeta$. For example, cohomology classes of monopole operators are expected to carry such position dependence in the presence of non-zero FI terms.
	
\subsection{Operators in the cohomology of $\cQ^H_i$}
\label{QHoperators}
	
In an SCFT, there was a state-operator map which allowed to identify the cohomology of local operators inserted at the origin with the state cohomology. Because of $\frac{1}{8}\{\cQ^H_1,\cQ^{H\dagger}_1\}=\frac{1}{8}\{\cQ^H_2,\cQ^{H\dagger}_2\}=D-R_1^{\,\, 1}$, one could completely describe cohomology by the equation $D=R_1^{\,\, 1}$. Unitarity also implied $D-R_1^{\,\, 1}\ge 0$, so states/operators with $D-R_1^{\,\, 1}=0$ had to be the highest weight states with respect to $\mathfrak{su}(2)_H$, i.e., they had the maximal eigenvalue of $R_1^{\,\, 1}$. This approach shows that the components $q_1$ and $\tq_1$ of the hypermultiplet scalars are examples of such operators in gauge theories built from hypermultiplets and vectormultiplets, and all other operators are constructed from them.
	
What should we do in our, generally non-conformal, case? One can check, by applying SUSY variations from the Section 2, that $q_1$ and $\tq_1=-\tq^2$, when inserted at the point $\theta=\pi/2$, $\varphi=0$, are still annihilated by $\cQ^H_1$ and $\cQ^H_2$. Twisted-translating them along the great circle parametrized by $\varphi$, we get twisted operators in the cohomology of $\cQ^H_i$:
\begin{align}
Q(\varphi)= q_1(\varphi)\cos\frac{\varphi}{2} + q_2(\varphi)\sin\frac{\varphi}{2}\,,
\qquad \tQ(\varphi) = \tq_1(\varphi)\cos\frac{\varphi}{2} + \tq_2(\varphi)\sin\frac{\varphi}{2} \ed \label{QQt}
\end{align}
In the gauged case, one should be slightly more precise, as we are allowed to consider only gauge invariant operators. This means that $\cQ^H_i$-closed operators that we can insert at the origin are gauge invariant polynomials in $q_1$ and $\tq_1$, and twisting-translating them along the circle gives gauge invariant polynomials in $Q$ and $\tQ$. 
	
These are the interesting $\cQ^H_i$-closed observables. Because we do not have a Hodge theory argument like in the conformal case, it becomes harder to argue that these operators are not $\cQ^H_i$-exact, and that there are no other cohomology classes besides those represented by products of $Q$ and $\tQ$.  But there is a roundabout: in the next section we will localize our theory to a 1d theory on the circle, and non-trivial cohomology classes of local operators will give local observables in that theory. We will see that all local observables in the 1d theory are generated by $Q$ and $\tQ$. (This will be especially clear from the 1d gauge theory interpretation of Section \ref{TOPOLOGICALQM}) This allows to prove that a 3d operator constructed from $Q$ and $\tQ$ is not $\cQ^H_i$-exact, otherwise it would vanish in the 1d theory (because a correlator of $\cQ^H_i$-closed operators with a $\cQ^H_i$-exact operator is zero). We can say that localization provides a surjective map from $\cQ^H_i$-closed observables in 3d to all local observables in 1d, which are just gauge invariant polynomials in $Q$ and $\tQ$. It does not  prove that there are no additional operators in the cohomology of the 3d theory, i.e., that this map is also injective, but we assume it to be the case.\footnote{Had it not been the case, this map would have a non-empty kernel, in other words there would exist a non-trivial operator in 3d which vanishes under the correlators with arbitrary insertions of $\cQ^H_i$-closed observables.}
	
The cohomology of $\cQ_{1,2}^H$ also contains interesting loop operators. For instance, the $\frac{1}{2}$-BPS Wilson loop wrapping the $\theta=\frac{\pi}{2}$ circle is defined by:
\begin{align}
W_{\cR} \equiv \trace_{\cR}\cP e^{-i\int_{\theta=\frac{\pi}{2}} d\varphi \left(A_{\varphi} + ir\Phi_{\dot{1}\dot 2} \right)} \ed \label{wilson}
\end{align}
One can verify that $W_{\cR}$ preserves the supercharges $\cQ^{(\ell_+)}_1$, $\cQ^{(\ell_-)}_2$, $\cQ^{(r_+)}_2$ and $\cQ^{(r_-)}_1$, which generate an $\mathfrak{su}(1|1)\oplus\mathfrak{su}(1|1)$ sub-algebra of $\mathfrak{su}(2|1)_{\ell}\oplus\mathfrak{su}(2|1)_r$. Our supercharges $\cQ_{1,2}^H$ are both part of this sub-algebra.
	
\subsection{Operators in the cohomology of $\cQ^C_i$}
\label{QCoperators}
Studying the Coulomb branch is not the central topic of this paper, but we nevertheless give some details on it. Again, in conformal theories one can use state-operator map and the equation:
\begin{equation}
\frac{1}{8}\{\cQ^C_1,\cQ^{C\dagger}_1\}=\frac{1}{8}\{\cQ^C_2,\cQ^{C\dagger}_2\}=D - \frac{1}{2}(\bar{R}_{\dot 1}^{\,\,\dot2} + \bar{R}_{\dot2}^{\,\,\dot1}).
\end{equation}
So, every state in the cohomology of an SCFT has to be the highest-weight state with respect to $(\sigma_1)_{\dot a}^{\,\,\dot b}\bar{R}_{\dot b}^{\,\,\dot a}$ and satisfy $D=\frac{1}{2}(\sigma_1)_{\dot a}^{\,\,\dot b}\bar{R}_{\dot b}^{\,\,\dot a}$. If we define
\begin{equation}
v^{\dot a}=\left( \begin{matrix}
1\cr 
1
\end{matrix} \right),
\end{equation}
then every operator ${\cal O}_{\dot{a}_1\cdots \dot{a}_n}$ of $\mathfrak{su}(2)_C$ spin $n/2$ has the highest weight component ${\cal O}_{\dot{a}_1\cdots \dot{a}_n}v^{\dot{a}_1}\cdots v^{\dot{a}_n}$. In gauge theories constructed from vectors and hypers, there is one obvious operator which satisfies this condition:
\begin{equation}
\Phi_{\dot a \dot b}v^{\dot a}v^{\dot b}=\Phi_{\dot1\dot1} + \Phi_{\dot2\dot2} + 2\Phi_{\dot1\dot2} \,,
\end{equation}
because $\Phi_{\dot a\dot b}$ at the conformal point has dimension $\Delta_\Phi=1$. For non-conformal theories, $\Phi_{\dot a \dot b}v^{\dot a}v^{\dot b}$ is also annihilated by $\cQ^C_i$ at the origin $\theta=\pi/2,\, \varphi=0$, as one can check using transformation rules from the Section 2. The corresponding twisted-translated operator is:
\begin{equation}
\Phi(\varphi)=e^{i\varphi}\Phi_{\dot1\dot1} + e^{-i\varphi}\Phi_{\dot2\dot2} + 2\Phi_{\dot1\dot2} \,. \label{hatPhi}
\end{equation}
Later we will be able to easily compute correlation functions of such operators. However, this is not the whole story for the Coulomb branch. The Coulomb branch chiral ring also contains monopole defect operators that contribute to the protected algebra in the cohomology of $\cQ^C_i$. Moreover, this cohomology contains line defect operators, known as vortex loops, which map to the Wilson loops \eqref{wilson} under mirror symmetry. As shown in \cite{Assel:2015oxa}, the vortex loop preserves an $\mathfrak{su}(1|1)\oplus\mathfrak{su}(1|1)$ sub-algebra of $\mathfrak{su}(2|1)_{\ell}\oplus\mathfrak{su}(2|1)_r$, which in our language is generated by $\cQ^{(\ell_+)}_1$, $\cQ^{(\ell_-)}_2$, $\cQ^{(r_+)}_1$ and $\cQ^{(r_-)}_2$. Both of the supercharges $\cQ^C_{1,2}$ are part of that sub-algebra. We postpone a detailed study of defect operators to a future publication.
	
\section{Localization}
\label{LOCALIZATION}
	
In this section we describe how to localize the theories on $S^3$ defined in Section \ref{THEORIES} to the 1d Higgs branch cohomological sector described in Sections \ref{COHOMOLOGY1} and \ref{COHOMOLOGY2}. We also provide some preliminary results on the Coulomb branch.  
	
Let us first briefly review how supersymmetric localization works. Given a supercharge $\cQ$ which generates a symmetry of our theory, we would like to calculate the path integral
\begin{align}
\cI = \int D\cV D\cH e^{-S[\cV,\cH]} (\cdots) \ecq  S \equiv S_{\text{YM}}[\cV] + S_{\text{hyper}}[\cH,\cV] \ec \label{Ischematic}
\end{align}
with $S_{\text{YM}}$ and $S_{\text{hyper}}$ defined in \eqref{SYM} and \eqref{Shyper}, respectively, and $(\cdots)$ representing some $\cQ$-closed insertions.\footnote{In most of this section we will not include FI terms and real masses to avoid clutter, but they can be very easily incorporated.} The first step is to deform the action in \eqref{Ischematic} by a $\cQ$-exact operator:
\begin{align}
\cI &\to \cI(t) = \int D\cV D\cH e^{-S_t[\cV,\cH]}(\cdots) \ec\notag\\
S_t &\equiv S + t V = S + t \int d^3x\sqrt{g}\{\cQ,\Psi(x)\} \ed \label{It}
\end{align}
If $[\cQ, V]=0$, then the path-integral \eqref{It} can be argued to be independent of $t$. Therefore, $\cI=\cI(0)$ is equal to the limit $\lim_{t\to\infty}\cI(t)$. In order for the path integral to converge as we take this limit, the bosonic part of $V$ is assumed to be non-negative. Then in the $t\to\infty$ limit, the path integral reduces to a sum over zeros of $V$, which are also its saddles. Those zeros include $\cQ$-invariant field configurations, and if $V$ is chosen properly, they are precisely identified with such configurations. Each of the saddles gives rise to two distinct contributions to $\cI$.  The first contribution comes from the classical action and insertions evaluated on the saddle point. The second contribution is the 1-loop determinant arising from integrating over the quadratic fluctuations of the fields around the saddle.
	
Our first task is then to choose the supercharge $\cQ$ in \eqref{It}. The twisted Higgs branch operators constructed from \eqref{QQt} are in the cohomology of $\cQ^H_{1,2}$ defined in \eqref{QH12}. To calculate their correlators we can therefore contemplate localizing with either $\cQ^H_1$ or $\cQ^H_2$. Similarly, to calculate correlators of twisted Coulomb branch operators, such as those constructed from \eqref{hatPhi}, we can consider localizing with $\cQ^C_1$ or $\cQ^C_2$, which were defined in \eqref{QC12}. In either case, as discussed in Section \ref{COHOMOLOGY1}, it is actually advantageous to localize with a linear combination
\begin{align}
\cQ_{\beta}^H &= \cQ_1^H + \beta \cQ_2^H \ec \label{QHloc}\\
\cQ_{\beta}^C &= \cQ_1^C + \beta \cQ_2^C \ec \label{QCloc}
\end{align} 
keeping $\beta\neq 0$ arbitrary. In what follows, we will describe the details of localizing our theories with respect to \eqref{QHloc} or  \eqref{QCloc}, starting with the vectormultiplet and then proceeding with the hypermultiplet.

\subsection{Vectormultiplets and a non-renormalization theorem}
	
As explained in \cite{Hama:2010av}, previous supersymmetric localization computations on $S^3$ may be simplified by taking the $\cN=2$ Yang-Mills and chiral superfield actions themselves as localizing terms. Indeed, these actions have non-negative bosonic parts, and are exact with respect to supercharges in the $\mathfrak{su}(2|1)\oplus\mathfrak{su}(2)$ symmetry algebra of $\cN=2$ theories on $S^3$. Moreover, they are symmetric under the full $\mathfrak{su}(2|1)\oplus\mathfrak{su}(2)$ algebra, which simplifies the evaluation of 1-loop determinants. 
	
It follows that our $\cN=4$ Yang-Mills action $S_{\text{YM}}$ defined in \eqref{SYM} is also exact with respect to supercharges in all of the  $\mathfrak{su}(2|1)\oplus\mathfrak{su}(2)$  sub-algebras of $\mathfrak{su}(2|1)_{\ell}\oplus\mathfrak{su}(2|1)_r$. In fact, an explicit calculation shows that $S_{\text{YM}}$ is also exact under both $\cQ_{\beta}^H$ and $\cQ_{\beta}^C$, even though they do not lie in any such $\cN=2$ sub-algebra.
	
Indeed, the Killing spinor generating $\cQ_{\beta}^H$ can be inferred from \eqref{Q12} and \eqref{delta2QS} to be
\begin{align}
(\xi_{\beta}^H)_{\alpha a \dot{a}} &= - e^{\Omega/2} \left[ \begin{pmatrix}
\beta\\0
\end{pmatrix}\otimes \begin{pmatrix}
0\\1
\end{pmatrix}\otimes \begin{pmatrix}
0\\1
\end{pmatrix} + \begin{pmatrix} 0\\1
\end{pmatrix} \otimes \begin{pmatrix}
0\\1
\end{pmatrix}\otimes \begin{pmatrix}
1\\0
\end{pmatrix} \right.\notag\\
&\left.+ \frac{1}{2r}\begin{pmatrix} x_1 - i x_2\\ -x_3
\end{pmatrix} \otimes \begin{pmatrix}
1\\0
\end{pmatrix}\otimes \begin{pmatrix}
1 \\ 0
\end{pmatrix} - \frac{\beta}{2r} \begin{pmatrix} x_3 \\ x_1 + i x_2
\end{pmatrix} \otimes \begin{pmatrix}
1\\0
\end{pmatrix}\otimes \begin{pmatrix}
0 \\ 1
\end{pmatrix} \right] \ed \label{xiQ1Q2}
\end{align}
One can then check using \eqref{Avar}--\eqref{dvar} that
\begin{align}
\delta_{\xi_{\beta}^H}\delta_{\xi_{-\beta}^H}\left(\frac{1}{2g^2_{\text{YM}}}h^{ab}\bar{h}^{\dot{a}\dot{b}}\int d^3x\sqrt{g}\trace\left(\lambda_{a\dot a}\lambda_{b\dot b} - 2 D_{ab}\Phi_{\dot{a}\dot b}\right)\right) = -i \beta  S_{\text{YM}} \ed \label{QexactYM}
\end{align}
Equation \eqref{QexactYM} also holds after the replacement $\xi_{\pm\beta}^H \to \xi_{\pm\beta}^C$, where $\xi_{\beta}^C$ is the Killing spinor generating $\cQ_{\beta}^C$. We conclude that $S_{\text{YM}}$ can be used as a localizing term for the vectormultiplet, whether we choose to localize with $\cQ^H_{\beta}$, $\cQ^C_{\beta}$, or the supercharge $\cQ_{\text{KWY}}$ that was used by \cite{Kapustin:2009kz} and lies in an $\cN=2$ sub-algebra of $\mathfrak{su}(2|1)_{\ell}\oplus\mathfrak{su}(2|1)_r$.\footnote{For example, in our notations we can take $\cQ_{\text{KWY}}= \cQ^{(\ell_+)}_1$.}
	
When calculating the variations in \eqref{QexactYM} we have set to zero total derivatives under the integral sign. Those total derivatives could give additional contributions in the presence of defect operators, such as monopoles, which introduce non-trivial boundary conditions for the fields at their insertion points. Consequently, the above result may have to be modified when defect operators are inserted in the path integral.
	
An immediate consequence of \eqref{QexactYM} is that for any $\cN=4$ theory on $S^3$, correlators of operators in the cohomology of $\cQ_{\beta}^C$ or $\cQ^H_{\beta}$ are independent of $g_{\text{YM}}$ in the absence of defect operators. In particular, for our theories, all correlators of the twisted Higgs branch operators constructed from \eqref{QQt}, or of the twisted Coulomb branch operators constructed from \eqref{hatPhi}, are independent of $g_{\text{YM}}$. The non-renormalization theorems of \cite{Intriligator:1996ex, Aharony:1997bx, Bullimore:2015lsa} make similar statements at the level of the chiral ring.
	
Let us now summarize the details of the localization of the $\cN=4$ vectormultiplet. Since we established that $S_{\text{YM}}$ can be used as a localizing term, the result can be entirely migrated from \cite{Kapustin:2009kz,Hama:2010av}. In our language, the fields \eqref{Vmul} in the vectormultiplet $\cV$ localize to 
\begin{align}
\cV\to \cV_{\text{loc}} = \{A_{\mu}^{\text{loc}}, \lambda_{a\dot b}^{\text{loc}}, \Phi_{\dot{a}\dot b}^{\text{loc}}, D_{ab}^{\text{loc}}\} \ec \label{VecLoc1}
\end{align}
where
\begin{align}
\Phi_{\dot{1}\dot 2}^{\text{loc}} = irD_{11}^{\text{loc}} = irD_{22}^{\text{loc}} = \frac{1}{r}\sigma \ecq A_{\mu}^{\text{loc}} = D_{12}^{\text{loc}}=\Phi_{\dot{1}\dot 1}^{\text{loc}} = \Phi_{\dot{2}\dot 2}^{\text{loc}} = \lambda_{a\dot b}^{\text{loc}} = 0 \ec \label{VecLoc2}
\end{align}
and $\sigma$ is a constant in the Cartan of the lie algebra $\mathfrak{g}$ parameterizing the different saddles of $S_{\text{YM}}$.  The action $S_{\text{YM}}$ itself, of course, vanishes when evaluated on the vectormultiplet localization locus \eqref{VecLoc2}:
\begin{align}
S_{\text{YM}}[\cV_{\text{loc}}(\sigma)] = 0 \ed
\end{align}
On the other hand, the hypermultiplet action \eqref{Shyper} becomes
\begin{align}
S_{\text{hyper}}[\cH,\cV_{\text{loc}}(\sigma)] &= \int d^3x \sqrt{g} \left[ \partial^{\mu}\tq^{a} \partial_{\mu} q_a - i\tpsi^{\dot{a}}\slashed{\nabla}\psi_{\dot{a}} + \frac{1}{r^2} \tq^{a} \left(\sigma^2+\frac{3}{4}\right)q_a + \frac{1}{r^2}\left(\tq^1\sigma q_2 - \tq^2\sigma q_1\right) \right.\notag\\
&\left. -\frac{i}{r}\left(\tpsi^{\dot{1}}\sigma\psi_{\dot{1}} -\tpsi^{\dot 2}\sigma\psi_{\dot 2}\right)\right] \ed \label{ShyperLoc3d}
\end{align}
Together with the contribution from the vectormultiplet 1-loop determinant, the path integral $\cI$ in \eqref{Ischematic} reduces to\footnote{Given a choice of a Cartan sub-algebra for the gauge group $G$ with generators $H^i$, we can decompose $\sigma=\sigma_i H^i$. If $\rho$ is a weight vector in a representation $\cR$ of $G$, then $\rho(\sigma)\equiv \rho^i\sigma_i$, and $\det_{\cR}f(\sigma) \equiv \Pi_{\rho\in\cR}f(\rho(\sigma))$. In the vectormultiplet determinant $\det{}'_{\text{adj}}\left[2\sinh(\pi\sigma)\right]$ in \eqref{vecloc} the product is only over the roots of $G$.}
\begin{align}
\cI = \frac{1}{|\cW|} \int_{\text{Cartan}} d\sigma \det{}'_{\text{adj}}\left[2\sinh(\pi\sigma)\right]\int D\cH e^{-S_{\text{hyper}}[\cH,\cV_{\text{loc}}(\sigma)]}(\cdots) \ec \label{vecloc}
\end{align}
where $|\cW|$ is the order of the Weyl group of the gauge group.  The prime over the determinant sign means we should restrict the action of $\sigma$ to the non-zero weights of the adjoint before taking the determinant.  Note that the determinant factor in \eqref{vecloc} is actually equal to the contribution of only an $\cN=2$ vectormultiplet. This is because the $\cN=4$ vectormultiplet decomposes into an $\cN=2$ vectormultiplet plus an adjoint chiral multiplet of R-charge $1$, and the 1-loop determinant of a chiral with precisely this R-charge is equal to $1$.
	
For each $U(1)$ factor in $G$, we can introduce an FI term \eqref{FIAction}, which leads to additional insertions in \eqref{vecloc} of
\begin{align}
e^{-S_{FI}} \to e^{-8\pi^2i r \zeta\sigma} \ec \label{FIloc}
\end{align}
where $\sigma$ in \eqref{FIloc} is understood to be the real scalar in the vectormultiplet that gauges the corresponding $U(1)$ factor. Real masses can be treated as follows.  For every Cartan generator of the flavor symmetry group we can couple the corresponding Abelian flavor current multiplet to a background vectormultiplet.  With our choice of matrices $h$ and $\bar{h}$, introducing a real mass parameter means giving an expectation value equal to $m$ to the background $\Phi_{\dot1\dot2}$ and to the corresponding components of $D_{ab}$ according to \eqref{PhiDBack}.  One can introduce as many real mass parameters as Cartan generators of the flavor symmetry algebra thus breaking the flavor symmetry to its Cartan subalgebra. At the level of localized action, turning on real masses corresponds to replacing:
 \es{sigmaReplacementMassive}{
  \sigma \to \sigma + r m \,,
 }
where $m$ is a mass matrix in the Cartan of the flavor symmetry group acting on $Q$ in the corresponding representation.  To be more precise, in the above expression, $\sigma$ acts only on gauge indices, while $m$ acts only on flavor indices.  In the notation of footnote~\ref{FlavorFootnote}, where ${\cal R}$ is viewed as a map from the gauge algebra into $\dim {\cal R} \times \dim {\cal R}$ hermitian matrices and ${\cal F}$ is defined as a map from the hypermultiplet flavor algebra into $\dim {\cal R} \times \dim {\cal R}$ hermitian matrices, one would write ${\cal R}(\sigma) \to {\cal R}(\sigma) + r {\cal F}(m)$ instead of \eqref{sigmaReplacementMassive}.  
	
\subsection{3d Gaussian theory coupled to a matrix model}
	
The expression \eqref{vecloc} for the path integral could be viewed as our final result.  Indeed, at fixed $\sigma$, the remaining path integral over the matter fields in $\cH$ is now easily calculable because the hypermultiplet action is quadratic.  Performing this integral gives a matrix model that depends on the precise operator insertions in \eqref{vecloc}.  We conclude that no further localization of the hypermultiplet is necessary for reducing the path integral \eqref{vecloc}, with any supersymmetric insertions, to a matrix integral.  (However, as we will see soon, we can do even better and replace the 3d Gaussian theory at fixed $\sigma$ by a very simple 1d Gaussian theory.)
	
Without any insertions, \eqref{vecloc} gives the $S^3$ partition function.  After integrating out the hypermultiplet fields using 
\begin{align}
Z_\sigma \equiv \int D\cH \,e^{-S_{\text{hyper}}[\cH,\cV_{\text{loc}}(\sigma)]} = \frac{1}{\det_{\cR}\left[2\cosh(\pi\sigma)\right]} \ec \label{Zsigma}
\end{align}
one obtains the KWY matrix model
\es{ZS3KWY}{
	Z_{S^3} = \frac{1}{|\cW|} \int_{\text{Cartan}} d\sigma \frac{\det{}'_{\text{adj}}\left[2\sinh(\pi\sigma)\right]}{\det_{\cR}\left[2\cosh(\pi\sigma)\right]} \ed 
}
	
The types of insertions $(\cdots)$ allowed in \eqref{vecloc} depend on the supercharge one chooses to localize with. As we saw, we can consider three possibilities: the supercharge $\cQ_{\text{KWY}}$ of \cite{Kapustin:2009kz}, the supercharge $\cQ^H_{\beta}$ in \eqref{QHloc}, or $\cQ^C_{\beta}$ defined in \eqref{QCloc}. The insertions $(\cdots)$ then must all lie either in the cohomology of $\cQ_{\text{KWY}}$, or in the cohomology of $\cQ^H_{\beta}$, or in the one of $\cQ^C_{\beta}$. Let us now discuss each of the three cases separately.

\subsubsection{Localizing with $\cQ_{\text{KWY}}$} 
	
The cohomology of $\cQ_{\text{KWY}}$ does not contain any local operators.  It contains, however, non-local operators such as the Wilson loop operators \eqref{wilson}.  Such a Wilson loop in representation $\cR_L$ of $G$ localizes to
\begin{align}
W_{\cR_L}\biggr|_{\text{loc}} \!\!\!\!= \tr_{\cR_L} e^{2\pi \sigma} \ed
\end{align}
These operators are not constructed from the hypermultiplet fields, so we can safely integrate those out using \eqref{Zsigma}, 
which gives 
\begin{align}
\langle W_{\cR_L}\rangle = \frac{1}{|\cW|} \int_{\text{Cartan}} d\sigma \frac{\det{}'_{\text{adj}}\left[2\sinh(\pi\sigma)\right]}{\det_{\cR}\left[2\cosh(\pi\sigma)\right]} \tr_{\cR_L} e^{2\pi \sigma}  \ed \label{ZKWY}
\end{align}
	
The modification of \eqref{ZS3KWY} that results from including real masses also allows for the calculation of integrated correlators of scalar operators in $\cN=2$ current multiplets associated with flavor symmetries \cite{Closset:2012vg}.

\subsubsection{Localizing with $\cQ^C_{\beta}$}
\label{LOCQC}

If we localize with $\cQ^C_{\beta}$, the allowed insertions include the twisted Coulomb branch operators discussed in Section \ref{QCoperators}. As shown there, with the exception of monopole operators, these are gauge invariant polynomials in the twisted field ${\Phi}(\varphi)$ defined in \eqref{hatPhi}.  On the localization locus \eqref{VecLoc2}, we have
\begin{align}
{\Phi}(\varphi) \biggr|_{\text{loc}} =  \frac{2 \sigma}{r} \ed
\end{align}
Because these operators are independent of $\cH$, we can again integrate out the hypermultiplet fields using \eqref{Zsigma}. We conclude that
\es{PolynCoulomb}{
	&\left\langle P_{1}\left({\Phi}(\varphi_1)\right) \cdots P_{n}\left({\Phi}(\varphi_n)\right) \right\rangle \\
	&\qquad\qquad\qquad\qquad= \frac{1}{Z_{S^3}|\cW|} \int_{\text{Cartan}} d\sigma \frac{\det{}'_{\text{adj}}\left[2\sinh(\pi\sigma)\right]}{\det_{\cR}\left[2\cosh(\pi\sigma)\right]} P_{1}\left(\frac{2\sigma}{r} \right)\cdots P_{n}\left(\frac{2\sigma}{r}\right) \ec
}
where $P_{i_n}$ are any polynomials and where we divided by $Z_{S^3}$ so that $\langle 1 \rangle = 1$. The treatment of defect operators in the cohomology of $\cQ^C_{\beta}$ is left for future work.
	
\subsubsection{Localizing with $\cQ^H_{\beta}$}
	
Finally, let us discuss the possible insertions when localizing with $\cQ^H_{\beta}$. They are gauge invariant polynomials constructed from $Q(\varphi)$ and $\widetilde{Q}(\varphi)$, as discussed in Section \ref{QHoperators}.  We conclude that the correlators of such operators ${\cO}_i(\varphi)$ can be calculated using \eqref{vecloc}:
\begin{align}
\langle  {\cO}_1(\varphi_1) \cdots {\cO}_n(\varphi_n) \rangle 
&= \frac{1}{Z_{S^3}|\cW|} \int_{\text{Cartan}} d\sigma \frac{\det{}'_{\text{adj}}\left[2\sinh(\pi\sigma)\right]}{\det_{\cR}\left[2\cosh(\pi\sigma)\right]} \langle {\cO}_1(\varphi_1) \cdots {\cO}_n(\varphi_n) \rangle_{\sigma} \ec  \label{higgscorr}
\end{align}
where 
\begin{align}
\langle {\cO}_1(\varphi_1) \cdots {\cO}_n(\varphi_n) \rangle_{\sigma} \equiv \frac{1}{Z_\sigma}\int D\cH e^{-S_{\text{hyper}}[\cH,\cV_{\text{loc}}(\sigma)]} \, {\cO}_1(\varphi_1) \cdots {\cO}_n(\varphi_n)  \ed \label{corrsigma}
\end{align}
	
Let us now explain how \eqref{higgscorr} is to be evaluated. The correlation functions \eqref{corrsigma} in the theory governed by the action $S_{\text{hyper}}[\cH,\cV_{\text{loc}}(\sigma)]$ are calculable since this action is quadratic.  Indeed, they are given by simply summing over all Wick contractions with the Green's function $G_{\sigma}(\vphi_1 - \vphi_2) \equiv \langle Q (\varphi_1) \tQ (\varphi_2) \rangle_{\sigma}$.  This Green's function can be calculated explicitly from \eqref{ShyperLoc3d}, and as shown in Appendix \ref{propagator}: 
\es{GotG}{
	G_{\sigma}(\vphi_1 - \vphi_2) \equiv  \langle Q (\varphi_1) \tQ (\varphi_2) \rangle_{\sigma} =   -\frac{\sgn(\vphi_1 - \vphi_2) + \tanh(\pi \sigma)}{8\pi r}  e^{-\sigma (\vphi_1 - \vphi_2)} \ec
}
where $\sigma$ is taken to be in the representation ${\cal R}$.  In the limit of coincident points, we can take as a definition that $\sgn(0) = 0$ in \eqref{GotG}.

To summarize, the correlation functions \eqref{higgscorr} reduce to the KWY matrix model with products of the propagator \eqref{GotG} inserted according to Wick's theorem applied to \eqref{corrsigma}. As discussed in Section \ref{QHoperators}, the Wilson loops \eqref{wilson} are also $\cQ^H_{\beta}$-closed so we can insert them as well. Correlators of only Wilson loops will of course be the same as in the KWY model \eqref{ZKWY}. Now, however, we have the possibility of calculating correlators of both Wilson loops and local operators, similarly to what was done in \cite{Giombi:2009ds,Giombi:2012ep} for 4d $\cN=4$ Yang-Mills theory. 
	
\subsection{1d Gaussian theory for twisted Higgs branch operators}\label{NoLocArg}
	
As we have seen, the solution for correlators of twisted Higgs branch operators can be obtained from a 3d Gaussian theory coupled to a matrix model. One may wonder, however, if, because these operators are restricted to lie on an $S^1$, their correlators can be described by a 1d quantum field theory that is coupled to the same matrix model. As advertised in the title of this paper and briefly reviewed in the Introduction, the answer is yes. In particular, the Green's function \eqref{GotG} can be obtained from the 1d quadratic action
\begin{align}
	S_\sigma [Q, \tQ] = - 4 \pi r \int_{-\pi}^{\pi} d\varphi \left( \tQ\partial_{\varphi} Q + \tQ \sigma Q \right)\ec \label{Shyper1d}
\end{align}
and so one can alternatively represent the 1d correlator given in \eqref{corrsigma} as
\begin{align}
\langle {\cO}_1(\varphi_1) \cdots {\cO}_n(\varphi_n) \rangle_{\sigma} = \frac{1}{Z_\sigma}\int DQ D\tQ e^{-S_\sigma[Q,\tQ]} \, {\cO}_1(\varphi_1) \cdots {\cO}_n(\varphi_n) \ed \label{3dTo1d}
\end{align} 
To check that the path integral \eqref{3dTo1d} with the 1d action \eqref{Shyper1d} can be used to calculate the correlators \eqref{corrsigma}, we need to check two things:
\begin{enumerate}
	\item We should check that \eqref{3dTo1d} is normalized so that $\langle 1 \rangle = 1$.  In other words, we must have
	\es{GaussianIntegral}{
		\int DQ D\tQ e^{-S_\sigma[Q,\tQ]} = Z_\sigma  \,. 
	}
	Indeed, we can check that if we assume that $Q$ and $\tQ$ are related by a reality condition (to be discussed in more detail later), such that the path integral \eqref{GaussianIntegral} is over half the number of complex integration variables than given by arbitrary complex fields $Q(\vphi)$ and $\tQ(\vphi)$, we have:
	\es{detRHS}{
		\int DQ D\tQ\, e^{-S_\sigma[Q,\tQ]} = \frac{1}{\det (\partial_\vphi + \sigma)} 
		= \frac{1}{\det_{\cal R} \prod_{n \in \Z+ \frac 12} (i n + \sigma)}  = \frac{1}{\det_{\cal R} \left[ 2 \cosh (\pi \sigma) \right]} \,,
	}
	where in evaluating the product over $n$ we used zeta-function regularization.  This expression indeed matches the formula for $Z_\sigma$ given in \eqref{Zsigma}.  
		
	The fact that the path integral in \eqref{3dTo1d} is over a middle-dimensional integration cycle in the space of complex fields $Q$ and $\tQ$ may seem mysterious at this point, but we should point out that it is absolutely necessary if \eqref{3dTo1d} were to make sense:  without such a choice of integration contour the path integral in \eqref{3dTo1d} would not converge.

	\item The second thing we need to check is that \eqref{GotG} is indeed the Green's function following from \eqref{Shyper1d}.  Such a Green's function would have to obey the differential equation   
	\es{GDiff}{
		(\partial_\varphi + \sigma) G_{\sigma}(\varphi) = -\frac{1}{4\pi r} \delta(\varphi) \ecq
	}
	and it should be an anti-periodic function of $\vphi$, as required by the anti-periodicity of the twisted fields $Q(\vphi)$ and $\tQ(\vphi)$. Indeed, it is easy to see that \eqref{GotG} obeys these properties.  
	\end{enumerate}

While so far we simply guessed the 1d action \eqref{Shyper1d}, let us now explain how to obtain it directly from a supersymmetric localization computation.  Additionally, we will provide a derivation of the middle-dimensional integration cycle in \eqref{3dTo1d}.

\subsection{1d theory from localization of hypermultiplet}

Let us now provide a derivation of \eqref{3dTo1d} using supersymmetric localization of the hypermultiplet.  Because  we have already argued for \eqref{Shyper1d} in a round-about way, the reader with applications in mind can safely skip to Section~\ref{PROPERTIES}. 
	
Our starting point is the remaining path integral over the hypermultiplet in \eqref{vecloc} 
\begin{align}
\cI_{\cH} = \int D\cH e^{-S_{\text{hyper}[\cH,\sigma]}} (\cdots)\ec \label{IH}
\end{align}
obtained after the vectormultiplet has been localized, and where $(\cdots)$ are $\cQ_{\beta}^H$-closed operators. We now wish to localize \eqref{IH}, by deforming $S_{\text{hyper}}[\cH,\cV_{\text{loc}}(\sigma)]$ with a $\cQ^H_{\beta}$-exact term constructed from the fields in $\cH$.

\subsubsection{Off-shell closure}\label{OFFSHELLCL}
	
For localization it is important that the algebra generated by the localizing supercharge $\cQ$ closes off-shell. Otherwise, a $\cQ$-exact deformation $\delta S_t = t V$ will generally not be $\cQ$-closed. Instead, $[\cQ, V]$ will contain non-vanishing factors that include equations of motion operators, thus making the path integral depend on the deformation parameter $t$ and spoiling the localization argument. In our case, while it may not be possible to close the full $\cN=4$ algebra off-shell, the sub-algebra generated by $\cQ^H_{\beta}$ can certainly be made to do so. We will first describe the general procedure and then apply it to $\cQ^H_{\beta}$.
	
Let us first define a new hypermultiplet $\cH'$ by supplementing $\cH$ with new auxiliary fields:
\begin{align}
\cH' = (q_a , \tq^a, \psi_{\alpha\dot a}, \tpsi_{\alpha\dot a}, G_a, \widetilde{G}^a) \ed \label{Hpmul}
\end{align}
In \eqref{Hpmul}, $G_a$ and $\widetilde{G}^a$ are new complex scalar fields in the fundamental and anti-fundamental of $\mathfrak{su}(2)_H$ respectively (as well as $\cR$ and $\overline\cR$ representations of $G$ in the gauged case), and whose reality condition is: 
\begin{equation}
\label{GReal}
\widetilde{G}^a = (G_a)^*\, .
\end{equation}
The action of $\cH'$ is defined as
\begin{align}
S'_{\text{hyper}}[\cH', \cV] = S_{\text{hyper}}[\cH, \cV] + \int d^3x\sqrt{g} \widetilde{G}^a G_a \ed \label{Sphyper}
\end{align}
The modifications \eqref{Hpmul} and \eqref{Sphyper} are harmless, since by integrating out $G^a$ and $\widetilde{G}^a$ we recover the original theory. The new action \eqref{Sphyper}, however, admits additional supersymmetries that act only on the fermion matter fields and the new auxiliary fields. The modified SUSY transformations are given by
\begin{alignat}{3}
\delta_{\xi,\nu} \psi_{\dot{a}} &= i\gamma^{\mu}\xi_{a\dot{a}} \cD_{\mu}q^a + i\xi'_{a\dot{a}}q^a - i\xi_{a\dot{c}}\Phi^{\dot{c}}{}_{\dot{a}}q^a + i\nu^a{}_{\dot a} G_a \ecq\,\, &\delta_{\xi,\nu} G^a &= i\nu^{a\dot a}\Psi^{\text{eom}}_{\dot a} \ec \label{xinuTrans1}\\
\delta_{\xi,\nu} \tpsi_{\dot{a}} &= i\gamma^{\mu}\xi_{a\dot{a}} \cD_{\mu}\tq^a + i\tq^a\xi'_{a\dot{a}} + i\xi_{a\dot{c}}\tq^a\Phi^{\dot{c}}{}_{\dot{a}} + i\nu^a{}_{\dot{a}} \widetilde{G}_a \ecq\,\,  &\delta_{\xi,\nu} \widetilde{G}^a &= -i\nu^{a\dot a}\widetilde{\Psi}^{\text{eom}}_{\dot a} \ec \label{xinuTrans2}
\end{alignat}
where $\Psi^{\text{eom}}$ and $\widetilde{\Psi}^{\text{eom}}$  were defined in \eqref{eompsi} and \eqref{eompsit}. In \eqref{xinuTrans1} and  \eqref{xinuTrans2} we introduced an arbitrary auxiliary spinor $\nu_{\alpha a\dot a}$ parameterizing the new symmetry, and it is taken to transform in the bi-fundamental of $\mathfrak{su}(2)_H\oplus\mathfrak{su}(2)_C$. 
	
Given a Killing spinor $\xi_{a\dot a}$, it is then sometimes possible to close the sub-algebra generated by $\delta_{\xi,\nu}$  off-shell on $\cH'$. This is done by tuning the value of $\nu_{a\dot a}$ as a function of $\xi_{a\dot a}$ to cancel the equations of motion that appear in $\delta_{\xi,\nu}^2\cH'$, as written in \eqref{psiclosure} and \eqref{tpsiclosure}. The resulting constraints on $\nu_{a\dot a}$ that ensure this cancellation are given by \footnote{To close more than one supersymmetry off-shell, say $\delta_{\xi, \nu}$ and $\delta_{\tilde{\xi}, \tilde{\nu}}$, we would have additional constraints on $\nu$ and $\tilde{\nu}$ from imposing the closure of the $\{\delta_{\xi, \nu},\delta_{\tilde{\xi}, \tilde{\nu}}\}$ transformation. The entire $\cN=4$ algebra cannot be closed off-shell in this way.}
\begin{align}
\xi^{\alpha c}{}_{\dot a}\xi_{\beta c \dot b} = \nu^{\alpha c}{}_{\dot b}\nu_{\beta c \dot a} \ecq \xi_a{}^{\dot c}\nu_{b \dot c} = 0 \ecq \xi_{(a}{}^{\dot c}\slashed{\nabla}\xi_{b)\dot c} = \frac{3i}{2} \nu_{(a}{}^{\dot c}\slashed{\nabla}\nu_{b)\dot c} \ed \label{nucond}
\end{align}
The constraints \eqref{nucond} on $\nu_{a\dot a}$ have generally many solutions. For the Killing spinor $\xi=\xi^H_{\beta}$, which was defined in \eqref{xiQ1Q2} and generates the supersymmetry we use for localization, a convenient solution of \eqref{nucond} is 
\begin{align}
\nu_{a\dot a} = (\xi^H_{-\beta})_{a\dot a} \ed \label{nuchoice}
\end{align}

\subsubsection{1d action from BPS Equations}
	
We are now ready to show that the 1d action $S_\sigma[Q,\tQ]$ follows from evaluating the 3d action $S_{\text{hyper}}'[\cH', \cV_\text{loc}(\sigma)]$ on bosonic $\cQ^H_{\beta}$-invariant field configurations. These configurations are obtained by setting the fermions and their $\cQ^H_{\beta}$ variations to zero. We will refer to them as the BPS locus and the corresponding equations as the BPS equations. Supplementing the BPS equations by reality conditions of the bosonic fields is equivalent to intersecting the BPS locus with the real middle-dimensional contour in the space of 3d bosonic fields. We call this intersection the bosonic localization locus throughout this paper. Note that, in general, the bosonic localization locus may not be the same as the full localization locus defined as the space of zero modes of $V$, which may also contain fermionic directions.   
	
An interesting fact, which is not crucial for our derivation, is that the BPS equations alone, without the reality conditions, reduce the 3d action to 1d. Namely, the full 3d action on $S^3$ evaluated at the BPS locus $\lambda_{a\dot a}=\psi_{\dot a}=\tpsi_{\dot a}=\delta_{\xi}\lambda_{a\dot a}=\delta_{\xi,\nu}\psi_{\dot a}= \delta_{\xi,\nu}\tpsi_{\dot a}=0$ is given by a 1d action on a great circle of $S^3$. This reduction is shown in full generality in Appendix \ref{BPSEQUATIONS}. Since we have already localized the vectormultiplet, in this section we show a slightly simpler result that the hypermultiplet action on the $\cV_\text{loc}(\sigma)$ background reduces to the 1d action after imposing the BPS equations. The reality conditions, however, will be absolutely crucial for us later. In the next section they will be used to achieve two goals. One is that only the intersection of the BPS locus by the real cycle is parametrized by fields $Q(\varphi)$ and $\tQ(\varphi)$ appearing in the 1d action, thus ensuring that there are no bosonic flat directions. Another is that the hypermultiplet reality conditions determine a middle-dimensional contour of integration in the space of complex fields of this 1d theory.
	
Note also that the results of this subsection do not really depend on the precise choice of the localizing term $V$. Since $V$ is $\cQ^H_\beta$-exact, it automatically vanishes on the BPS locus, so the localization answer includes an integral over the BPS locus. By properly choosing $V$, one can actually ensure that it does not have any other zeros besides the BPS locus.
	
The fact that we take $\beta\neq 0$ now plays an important role. Indeed, the $\cQ^H_{\beta}$-invariant field configurations must also be invariant under $(\cQ^H_{\beta})^2$, where
\begin{align}
(\cQ^H_{\beta})^2 = \beta\{\cQ_1^H,\cQ_2^H\} = \frac{4i\beta}{r} (P_\tau + R_C) \equiv \frac{4i\beta}{r}\cZ \ed \label{cZS3def}
\end{align}
It follows that operators with zero $R_C$ charge, such as the action, must be $\tau$-independent.\footnote{Generally, $\tau$-independence would only follow from \eqref{cZS3def} up to a gauge transformation. However, in the treatment of this section, even non-gauge invariant fields are $\tau$-independent since we have already localized the vectormultiplet. This is because the gauge parameter $\Lambda$ in \eqref{GotLambda} evaluated for $\xi=\tilde\xi=\xi_\beta^H$ localizes to zero on \eqref{VecLoc2}.} The action can therefore be dimensionally reduced in the $\tau$ direction leaving us with a theory on the disk $D^2$ with metric
\begin{align}
ds^2(D^2) = r^2\left(d\theta^2+ \sin^2 \theta \,d\varphi^2\right) \ed
\end{align} 
The second ingredient in the derivation comes from  solving the BPS equations in the background \eqref{VecLoc2} of the localized vectormultiplet
\begin{align}
\delta_{\xi^H_{\beta}, \nu} \psi_{\dot a}\biggr|_{\cV=\cV_{\text{loc}}} = \delta_{\xi^H_{\beta}, \nu} \tpsi_{\dot a}\biggr|_{\cV=\cV_{\text{loc}}} = 0 \ed \label{BPSequationsLoc}
\end{align}
As evident from \eqref{xinuTrans1} and \eqref{xinuTrans2}, one can solve \eqref{BPSequationsLoc} for the auxiliary fields $G_a$ and $\widetilde{G}_a$ in terms of functions linear in $q_a$ and $\tq_a$,
\begin{align}
G_a\bigr|_{(\text{BPS})} &= f_{\nu}(q_a, \sigma) \ecq  \widetilde{G}_a\bigr|_{(\text{BPS})} = \tilde{f}_{\nu}(\tq_a,\sigma) \ec \label{Gsolschematic}
\end{align}
where explicit expressions for $f_{\nu}$ and $\tilde{f}_{\nu}$ are given in Appendix \ref{BPSEQUATIONS}, and we have indicated that they depend on the choice of auxiliary spinor $\nu_{a\dot a}$. Without imposing reality conditions on the fields,  $\tau$-independence and the solutions \eqref{Gsolschematic} constitute the full set of constraints that follow from the BPS equations \eqref{BPSequationsLoc}. 
	
To summarize, the action $S_{\text{hyper}}'[\cH',\cV_\text{loc}(\sigma)]$ evaluated on $\cQ^H_{\beta}$-invariant configurations is obtained by first performing dimensional reduction on $\tau$, and then plugging in the solutions \eqref{Gsolschematic}. Using \eqref{nucond}, the result can be shown to be independent of the choice of $\nu_{a\dot a}$, even though the solutions for the auxiliary fields \eqref{Gsolschematic} depend on it. After some algebra, one finds that the action evaluates to a total derivative on $D^2$:
\begin{align}
S_{\text{hyper}}'[\cH']\biggr|_{\cQ_{\beta}^H-\text{BPS}} = 2\pi\int_{D^2} d^2x\sqrt{g_{D^2}} \nabla_{\mu} K^{\mu} \ec \label{DKloc}
\end{align}
where 
\begin{align}
K^{\theta} = \tq_a \cA^{ab}_{1}\partial_{\varphi} q_b + \tq_a \cB^{ab}_{1} q_b \ecq K^{\varphi} = \tq_a \cA^{ab}_{2}\partial_{\theta} q_b + \tq_a \cB^{ab}_{2} q_b \ec
\end{align}
and
\begin{align}
\cA_1^{ab} &= \frac{1}{r\sin(\theta)}\begin{pmatrix}
-\cos(\varphi)\sin(\theta)-1 & -\sin(\varphi)\sin(\theta) \\ -\sin(\varphi)\sin(\theta) & \cos(\varphi)\sin(\theta)-1
\end{pmatrix} \ecq \cA_2^{ab} = - \cA_1^{ab} \ec \notag\\
\cB_1^{ab} &= \frac{1}{2r}\begin{pmatrix}
\sin(\varphi) - 2(\cos(\varphi)+\sin(\theta))\sigma & -\cos(\varphi)-\sin(\theta)-2\sin(\varphi)\sigma \\ -\cos(\varphi)+\sin(\theta)-2\sin(\varphi)\sigma & -\sin(\varphi) + 2(\cos(\varphi)-\sin(\theta))\sigma
\end{pmatrix} \ec\label{ABmatrices}\\
\cB_2^{ab} &= \frac{\cot(\theta)}{2r}\begin{pmatrix}
\cos(\varphi) + 2\sin(\varphi)\sigma & +\sin(\varphi)-2\cos(\varphi)\sigma \\ \sin(\varphi)-2\cos(\varphi)\sigma & -\cos(\varphi) - 2\sin(\varphi)\sigma
\end{pmatrix} \ed\notag
\end{align}
	
One can verify that the boundary term left from \eqref{DKloc} is precisely the 1d action \eqref{Shyper1d}, which completes the derivation. As already mentioned before, in Appendix \ref{BPSEQUATIONS} we show a slightly more general result. In particular, imposing $\delta_{\xi}\lambda_{a\dot a}=\delta_{\xi,\nu}\psi_{\dot a}= \delta_{\xi,\nu}\tpsi_{\dot a}=0$, without assuming that $\cV$ is initially set to its localization locus, is sufficient in order to reduce the full non-Gaussian hypermultiplet action $S_{\text{hyper}}[\cH,\cV]$ to
\begin{align}
\label{S1d}
S=-4\pi r\int d\varphi\, \tQ \widehat{D}_{\varphi} Q \ec
\end{align}
where $\widehat{D}_{\varphi} = \partial_{\varphi} - i (A_{\varphi}+ir\Phi_{\dot{1}\dot 2})$ is the $\frac{1}{2}$-BPS connection from which Wilson loops \eqref{wilson} are constructed. The procedure outlined in this section is analogous to the one in the work \cite{Pestun:2009nn} of Pestun on localization of 4d maximally supersymmetric Yang-Mills on $S^4$ to an $S^2$.\footnote{Note that the localization in \cite{Pestun:2009nn} also allows for insertions of local operators on $S^2$ \cite{Giombi:2009ds, Giombi:2012ep}. Those operators are ``twisted-translated" operators that were first defined by Drukker and Plefka \cite{Drukker:2009sf}. It would be fascinating to generalize \cite{Pestun:2009nn} to $\cN=2$ theories in a way that allows insertions of the more general class of twisted operators found in \cite{Beem:2013sza}.}

\subsubsection{Hypermultiplet 1-loop determinant}
\label{1LoopSec}

Under the assumption that a well-defined localizing term exists, we know by now that the Gaussian 3d theory \eqref{ShyperLoc3d} localizes to the Gaussian 1d theory \eqref{Shyper1d}, up to the presence of a possible 1-loop determinant. The well-definiteness of the localizing term means that this 1-loop determinant is finite and non-zero. We also loosely describe this situation by saying that there are no flat directions. In our case, this means that the localization locus has no fermionic directions, and the bosonic directions are parametrized by the 1d fields $Q(\varphi)$ and $\tQ(\varphi)$ appearing in the action $S_\sigma[Q,\tQ]$ given in \eqref{Shyper1d} and obeying an extra reality condition which relates $Q(\varphi)$ and $\tQ(\varphi)$. Therefore, the term $e^{-S_\sigma}$, multiplied by a possible 1-loop determinant, provides a good integration measure over the localization locus, which is what we mean by saying that there are no flat directions.
	
Being more precise, the statement is that:
\begin{align}
\label{LocEq}
\int D\cH\, e^{-S_{\text{hyper}}} \left( \text{$\cQ^H_\beta$-closed insertions} \right)=\int DQ D\tQ\, e^{-S_\sigma[Q,\tQ]} \Delta(\sigma, r)\left( \text{$\cQ^H_\beta$-closed insertions} \right),
\end{align}
where $\Delta(\sigma, r)$ is a possible 1-loop determinant for fluctuations around the localization locus. It is clear that it can only be a function of $\sigma$ and $r$, (as well as masses and FI parameters if they are present in the theory) because these are the only parameters appearing in either the 3d action \eqref{ShyperLoc3d} or the 1d action \eqref{Shyper1d}.   Since both 1d and 3d theories are Gaussian, we can find $\Delta(\sigma, r)$ by simply computing partition functions for theories on the left and on the right in \eqref{LocEq} with no $\cQ^H_\beta$-closed insertions whatsoever.  We performed this computation in \eqref{GaussianIntegral}--\eqref{detRHS} and noticed that the results agree.  We therefore conclude that 
\es{DeltaOne}{
	\Delta(\sigma,r)=1 \,.
	}
	
To finish the argument, we need to construct the localizing term and show that there are indeed no flat directions. We will construct the localizing term below and give some evidence that it has the required properties, in particular that there are no fermionic flat directions and the 1-loop determinant is non-zero and finite. Unfortunately, we do not have a completely rigorous proof of the last statement.

\textbf{The localizing term}
	
In the localization, it is useful to organize fields into the multiplets of $\cQ_\beta^H$. For that purpose, define fields: 
\begin{align}
\label{etaUps}
\eta_a&=(\xi_\beta^H)_a^{\,\,\dot a}\psi_{\dot a}\, ,\quad \Upsilon_a= i \mu_a^{\,\, b}\nu_b^{\,\,\dot a}\psi_{\dot a}\, ,\cr
\widetilde\eta^a&=(\xi_\beta^H)^{a\dot a}\widetilde\psi_{\dot a}\, ,\quad \widetilde\Upsilon^a =i \mu^{ab}\nu_b^{\,\,\dot a}\widetilde\psi_{\dot a}\, ,
\end{align}
where, as before, $\nu\equiv \xi_{-\beta}^H$, and the matrix $\mu_{ab}$ is defined as in Appendix~\ref{BPSEQUATIONS}:
\begin{equation}
\mu_{ab}=(\xi_\beta^H)_a^{\,\,\dot c}(\xi_\beta^H)_{b\dot c} = -\nu_a^{\,\,\dot c}\nu_{b\dot c}.
\end{equation}
The choice of $\xi_\beta^H$ breaks spacial and R-symmetries of the model. The bosonic symmetry algebra $\mathfrak{su}(2)_\ell\oplus \mathfrak{u}(1)_\ell \oplus \mathfrak{su}(2)_r \oplus \mathfrak{u}(1)_r$ is broken down to $\mathfrak{u}(1)_\varphi \oplus \mathfrak{u}(1)_\tau$. Here $\mathfrak{u}(1)_\varphi$ represents twisted rotations of the circle generated by $\widehat{P}^H_\varphi=P_\varphi + R_H$, where $R_H=\frac{1}{2}\left(R_\ell + R_r\right)$, and $R_\ell$ and $R_r$ are, respectively, the generators of $\mathfrak{u}(1)_\ell$ and $\mathfrak{u}(1)_r$. The $\mathfrak{u}(1)_\tau$ represents twisted $\tau$-rotations generated by $(\cQ_\beta^H)^2\propto P_\tau + R_C$, where $R_C=\frac{1}{2}(R_\ell- R_r)$. The new fermions $\eta_a, \Upsilon_a$ transform as scalars under $\mathfrak{u}_\tau(1)$, while for $\mathfrak{u}(1)_\varphi$:
\begin{equation}
[\widehat{P}^H_\varphi, \eta_a]= i\partial_\varphi \eta_a + \frac{1}{2}h_a^{\,\, b}\eta_b,
\end{equation}
and analogously for $\Upsilon_a$. For conformal theories, $\mathfrak{u}(1)_\varphi$ is enhanced to the algebra $\mathfrak{su}(2)_{S^1}\subset \mathfrak{so}(4)\oplus\mathfrak{sp}(4)$ that acts by twisted conformal transformations on the circle, and the fields $\eta_a, \Upsilon_a$ transform in the spin-$1/2$ representation of this $\mathfrak{su}(2)_{S^1}$.

The matrix $\mu_{ab}$ has the properties:
\begin{equation}
\det \mu_{ab}=\beta^2\cos^2\theta \equiv \mu\,,\quad \mu_{ab} \mu^{bc}= \mu\delta_a^c \,.
\end{equation}
The fields $q_a$ and $\eta_a$, as well as $\widetilde{q}^a$ and $\widetilde\eta^a$, form multiplets of $\delta_{\xi_\beta^H}$:
\es{qeta}{
	\delta_{\xi_\beta^H} q_a &= \eta_a\, ,\quad \delta_{\xi_\beta^H} \eta_a = \frac{\beta}{r}\partial_\tau q_a\,,\cr
	\delta_{\xi_\beta^H} \widetilde{q}^a &= \widetilde\eta^a\, ,\quad \delta_{\xi_\beta^H} \widetilde\eta^a = \frac{\beta}{r}\partial_\tau \widetilde{q}^a\,.\cr
	}
We then complete $\Upsilon_a$ and $\widetilde\Upsilon^a$ into multiplets:
\es{UpsilonH}{
\delta_{\xi_\beta^H} \Upsilon_a &= H_a\, ,\quad \delta_{\xi_\beta^H} H_a = \frac{\beta}{r}\partial_\tau\Upsilon_a\,,  \cr
	\delta_{\xi_\beta^H} \widetilde\Upsilon^a &= \widetilde{H}^a\, ,\quad \delta_{\xi_\beta^H} \widetilde{H}^a = \frac{\beta}{r}\partial_\tau\widetilde\Upsilon_a\,,\cr
}
where the new bosonic fields $H_a$ and $\widetilde{H}^a$ are explicitly given by:
\begin{align}
\label{Hs}
H_a &= - \mu G_a - \mu_a^{\,\, b}\nu_b^{\,\,\dot a}\gamma^\mu(\xi_\beta^H)_{c\dot a}\partial_\mu q^c - \mu_a^{\,\, b}\nu_b^{\,\,\dot a}(\xi'^H_\beta)_{c\dot a}q^c + \mu_a^{\,\, b}\nu_b^{\,\,\dot a}(\xi_\beta^H)_{c\dot c}\Phi^{\dot c}_{\ \dot a}q^c,\cr
\widetilde{H}^a &= -\mu\widetilde{G}^a + \mu^a_{\,\, b}\nu^{b\dot a}\gamma^\mu(\xi_\beta^H)_{c\dot a}\partial_\mu \widetilde{q}^c + \mu^a_{\,\, b}\nu^{b\dot a}(\xi'^H_\beta)_{c\dot a}\widetilde{q}^c - \mu^{ab}\nu_b^{\ \dot a}(\xi_\beta^H)_{c\dot c}\Phi^{\dot c}_{\ \dot a}\widetilde{q}^c.
\end{align}
Recall that $\Phi_{\dot a\dot b}$ has only one non-zero component $\Phi_{\dot 1\dot2}=\Phi_{\dot2\dot1}=\sigma/r$.
	
We can group the fields into four sets: $X_0=(q_a,\, \widetilde{q}^a)^t$, $X_1=(\Upsilon_a,\, \widetilde\Upsilon^a)^t$, $X_0' = (\eta_a,\, \widetilde\eta^a)^t$ and $X_1'=(H_a,\, \widetilde{H}^a)^t$. Now the multiplet structure is very simple:
\begin{align}
\delta_{\xi_\beta^H} X_0 &= X_0'\, ,\cr
\delta_{\xi_\beta^H} X_1 &= X_1'\, .
\end{align}
However, the conjugation property of the new bosonic fields $X_0=(q_1, q_2, \widetilde{q}^1, \widetilde{q}^2)^t$ and $X_1'=(H_1, H_2, \widetilde{H}^1, \widetilde{H}^2)^t$ (we suppress gauge and flavor indices) is more complicated, as it follows from \eqref{RealqAPhD} and \eqref{GReal}:
\begin{align}
\label{conjH}
(X_1')^* &= D_{11}X_1' + D_{10}X_0\,,\cr
(X_0)^* &= D_{11}X_0\,.
\end{align}
Here,
\begin{align}
\label{operatorsD}
D_{11}=\left(\begin{matrix}
0_2 & 1_2\cr
1_2 & 0_2
\end{matrix} \right)\, ,\quad D_{10}=\left(\begin{matrix}
0_2 & \widehat{D}\cr
-\widehat{D} & 0_2
\end{matrix} \right)\, ,\quad \widehat{D}=2\frac{\beta^2}{r}\left(\begin{matrix}
-\dd_- & \dd_+\cr
\dd_+ & \dd_-
\end{matrix} \right)-2\mu\frac{\sigma}{r}\left(\begin{matrix}
1 & 0\cr
0 & 1
\end{matrix} \right)\, ,
\end{align}
where $0_2$ and $1_2$ represent zero and unit $2\times 2$ matrices respectively, and the differential operators $\dd_\pm$ on the disk parameterized by $z_2 = \sin\theta e^{i\varphi}$ are given by:
\begin{align}
\dd_+ =\cos^2\theta\left(\frac{\cos\varphi}{\cos\theta}\frac{\partial}{\partial\theta}-\frac{\sin\varphi}{\sin\theta}\frac{\partial}{\partial\varphi}\right)=\cos^2\theta \left(\frac{\partial}{\partial z_2}+\frac{\partial}{\partial\bar z_2}\right)\, ,\\ 
\dd_- =\cos^2\theta\left(\frac{\sin\varphi}{\cos\theta}\frac{\partial}{\partial\theta}+\frac{\cos\varphi}{\sin\theta}\frac{\partial}{\partial\varphi}\right)= i\cos^2\theta\left(\frac{\partial}{\partial z_2}-\frac{\partial}{\partial\bar z_2} \right)\, .
\end{align}
These differential operators are globally well-defined on $S^3$ as well.
	
Let us define a quantity which we call the canonical localizing fermion:
\begin{equation}
\label{LocPsi}
\Psi =\int_{S^3}{\rm Vol} \sum_{a=1,2}\left[\eta_a \left(\delta_{\xi_\beta^H}\eta_a\right)^* + \tilde\eta^a \left(\delta_{\xi_\beta^H}\tilde\eta^a\right)^* + \Upsilon_a \left(\delta_{\xi_\beta^H}\Upsilon_a\right)^* + \tilde\Upsilon^a \left(\delta_{\xi_\beta^H}\tilde\Upsilon^a\right)^*\right].
\end{equation}
This can be written as:
\begin{align}
\Psi = \int_{S^3}{\rm Vol}\, \left(\begin{matrix}X_0^{'t} & X_1^t\end{matrix} \right) \left(\begin{matrix}
D_{00} & 0\cr
D_{10} & D_{11}
\end{matrix} \right)\left(\begin{matrix}
X_0\cr
X_1'
\end{matrix}\right),
\end{align}
where we introduced one more notation:
\begin{equation}
D_{00}=\frac{\beta}{r}D_{11}\partial_\tau\, .
\end{equation}
The canonical localizing term is then defined as
\begin{equation}
\label{LocV}
V = \delta_{\xi_\beta^H} \Psi = \int_{S^3}{\rm Vol}\, \left(\begin{matrix}X_0^t & X_1^{'t}\end{matrix} \right) \widehat\Delta_b\left(\begin{matrix}
X_0\cr
X_1'
\end{matrix}\right)+\int_{S^3}{\rm Vol}\,\left(\begin{matrix}X_0^{'t} & X_1^t\end{matrix} \right) \widehat\Delta_f\left(\begin{matrix}
X_0'\cr
X_1
\end{matrix}\right).
\end{equation}
We could also introduce a notation:
\begin{equation}
\langle A, B \rangle \equiv \int_{S^3}{\rm Vol}\, A^t B.
\end{equation}
In terms of this inner product, the localizing term is:
\begin{equation}
V=\left\langle \left(\begin{matrix}
X_0\cr
X_1'
\end{matrix}\right), \widehat\Delta_b\left(\begin{matrix}
X_0\cr
X_1'
\end{matrix}\right) \right\rangle + \left\langle \left(\begin{matrix}
X_0'\cr
X_1
\end{matrix}\right), \widehat\Delta_f\left(\begin{matrix}
X_0'\cr
X_1
\end{matrix}\right) \right\rangle\, .
\end{equation}
	
If we naively compute the above expression, the bosonic/fermionic operators $\widehat\Delta_{b,f}$ will appear neither symmetric nor anti-symmetric with respect to $\langle \cdot,\cdot\rangle$. Therefore, one should symmetrize/anti-symmetrize them first using integration by parts, because correct application of Gaussian integration formulas requires them to have such properties. We obtain: 
\begin{align}
\label{Deltas}
\widehat\Delta_b =  \left(\begin{matrix}
-(\beta/r)^2 D_{11}\partial_\tau^2 & \frac{1}{2}D_{10}^\dagger\cr
\frac{1}{2}D_{10} & D_{11}
\end{matrix} \right),\quad \widehat\Delta_f = \left(\begin{matrix}
-D_{00} & \frac{1}{2}D_{10}^\dagger\cr
-\frac{1}{2}D_{10} & -D_{00}
\end{matrix} \right).
\end{align}
Here we wrote Hermitian conjugate on $D_{10}$ instead of transpose, because this operator is real. Note that the operator $\widehat\Delta_b$ is real symmetric and $\widehat\Delta_f$ is real and anti-symmetric. We claim that $V$ is a well-defined localizing term. Below we will give some evidence to support this claim.
	
\textbf{The localization for bosons}
	
We find no difficulties studying the localization locus for bosons. The bosonic part of $V$ can be written as $\sum_a( |\delta_{\xi_\beta^H} \eta_a|^2 + |\delta_{\xi_\beta^H} \widetilde\eta^a|^2 + |\delta_{\xi_\beta^H} \Upsilon_a|^2 + |\delta_{\xi_\beta^H} \widetilde\Upsilon^a|^2)$. It has a global minimum and actually vanishes at the localization locus described by:
\begin{align}
\delta_{\xi_\beta^H}\eta_a &= (\delta_{\xi_\beta^H}\eta_a)^* =0\, ,\quad \delta_{\xi_\beta^H}\widetilde\eta^a = (\delta_{\xi_\beta^H}\widetilde\eta^a)^* =0\, ,\cr
\delta_{\xi_\beta^H}\Upsilon_a &= (\delta_{\xi_\beta^H}\Upsilon_a)^* =0\, ,\quad \delta_{\xi_\beta^H}\widetilde\Upsilon^a = (\delta_{\xi_\beta^H}\widetilde\Upsilon^a)^* =0\, ,
\end{align}
which are nothing but the BPS equations supplemented by the reality conditions. The equations in the first line imply that $q_a$ and $\widetilde{q}^a$ are $\tau$-independent, something we already saw before from the BPS equations. In other words, $q_a$ and $\tilde{q}^a$ are functions on the disk $D^2$, which is a base of the $U(1)$ fibration with fibers parametrized by $\tau$. The equations in the second line imply $H_a = \widetilde{H}^a=0$, i.e., $X_1'=0$, as well as $(X_1')^*=0$. The equations $H_a = \widetilde{H}^a=0$ are again the BPS equations, they simply express $G_a$ and $\widetilde{G}^a$ in terms of $q_a$ and $\widetilde{q}^a$ (as in \eqref{Gsol} and \eqref{Gtsol}), while $(X_1')^*=0$, due to the unusual conjugation property \eqref{conjH}, gives one more equation on $q$'s concisely written as:
\begin{equation}
D_{10}X_0=0 \,.
\end{equation}
Introducing notation
\begin{equation}
q_\pm = q_1 \pm i q_2\, ,\quad \widetilde{q}^\pm = \widetilde{q}^1 \pm i \widetilde{q}^2\, ,
\end{equation}
this equation is equivalent to the following system of equations:
\es{Systemqpqm}{
	\frac{\partial q_+}{\partial z_2} &= i\frac{\sigma}{2} q_-\,,\quad \frac{\partial q_-}{\partial \bar{z}_2} = -i\frac{\sigma}{2} q_+\,,\\
	\frac{\partial \widetilde{q}^+}{\partial z_2} &= i\frac{\sigma}{2} \widetilde{q}^-\,,\quad \frac{\partial \widetilde{q}^-}{\partial \bar{z}_2} = -i\frac{\sigma}{2} \widetilde{q}^+\,.
}
Because $\widetilde{q}^a$ are complex conjugates of $q_a$, the above equations on $\widetilde{q}^a$ are complex conjugates of those for $q_a$, so we only need to solve for $q_a$.

The case $\sigma = 0$ (the free hyper case) is simpler, so let us discuss it first.  In this case, $q_-$ is simply a holomorphic function of $z_2$ on $D^2$, and $q_+$ is anti-holomorphic. The holomorphy of $q_-$ on the disk implies that it can be written as a convergent power series $q_-=\sum_{n=0}^\infty a_n (z_2)^n$, and similarly $q_+=\sum_{n=0}^\infty b_n (\bar{z}_2)^n$.   Such functions are uniquely determined by their values at the boundary of the disk, where their Taylor expansions turn into the Fourier expansions: $q_-|_{\partial D^2}=\sum_{n=0}^\infty a_n e^{in\varphi}$ and $q_+|_{\partial D^2}=\sum_{n=0}^\infty b_n e^{-in\varphi}$. We see that equations at $\sigma=0$ imply that the functions $q_a$ are determined uniquely by their values at the boundary $\partial{D}^2= S^1$ (which is where our 1d theory lives), and moreover, $q_-$ at the boundary has only Fourier modes $e^{in\varphi}$ with $n\ge0$, while $q_+$ at the boundary has only Fourier modes $e^{-in\varphi}$ with $n\ge0$. If we now look at the anti-periodic linear combination:
\es{QModes}{
	Q(\varphi)&=q_1(\varphi)\cos\frac{\varphi}{2} + q_2(\varphi)\sin\frac{\varphi}{2}=\frac{1}{2}q_+(\varphi) e^{-i \varphi/2} + \frac{1}{2}q_-(\varphi) e^{i \varphi/2}\\
	&=\frac{1}{2}\sum_{n=0}^\infty a_n e^{i\varphi\left(n+\frac{1}{2}\right)} + \frac{1}{2}\sum_{n=0}^\infty b_n e^{-i\varphi\left(n+\frac{1}{2}\right)}\, ,
}
we see that the most general Fourier series for $Q(\varphi)$ completely encodes boundary values for $q_+$ and $q_-$. This shows that $Q(\varphi)$ parametrizes the bosonic part of the localization locus at $\sigma=0$.  An analogous calculation can also be performed for $\widetilde{Q}(\varphi)$ as defined in \eqref{QQt}.   Alternatively, we could start from $Q(\vphi)$, from which by \eqref{QModes} we can extract $q_a$, then determine $\tq^a$ by complex conjugation, and consequently determine $\tQ(\vphi)$ using \eqref{QQt}.  This procedure shows that $\widetilde{Q}(\varphi)$ is not simply the complex conjugate of $Q(\varphi)$; the relation between $Q$ and $\tQ$ is more complicated and will be discussed later.
	
It is straightforward to extend this analysis to $\sigma\ne0$, with the equations becoming slightly more complicated and holomorphy lost. In this case, the solution for $q_\pm$ is:
\begin{align}
\label{qpmSol}
q_- &= \sum_{n\in\bZ} a_n I_n(\sigma\sin\theta)e^{in\varphi}\, ,\cr
q_+ &= \sum_{n\in\bZ} i a_{n-1}I_n(\sigma\sin\theta)e^{in\varphi}\, ,
\end{align}
where $I_n$ are modified Bessel functions that are regular at zero. Just as for $\sigma = 0$, the solutions \eqref{qpmSol} are completely determined by the values of $q_\pm$ at the boundary $\partial D^2$. The expressions for $\widetilde{q}^a$ are obtained by complex conjugation. The fields $Q(\varphi)$ and $\widetilde{Q}(\varphi)$ at the boundary are:
\begin{align}
\label{QandIn}
Q(\varphi) &= q_1(\varphi)\cos\frac{\varphi}{2} + q_2(\varphi)\sin\frac{\varphi}{2} = \sum_{n\in\bZ}ia_n\frac{I_{n+1}(\sigma)-i I_n(\sigma)}{2}e^{i\left(n+\frac{1}{2}\right)\varphi}\, ,\cr
\widetilde{Q}(\varphi) &= \widetilde{q}_1(\varphi)\cos\frac{\varphi}{2} + \widetilde{q}_2(\varphi)\sin\frac{\varphi}{2} = -\sum_{n\in\bZ}a_n^*\frac{I_{n+1}(\sigma)-i I_n(\sigma)}{2}e^{-i\left(n+\frac{1}{2}\right)\varphi}\, .
\end{align}
We see that, again, all information about the fields $q_a$ and $\widetilde{q}^a$ that solve $D_{10}X_0=0$ in the bulk of $D^2$ is encoded in the most general anti-periodic fields $Q(\varphi)$ and $\widetilde{Q}(\varphi)$ at the boundary $\partial D^2$, subject to a certain reality constraint relating $Q(\varphi)$ and $\widetilde{Q}(\varphi)$. This completes the proof that the bosonic part of the localization locus is completely parametrized by $Q(\varphi)$ and $\widetilde{Q}(\varphi)$, the fields present in the 1d action \eqref{Shyper1d}, subject to the reality constraint.
	
Knowing the localization locus for bosons, there are no further subtleties with the 1-loop determinant for fluctuations of bosons in the transverse directions of the field space. Since the localizing term is quadratic, any zero mode would correspond to moving along the localization locus, so the determinant for transverse directions is well-defined and non-zero. In fact, one can easily check that the bosonic localization locus, as found above, coincides precisely with the space of zero modes of $\widehat\Delta_b$. These are a discrete series of normalizable zero modes of $\widehat\Delta_b$, and there are no other zero modes of $\widehat\Delta_b$ (which could be part of continuous spectrum, for example).  Concretely, one can check that there is a gap in the spectrum of $\widehat\Delta_b$ which separates discrete zero modes from the rest of the spectrum. The existence of this gap gives an even better evidence that everything works well with the localization of bosons using $e^{-tV}$ as the localizing term.
	
\textbf{The localization of fermions}
	
To understand what the localization by $e^{-tV}$ does to fermions, we have to study the spectrum of the operator acting on fermions that appear in $V$, that is of $\widehat\Delta_f$. To do that, we have to be precise about the space that $\widehat\Delta_f$ acts on, and this is where the subtleties begin.
	
\textbf{Spaces of fields and $L^2$ and $L'^2$ structures}
	
The fields that initially appear in the path integral are $\psi_{\alpha\dot a},\, \widetilde\psi_{\alpha\dot a}$. It is natural to postulate that they are square-integrable, and that the path integral is taken over the corresponding Hilbert space. Note that when we describe the space of fields, their statistics does not matter, so we might as well assume that they take values in the ordinary complex numbers. There is a norm which we refer to as an $L^2$ structure:
\begin{align}
||(\psi,\widetilde\psi)||_{L^2}^2&=\int_{S^3}{\rm Vol}\, \sum_{\alpha,\dot a}\left[(\psi_{\alpha\dot a})^*\psi_{\alpha\dot a} + (\widetilde\psi_{\alpha\dot a})^*\widetilde\psi_{\alpha\dot a} \right]\, .
\end{align}
	
One then chooses some convenient operator on $S^3$ that is self-adjoint with respect to this norm, e.g., the Dirac operator, expands fields into their eigenmodes, and uses this expansion to define the path integral. At this point, statistics becomes relevant since the modes of spinors are defined to be Grassmann numbers.
	
A key step in our discussion of the localizing term is to introduce new fields $\eta_a, \widetilde\eta^a, \Upsilon_a, \widetilde\Upsilon^a$. We want to pass to integration over these new fields, but this is a subtle point. Even though the expressions for $\eta_a, \Upsilon_a$ in terms of the old fields are perfectly smooth, they degenerate at the $\theta=\pi/2$ circle. The inverse transforms blow up there.  Indeed, we have
\begin{align}
\label{etaUps_inv}
\psi_{\dot a} &=\frac{1}{\mu}\left( (\xi_\beta^H)_{a\dot a}\mu^{ab}\eta_b - i\nu^a_{\ \ \dot a}\Upsilon_a \right),\cr
\widetilde\psi_{\dot a} &= \frac{1}{\mu}\left( (\xi_\beta^H)^a_{\ \ \dot a}\mu_{ab}\widetilde\eta^b + i\nu_{a\dot a}\widetilde\Upsilon^a \right) \,,
\end{align}
which blows up at $\theta=\pi/2$.  Since we want $\psi$ and $\tilde\psi$ to be $L^2$-normalizable, the fields $\eta_a$ and $\Upsilon_a$ cannot behave arbitrarily in the vicinity of $\theta=\pi/2$, where $\mu$ vanishes. They should obey certain ``boundary'' conditions in the vicinity of the $\theta=\pi/2$ circle that ensure that
\begin{align}
\label{L2norm}
\sum_{\alpha,\dot a}(\psi_{\alpha\dot a})^*\psi_{\alpha\dot a}=-\frac{1+\beta^2}{2\beta}\left(\frac{1}{\mu}\mu^{ab}\eta_a^*\eta_b + \frac{1}{\mu^2}\mu^{ab}\Upsilon_a^*\Upsilon_b \right) - \frac{1-\beta^2}{2\beta}\frac{1}{\mu}i\varepsilon^{ab}\left(\Upsilon_a^*\eta_b - \eta_b^*\Upsilon_a \right)
\end{align}
is integrable on $S^3$.

We can introduce another norm, which we refer to as an $L'^2$ structure. If we write $A=(\eta_a\, \widetilde\eta^a\, \Upsilon_a\, \widetilde\Upsilon^a)^t$, then the $L'^2$ structure is characterized by the norm:
\begin{align}
||(\eta,\widetilde\eta,\Upsilon,\widetilde\Upsilon)||_{L'^2}^2 &= \langle A^*, A \rangle = \int_{S^3} {\rm Vol}\, \sum_a \left[ (\eta_a)^*\eta_a + (\widetilde\eta^a)^*\widetilde\eta^a + (\Upsilon_a)^*\Upsilon_a + (\widetilde\Upsilon^a)^*\widetilde\Upsilon^a \right]\, .
\end{align}
Our real operator $\widehat\Delta_f$, being anti-symmetric with respect to $\langle\cdot,\cdot\rangle$, becomes anti-Hermitian in the $L'^2$ structure. If we study its spectrum in the $L'^2$ structure, all its discrete spectrum eigenfunctions have to be $L'^2$ normalizable. But, because of the singular relations like \eqref{L2norm}, they won't necessarily be $L^2$ normalizable. 
	
For the Hilbert space with the $L^2$ inner product, we will use the notation $\mathcal{H}$, while $\mathcal{H}'$ will be used to denote the $L'^2$ Hilbert space. The definitions \eqref{etaUps} of $\eta, \Upsilon$ and $\widetilde\eta, \widetilde\Upsilon$ provide an embedding of $\mathcal{H}$ as a linear subspace of $\mathcal{H}'$. Linearity simply follows from the linearity of \eqref{etaUps}.
	
When we write the path integral in terms of $\psi, \widetilde\psi$, we simply think of integration over $\mathcal{H}$ (with some modes cut-off used to make this precise). When we write the path integral in terms of $\eta,\Upsilon,\widetilde\eta,\widetilde\Upsilon$ though, because these fields live in a bigger space $\mathcal{H}'$, we should think of integration over $\mathcal{H}$ as a subspace of $\mathcal{H}'$. This is important, in particular, because the operator $\widehat\Delta_f$, being anti-Hermitian in $\mathcal{H}'$, is not anti-Hermitian when restricted to $\mathcal{H}$.
	
\textbf{No fermionic zero modes in the discrete spectrum}
	
The above discussion shows that the operator that enters the localization procedure is $\widehat\Delta_f$ restricted to $\mathcal{H}\subset \mathcal{H}'$, where by ``restricted'' we mean the orthogonal projection with respect to $\langle\cdot,\cdot\rangle$. It would be difficult to study such a projected operator, so instead we do the following.
	
Let us introduce an eight-column $\Gamma=(\psi_{1\dot1}\, \psi_{1\dot2}\, \psi_{2\dot1}\, \psi_{2\dot2}\, \widetilde\psi_{1\dot1}\, \widetilde\psi_{1\dot2}\, \widetilde\psi_{2\dot1}\, \widetilde\psi_{2\dot2})^t$. We already introduced notations $X_0'$ and $X_1$ for the columns with $\eta,\widetilde\eta$ and $\Upsilon,\widetilde\Upsilon$ before. The relation \eqref{etaUps} can be written as:
\begin{equation}
\left(\begin{matrix}
X_0'\\
X_1
\end{matrix} \right) = T_f \Gamma\,,
\end{equation}
where $T_f$ is a coordinate dependent 8 by 8 matrix. Its determinant is $\beta^8\cos^8\theta$, which of course vanishes at the $\theta=\pi/2$ circle and is the reason why $\mathcal{H}$ and $\mathcal{H}'$ are not the same. The localizing term for fermions is:
\begin{equation}
\int_{S^3} {\rm Vol}\, \left(\begin{matrix}X_0^{'t} & X_1^t\end{matrix} \right) \widehat\Delta_f\left(\begin{matrix}
X_0'\cr
X_1
\end{matrix}\right) = \int_{S^3} {\rm Vol}\, \Gamma^t T_f^t \widehat\Delta_f T_f\Gamma \,.
\end{equation}
The matrix $T_f$ is complex, so even though $T_f^t \widehat\Delta_f T_f$ is anti-symmetric, it is not real, so we cannot apply the spectral theorem. However, it is possible to find another complex and everywhere non-degenerate matrix $M_f$, such that:
\begin{equation}
T_f = R_f M_f \,,
\end{equation}
where $R_f$ still degenerates at $\theta=\pi/2$, but now it is real. There is a lot of freedom in choosing such a matrix $M_f$, and we fix it to be: 	\begin{equation}
	M_f = \left(\begin{matrix}
	\mathbb{B} & 0_4\cr
	0_4 & \mathbb{B}
	\end{matrix} \right)\, ,\quad \mathbb{B} = \left(\begin{matrix}
	1 & 0 & 0 & -\frac{1}{\beta}\cr
	0 & 1 & \beta & 0\cr
	i & 0 & 0 & \frac{i}{\beta}\cr
	0 & i & -i\beta& 0
	\end{matrix} \right)\, .
\end{equation}
The localizing term for fermions becomes:
\begin{equation}
\int_{S^3} {\rm Vol}\, \Gamma^t M_f^t R_f^t \widehat\Delta_f R_f M_f\Gamma.
\end{equation}
The change of variables $\chi= M_f \Gamma$ is a non-degenerate field redefinition, and we could imagine defining the path integral over $\chi$ rather than $\Gamma$. We could define yet another norm:
\begin{equation}
\int_{S^3} {\rm Vol}\, \chi^\dagger\chi\, ,
\end{equation}
but because $\chi^\dagger\chi = \Gamma^\dagger M_f^\dagger M_f \Gamma = 2(|\psi_{1\dot1}|^2+|\psi_{1\dot2}|^2+\beta^2|\psi_{2\dot1}|^2+\frac{1}{\beta^2}|\psi_{2\dot2}|^2+|\widetilde\psi_{1\dot1}|^2+|\widetilde\psi_{1\dot2}|^2+\beta^2|\widetilde\psi_{2\dot1}|^2+\frac{1}{\beta^2}|\widetilde\psi_{2\dot2}|^2)$, this $\chi^\dagger\chi$ is integrable if and only if $\Gamma^\dagger\Gamma=\sum_{\alpha\dot a}\left(|\psi_{\alpha\dot a}|^2 + |\widetilde\psi_{\alpha\dot a}|^2\right)$ is integrable. So this norm is equivalent to the $L^2$ structure we had before, and integrating over $\chi$ with this norm is equivalent to integrating over the same Hilbert space $\mathcal{H}$.
	
The localizing term now takes the form $\int_{S^3} {\rm Vol}\, \chi^t R_f^t \widehat\Delta_f R_f \chi$, and the relevant operator is $R_f^t \widehat\Delta_f R_f$. It is real and anti-symmetric, thus the spectral theorem applies to it now, and so it makes sense to look for its eigenfunctions. If it has any zero modes, they would be of the form ``$R_f^{-1}$ times the zero mode of $\widehat\Delta_f$''. If they are part of the discrete spectrum, they should be $L^2$ normalizable. Let us see if there are any. We try to solve:
\begin{equation}
\widehat\Delta_f\left(\begin{matrix}
X_0'\cr
X_1
\end{matrix} \right)=0\, \Rightarrow \Bigg\{ \begin{matrix}
D_{00}X_0' - \frac{1}{2}D_{10}^\dagger X_1 &=0\, ,\cr
\frac{1}{2}D_{10}X_0' + D_{00}X_1 &=0\, .
\end{matrix}
\end{equation}
where as before, $X_0'=(\eta_1,\, \eta_2,\, \widetilde\eta^1,\, \widetilde\eta^2)^t$ and $X_1=(\Upsilon_1,\,\Upsilon_2,\, \widetilde\Upsilon^1,\,\widetilde\Upsilon^2)^t$ are four-component columns. Acting on the second equation with $D_{00}$, using that it anticommutes with $D_{10}$ and using the first equation to eliminate $D_{00}X_0'$, we arrive at:
\es{DDB}{
	D_{10}D_{10}^\dagger X_1 = \left(\frac{2\beta}{r}\right)^2 \partial_\tau^2 X_1 \,.
}
The differential operator $\partial_\tau^2$ appearing on the right is non-positive definite: it can only have zero or negative eigenvalues. The differential operator $D_{10}D_{10}^\dagger$ appearing on the left is manifestly non-negative definite. Therefore, the above equation can only be satisfied if $\partial_\tau X_1=0$ (so the solutions of \eqref{DDB} are defined on the disk $D^2$) and $D_{10}D_{10}^\dagger X_1 =0\Rightarrow D_{10}^\dagger X_1=0$. Let us temporarily write $D^{(\sigma)}_{10}$ for $D_{10}$ to make $\sigma$-dependence of $D_{10}$ explicit. Then
\begin{equation}
\label{conjugation}
(D_{10}^{(\sigma)})^\dagger = \frac{1}{\cos^2\theta}D_{10}^{(-\sigma)}\cos^2\theta \,.
\end{equation}
So $(D^{(\sigma)}_{10})^\dagger X_1=0$ implies $D_{10}^{(-\sigma)}(\cos^2\theta X_1)=0$. Modes in the kernel of $D_{10}^{(-\sigma)}$ are completely determined by their value at the boundary of the disk, as we have seen in the analysis of bosonic localization locus. If $X_1$ is regular, then $X_1\cos^2\theta$ vanishes at $\theta=\pi/2$, that is at $\partial D^2$, and so $D_{10}^{(-\sigma)}(\cos^2\theta X_1)=0$ implies $X_1=0$. The only way to avoid this conclusion is to assume that $X_1$ is singular at the boundary, with the singularity being no weaker than $1/\cos^2\theta$.  However, this asymptotic behavior will give a mode which is not normalizable. So we infer that $X_1=0$. The remaining equations imply $D_{00}X_0'=0$, that is $\partial_\tau X_0'=0$, and:
\begin{equation}
D_{10}X_0'=0 \,.
\end{equation}
These are again the same equations we obtained before for the bosonic part of the localization locus. Therefore, we could use the same solution, which would lead us to a naive conclusion that there are fermionic zero modes: $\Upsilon_a=\widetilde\Upsilon^a=0$ and $\eta_a,\, \widetilde\eta^a$ are determined by $\eta_\pm =\eta_1\pm i\eta_2$ and $\widetilde\eta_\pm = \widetilde\eta_1\pm i\widetilde\eta_2$, which are in turn given by expressions like \eqref{qpmSol}.
	
We now impose $L^2$ normalizability of these zero modes. Using \eqref{L2norm}, the integral $\int_{S^3}\frac{1}{\mu}\mu^{ab}\eta_a^*\eta_b \times \sin\theta\cos\theta\, d\theta\, d\varphi\, d\tau$ has to be convergent. Since $\mu=\beta^2\cos^2\theta$, we are integrating over the disk the following:
\es{etaNorm}{
	-\frac{2\pi}{\beta^2}\int_{D^2} \mu^{ab}\eta_a^* \eta_b \frac{d(\cos\theta)}{\cos\theta}d\varphi.
}
For possible zero modes we have:
\begin{align}
\eta_- &= \sum_n \eta_n I_n(\sigma\sin\theta)e^{in\varphi}\, ,\cr
\eta_+ &= \sum_n i\eta_{n-1} I_n(\sigma\sin\theta) e^{in\varphi}\, .
\end{align}
If we insert these expressions into \eqref{etaNorm}, we obtain
\begin{equation}
\frac{2\pi^2}{\beta}\int_{\theta=0}^{\theta=\pi/2}\sum_n (\eta_{n-1}^*\eta_{n-1} + \eta_n^*\eta_n) I_n^2(\sigma\sin\theta)\frac{d(\cos\theta)}{\cos\theta}.
\end{equation}
The convergence of this integral requires that all $\eta_n=0$,  so we conclude that zero modes are not in the discrete spectrum of $R_f^t \widehat\Delta_f R_f$.
	
\textbf{Summary}
	
Let us summarize what has been done in the last couple of pages. We have rewritten the fermionic path integral as an integral over $\chi$. We were able to prove that the operator $R_f^t \widehat\Delta_f R_f$ that appears in the localizing term does not have any $L^2$ normalizable zero modes. If there were such zero modes, we would say that the localization locus is actually a supermanifold and they would be the fermionic coordinates on it. As we see, this does not happen, and the localization locus is purely bosonic, parametrized by $Q(\varphi)$, as we explained before.
	
However, the spectrum of $R_f^t \widehat\Delta_f R_f$ still might have continuous branches that pass through zero. They manifest themselves as ``non-normalizable'' zero modes we encountered above. In fact, a more detailed analysis of this operator and the asymptotic behavior of its eigenfunctions next to $\theta=\pi/2$ shows that this is indeed the case. We believe that for such operators, the determinant can still be defined after appropriate regularization and renormalization have been introduced. We are not going to study this here.
	
Let us also emphasize that even though we have not been able to prove rigorously that localization works well for fermions, our final result about the 1d theory \eqref{Shyper1d} is still reliable.  Indeed, in Section~\ref{NoLocArg} we showed that the 1d theory \eqref{Shyper1d} gives precisely the same correlators as the 3d Gaussian theory coupled to a matrix model \eqref{ShyperLoc3d}.  What we have shown in the discussion above is that the same result can be obtained from the localization of the hypermultiplet, up to an unresolved problem about the existence of the determinant of an operator with continuous spectrum.  Moreover, our analysis identified a relation between $Q$ and $\tQ$ that can be used to define a middle-dimensional integration cycle in the path integral \eqref{3dTo1d} that we now discuss.

\subsubsection{Integration cycle from localization}
\label{REALITY}

If we look at the linear combinations $Q = q_1 \cos\frac{\varphi}{2} + q_2 \sin\frac{\varphi}{2}$ and $\widetilde{Q} = \widetilde{q}_1 \cos\frac{\varphi}{2} + \widetilde{q}_2 \sin\frac{\varphi}{2}$ in the 3d theory, they are completely independent. They are also not related by the complex conjugation, since $Q^*=\widetilde{q}_2\cos\frac{\varphi}{2} - \widetilde{q}_1\sin\frac{\varphi}{2}$ and $\widetilde{Q} = \widetilde{q}_1 \cos\frac{\varphi}{2} + \widetilde{q}_2 \sin\frac{\varphi}{2}$ are linearly independent.
	
Things change once we pass to the 1d theory of the localization locus. The fields $Q(\varphi)$ and $\widetilde{Q}(\varphi)$ become related by a complicated reality condition. If we write Fourier expansions of these fields as:
\es{QExpansion}{
	Q(\varphi)&=\sum_{n\in\bZ + \frac 12} c_n e^{i n \varphi}\, ,\cr
	\widetilde{Q}(\varphi)&=\sum_{n\in\bZ + \frac 12} \widetilde{c}_n e^{-i n \varphi}\, ,
}
then, as one can see from \eqref{QandIn}, the reality condition is:
\es{ctilde}{
	\widetilde{c}_n  = e^{i \alpha_n(\sigma)}  c_n^* \,, \qquad
	e^{i \alpha_n(\sigma)} \equiv \frac{1}{i} \frac{I_{n+\frac 12}(\sigma)-i I_{n - \frac 12} (\sigma)}{I_{n+\frac 12}(\sigma)+i I_{n-\frac 12} (\sigma)} \,.
}

The relation $\tc_n = e^{i \alpha_n(\sigma)} c_n^*$ can be interpreted as a choice of integration cycle in the Gaussian integrals in \eqref{3dTo1d} or \eqref{detRHS}.  The choice \eqref{ctilde} is such that these integrals converge.  Indeed, the Gaussian integrals in \eqref{detRHS} can be written in terms of the mode coefficients as
\es{GaussianModes}{
	\prod_{n \in \Z + \frac 12} \int d c_n\, d\tc_n  \, e^{-2 \pi  (in + \sigma) c_n \tilde c_n}
	=  \prod_{n \in \Z + \frac 12}  \int d c_n\, dc_n^* \, e^{-2 \pi  (in + \sigma) e^{i \alpha_n(\sigma)}\abs{c_n}^2 }
}
where we used the fact that the Jacobian for the change of variables from $\tc_n$ to $c_n^*$ is equal to $1$ due to the property $\alpha_n(\sigma) + \alpha_{-n}(\sigma) = 0$ that can be easily inferred from \eqref{ctilde}.  The Gaussian integral in \eqref{GaussianModes} converges provided that
\es{ConvergenceCondition}{
	\Re \left[ (i n + \sigma) e^{i \alpha_n (\sigma)} \right]\geq 0 
}
for all $n$ and $\sigma$.  This condition holds, as can be checked explicitly using the expression for $\alpha_n(\sigma)$ given in \eqref{ctilde}.  
	
One can of course deform the integration cycle while preserving the convergence of the Gaussian integrals.  For instance, an alternative choice of integration cycle is given by
\es{ctildeSimple}{
	\tc_n = i \sgn(n) c_n^*
}
which has the advantage of being independent of $\sigma$.  This is in fact the $\sigma \to 0$ limit of \eqref{ctilde}.  While very simple in terms of the Fourier modes, the condition \eqref{ctildeSimple} is rather complicated in position space.  It takes the form
\es{SimplePosition}{
	\tQ(\vphi) = \frac{1}{2 \pi} \, \text{P.V.} \int d\vphi' \frac{1}{\sin \frac{\vphi - \vphi'}{2}} Q^*(\vphi') \,,
}
with P.V. denoting principal value integration.

\subsection{Integration cycle from Morse theory}\label{MORSE}

The 1d Gaussian theory \eqref{Shyper1d} is very similar to the one studied by Witten in~\cite{WittenQM}.  The two theories are given by the same Hamiltonian path integral for quantum mechanics, with the only exception that in \eqref{Shyper1d} the scalar fields are taken to be anti-periodic on the circle, while in \cite{WittenQM} they are periodic.  When $\sigma = 0$, the description of integration cycle used in \cite{WittenQM} involves holomorphic functions on an auxiliary two-dimensional disk, which is suggestively similar to the description we found in \eqref{Systemqpqm}!  

Let us thus explore the connection between our 1d Gaussian theory \eqref{Shyper1d} and that studied in~\cite{WittenQM} in more detail.  For simplicity, let us focus on the case of a free uncharged hypermultiplet and leave the generalization to a gauged hypermultiplet for future work. It was explained in \cite{WittenQM, WittenCS} that one can construct sensible integration cycles for the path integral using Morse theory.  Let us apply this formalism to our case.  While we will not give a comprehensive review of the Morse theory formalism developed in \cite{WittenQM, WittenCS}, we refer interested readers to these references for more details.  

We start with the holomorphic functional
\begin{equation}
\exp \left[ 4\pi r\int_{-\pi}^\pi \tQ \partial_\varphi Q d\varphi \right] \,,
\end{equation}
which we want to integrate over some middle-dimensional real cycle in the space of complex fields $Q(\varphi)$ and $\tQ(\varphi)$ that describe anti-periodic maps $S^1 \to \C^2$.  At every $\vphi$, it is convenient to parameterize the target space $\C^2$ by $Q=x+iy$ and $\tQ=\widetilde{x}+i\widetilde{y}$, where $x$, $y$, $\widetilde x$, and $\widetilde{y}$ are real coordinates.

The integration cycle in the path integral should be chosen in such a way that the functional $\exp \left[4 \pi r \int_{-\pi}^\pi \tQ \partial_\varphi Q d\varphi \right]$ vanishes at large field values in order for the path integral to converge. A way to construct such an integration cycle is to consider the real part of the holomorphic action as a Morse function:
\begin{equation}
h=\Re \int \tQ dQ = \int (\widetilde{x}dx - \widetilde{y}dy) \,,
\end{equation}
(we dropped the positive factor $4\pi r$ because it is irrelevant in this discussion) and construct the integration cycle using the gradient flow lines associated with this Morse function that start from the critical points of $h$. Since the Morse function $h=\Re \int \tQ dQ$ is strictly decreasing along any gradient flow, it is bounded from above by its value at the critical point at which the flow starts. As explained in \cite{WittenQM, WittenCS}, if the integration cycle is, roughly speaking, defined as a union of all gradient flow lines starting from the critical point, the functional $\exp \left[ \int_{-\pi}^\pi \tQ dQ \right]$ remains bounded along this cycle and vanishes at infinity, so it can be integrated.  

Since the fields are taken to be antiperiodic on the circle---which is the main difference between our theory and that in \cite{WittenQM}---the only critical point of $h$ is $x=\widetilde{x}=y=\widetilde{y}=0$, as one can find from setting $\delta h=0$. One can check that among the eigenvalues of the Hessian of $h$ at this critical point, precisely half are positive and half are negative.\footnote{For that, we can expand fields $x, y, \widetilde{x}, \widetilde{y}$ in Fourier modes on $S^1$ and introduce a UV cutoff on the modes in order to obtain a finite-dimensional vector space. Then $h$ becomes a real quadratic form in those modes, half of whose eigenvalues are positive and half are negative.} A gradient flow line can go in any of the negative directions, and so the union of all gradient flow lines indeed defines a middle-dimensional cycle, as is always the case for isolated critical points.

The definition of gradient flow requires a metric on target space.  Indeed, in order to define gradient flow, we would like to turn the one-form $\delta h$ into the vector field defining the flow line using a metric on the space of maps parameterized by $Q(\varphi)$ and $\tQ(\varphi)$.  To achieve this goal, we first choose a metric on $\C^2$, 
\begin{equation}
ds^2 = |d\tQ|^2 + |d Q|^2 = dx^2 + dy^2 + d\widetilde{x}^2 + d\widetilde{y}^2 \,.
\end{equation}
This metric on $\C^2$ induces a metric on the space of maps $Q(\varphi), \tQ(\varphi)$ given by $\int \left(|\delta Q|^2 + |\delta\tQ|^2 \right)d\varphi$, and this latter metric allows us to define the gradient flow lines of $h$.  If $s\in (-\infty,0]$ is the flow parameter, the flow equations for this metric are:
 \es{Flow}{
\frac{\partial x(s,\varphi)}{\partial s}&=\frac{\partial\widetilde{x}(s,\varphi)}{\partial \varphi}\, ,\quad \frac{\partial \widetilde{x}(s,\varphi)}{\partial s}=-\frac{\partial x(s,\varphi)}{\partial \varphi}\,,\\
\frac{\partial y(s,\varphi)}{\partial s}&=-\frac{\partial\widetilde{y}(s,\varphi)}{\partial \varphi}\, ,\quad \frac{\partial \widetilde{y}(s,\varphi)}{\partial s}=\frac{\partial y(s,\varphi)}{\partial \varphi}\,.
 }
The fields appearing in \eqref{Flow} have to be anti-periodic in $\varphi$, as we mentioned before. The boundary condition at $s=-\infty$ is that the flow line should start from the critical point of $h$, that is the fields $x, y, \widetilde{x}, \widetilde{y}$ should vanish at $s=-\infty$. With such a boundary condition, the space of all possible solutions of \eqref{Flow} restricted to $s=0$ provides a good integration cycle, as explained in \cite{WittenQM} and reviewed briefly above. Note that taking $s=0$ is just a convenient choice. Since we can go along the flow with different speeds, flowing for the time $(-\infty, 0]$ with all possible initial speeds sweeps out the whole range of possible flow lines.

What the equations \eqref{Flow} tell us is that $x+i\widetilde{x}$ and $\widetilde{y}+iy$ should be holomorphic functions in the complex coordinate $w=s+i\varphi$. Locally, this is the same as to say that they are holomorphic in the coordinate $z=\exp w$, that parameterizes the unit disk $D^2$. The locus $s=0$ corresponds to the boundary of this disk. Unlike in \cite{WittenQM}, fields $x+ i\widetilde{x},\ \widetilde{y}+iy$ are not holomorphic functions on this disk, but rather, because of their anti-periodicity in $\varphi$, are holomorphic functions on the double cover parametrized by $\sqrt{z}$. Being holomorphic in $\sqrt{z}$ and anti-periodic in $\varphi=2\arg\sqrt{z}$, the Taylor expansions of these functions can contain only odd positive powers of $\sqrt{z}$. Any function of this type can be written as $\sqrt{z}$ times an arbitrary holomorphic function of $z$. So we conclude that the most general solution of the flow equations satisfying the boundary condition at $s=-\infty$ is:
\begin{equation}
\label{Mcont}
x+i\widetilde{x} = \sqrt{z} a(z)\,,\qquad \widetilde{y} + iy = \sqrt{z} b(z) \,,
\end{equation}
where $a(z)$ and $b(z)$ are arbitrary holomorphic functions on the disk. Then the space of boundary values of these solutions provides a good integration cycle.

Let us make contact with the fields of our theory. Recall that in Section~\ref{1LoopSec} we defined the linear combinations of hypermultiplet scalars $q_\pm=q_1\pm iq_2$ and $\widetilde{q}^\pm = \widetilde{q}^1\pm i\widetilde{q}^2$, in terms of which $Q(\vphi)$ and $\tQ(\vphi)$ are 
 \es{Oldq}{
Q(\varphi)=\frac{1}{2}q_-(\varphi) e^{i\varphi/2} + \frac{1}{2}q_+(\varphi) e^{-i\varphi/2}\,,\\
\tQ(\varphi)=\frac{i}{2}\widetilde{q}^+(\varphi) e^{-i\varphi/2} - \frac{i}{2}\widetilde{q}^-(\varphi) e^{i\varphi/2}\,.
 }
Using \eqref{Mcont} restricted to the boundary $z=e^{i\varphi}$, and using $Q=x+iy$ and $\tQ=\widetilde{x}+i\widetilde{y}$, we obtain
 \es{QFromab}{
Q(\varphi)&=\frac{1}{2}(a(e^{i\varphi})+b(e^{i\varphi}))e^{i\varphi/2} + \frac{1}{2}(\bar{a}(e^{-i\varphi})-\bar{b}(e^{-i\varphi}))e^{-i\varphi/2}\,,\\
\tQ(\varphi)&=\frac{i}{2}(b(e^{i\varphi})-a(e^{i\varphi}))e^{i\varphi/2} + \frac{i}{2}(\bar{a}(e^{-i\varphi})+\bar{b}(e^{-i\varphi}))e^{-i\varphi/2}\,.
 }
Comparing \eqref{QFromab} with \eqref{Oldq}, we conclude that we can take:
\begin{align}
q_-(\varphi) = \left[a(z) + b(z) \right]\big|_{z=e^{i\varphi}}\, ,\quad q_+ = \left[\bar{a}(\bar{z}) - \bar{b}(\bar{z}) \right]\big|_{z=e^{i\varphi}}\, ,\cr
\widetilde{q}^+(\varphi) = \left[\bar{a}(\bar{z}) + \bar{b}(\bar{z}) \right]\big|_{z=e^{i\varphi}}\, ,\quad \widetilde{q}^- = \left[a(z) - b(z) \right]\big|_{z=e^{i\varphi}}\, ,
\end{align}
and therefore completely describe the integration cycle by saying that it is given by \eqref{Oldq}, with $q_-(\varphi)$ being the boundary value of a holomorphic function on a disk, and $q_+(\varphi)$ being the boundary value of an antiholomorphic function on a disk. For the tilded fields we have $\widetilde{q}^\pm = (q_\mp)^*$.  But these descriptions coincide \emph{precisely} with what we obtained in Section \ref{1LoopSec} from localization!  We therefore conclude that the integration cycle we find here using the Morse theory is the same as the integration cycle given in  \eqref{ctildeSimple} obtained before from localization. 	
	
\section{Properties of twisted Higgs branch theory}
\label{PROPERTIES}
	
Before we move on to applications, let us provide a brief summary of the results of the previous section and describe some important features of the 1d theory used to calculate correlation functions of twisted Higgs branch operators.   
	
\subsection{Brief summary}
	
The main result of our work is a concise formula representing a 1d Gaussian theory coupled to a matrix model that can be used to compute correlation functions of twisted Higgs branch operators.  The matrix degree of freedom, $\sigma$, has its 3d origin in the constant mode of one of the ${\cal N} = 4$ vectormultiplet scalars and is thus valued in the Lie algebra $\mathfrak{g}$ of the gauge group $G$.  By a gauge transformation, $\sigma$ can be taken to lie within the Cartan subalgebra, and one can write the $S^3$ partition function as
\es{ZS3}{
	Z = \frac{1}{\abs{{\cal W}}} \int_\text{Cartan} d\sigma \det{}^\prime_\text{adj}(2 \sinh( \pi \sigma)) \, Z_\sigma \,,
}
where $\abs{{\cal W}}$ is the order of the Weyl group of the gauge group, and $Z_\sigma$ is the partition function of the 1d Gaussian theory that couples to $\sigma$.  This 1d Gaussian theory is written in terms of anti-periodic scalars $Q(\varphi)$ and $\tQ(\varphi)$ living on a great circle of $S^3$ parameterized by $\varphi \in [-\pi, \pi]$.  They have their 3d origin in the twisted operators formed from the hypermultiplet scalars.  For a hypermultiplet transforming in the representation ${\cal R}$ of $\mathfrak{g}$, we have that $Q(\varphi)$ and $\tQ(\varphi)$ transform in ${\cal R}$ and $\overline{\cal R}$, respectively.  The partition function of the 1d theory is
\es{Z1dFinal}{
	Z_\sigma = \int DQ D\tQ e^{- S_\sigma[Q, \tQ]} \,, \qquad
	S_\sigma[Q, \tQ] = \ell \int d\varphi\, \left[ \tQ \partial_\varphi Q + \tQ (\sigma + m r) Q  - 2 \pi i \tr_\zeta \sigma \right] \,,
} 
where $\ell \equiv - 4 \pi r$, $r$ being the radius of $S^3$, $m$ is the real mass matrix, and $\tr_\zeta: \mathfrak{g} \to \C$ is a weight of $\mathfrak{g}$ corresponding to FI parameters, i.e., if $t_a$ are generators of abelian factors in $\mathfrak{g}$, then $\tr_\zeta \sigma = \tr_\zeta \sum_a \sigma_a t_a = \sum_a \sigma_a \tr_\zeta(t_a)$, with $\tr_\zeta(t_a)=\zeta_a$ the FI parameters.  The path integral in \eqref{Z1dFinal} is over a middle-dimensional integration cycle in the space of complex $Q$ and $\tQ$, as explained in Sections~\ref{REALITY} and \ref{MORSE}. However, for the applications we are interested in, the precise choice of the integration cycle will not be important.
	
The theory \eqref{ZS3}--\eqref{Z1dFinal} can be used to calculate correlation functions of twisted Higgs branch operators ${\cal O}_i$ via
\es{CorrelFunctions}{
	\langle {\cal O}_1(\varphi_1) \ldots {\cal O}_n(\varphi_n) \rangle
	&= \frac{1}{|\cW|Z} \int_\text{Cartan} d\sigma \det{}^\prime_\text{adj} [ 2 \sinh( \pi \sigma) ] \langle {\cal O}_1(\varphi_1) \ldots {\cal O}_n(\varphi_n) \rangle_\sigma \, Z_\sigma  \,,
}
where, $\langle\, \cdots \rangle$ represents an expectation value in the full 3d ${\cal N} = 4$ SCFT, while $\langle\, \cdots \rangle_\sigma$ represents the expectation value in the Gaussian 1d theory \eqref{Z1dFinal}.   The latter expectation value can be computed using Wick contractions with the propagator
\es{GSummary}{
	G_\sigma(\vphi_1 - \vphi_2) = \langle Q(\vphi_1) \tQ(\vphi_2) \rangle_\sigma 
	=  \frac{\sgn(\vphi_1 - \vphi_2) + \tanh(\pi (\sigma + m r))}{2 \ell}  e^{-(\sigma + m r) (\vphi_1 - \vphi_2)} \,.
}
At coincident points, this propagator reduces to 
\es{G0Again}{
	G_\sigma(0) =  \frac{\tanh(\pi (\sigma + mr))}{2 \ell} \,.
}
In the expressions above, $\sigma + m r$ is a shorthand notation for ${\cal R}(\sigma) + r {\cal F}(m)$, where ${\cal R}$ and ${\cal F}$ are maps form the gauge and flavor algebras, respectively, into $\dim {\cal R} \times \dim {\cal R}$ hermitian matrices, as introduced in Footnote~\ref{FlavorFootnote}.

\subsection{Topological gauged quantum mechanics}
\label{TOPOLOGICALQM}
	
Let us now discuss two important properties of the theory presented above, first in the conformal case $m = \zeta = 0$, and then in the case where $m$ and $\zeta$ are non-zero.
	
\subsubsection{The conformal case}
	
The first property of \eqref{ZS3} when $m = \zeta = 0$ is that correlation functions of gauge-invariant operators are topological.  Indeed, at fixed $\sigma$, the equations of motion for $Q$ and $\tQ$ give $\partial_\varphi Q = -{\cal R}(\sigma) Q$ and $\partial_\varphi \tQ = +{\cal R}(\sigma) \tQ$.  The theory \eqref{Z1dFinal} being Gaussian, it follows that any operator ${\cal O}$ built out of $Q$ and $\tQ$ obeys $\partial_\varphi {\cal O} = - {\cal R}_{\cal O}(\sigma) {\cal O}$, with ${\cal R}_{\cal O}$ being the gauge representation of ${\cal O}$.  Gauge-invariant operators have ${\cal R}_{\cal O}(\sigma) = 0$, and therefore they obey $\partial_\varphi {\cal O} = 0$.  Consequently, the correlation functions of gauge-invariant operators must be independent of the precise distance between the operator insertions.  However, these correlation functions could and, in general, do depend on the ordering of the insertions.  Note that gauge-non-invariant operators may have position-dependent correlation functions, as can be seen, for instance, in \eqref{GSummary} in the case of the gauge non-invariant operators $Q$ and $\tQ$.
	
Another property that is rather hard to see from \eqref{ZS3}, but that we have checked in detail in many examples, is that, when $m = \zeta = 0$, we have 
\es{Relation1d}{
	(\tQ {\cal R}(T) Q) (\vphi) = 0\,, \qquad \text{for all } T \in \mathfrak{g}\,,  \qquad \text{(up to contact terms)}\,.
}
This identity can be used in the correlation functions of gauge-invariant operators (up to contact terms).  From the point of view of the 3d theory, the relation \eqref{Relation1d} is a D-term relation.  Indeed, at the SCFT fixed point attained by sending $g_\text{YM} \to \infty$, the $D^{ab}$ equation of motion gives the D-term relations $\tq_{(a} {\cal R}(T) q_{b)} = 0$ for all $T \in \mathfrak{g}$.  Taking appropriate linear combinations of these relations and using the definitions of $Q$ and $\tQ$ in \eqref{QQt} yields \eqref{Relation1d}.  Note that when $\tQ \cR(T) Q$ appears in a composite operator, for instance in a gauge-invariant operator $\tQ \cR(T) Q {\cal O}$, this operator need not vanish exactly;  instead, it can be written as a linear combination of operators that do not involve $\tQ \cR(T) Q$.  The precise form of this linear combination depends on the precise definition of the composite operator $\tQ \cR(T) Q {\cal O}$, as we will see explicitly in examples.
	
It is possible to provide a more formal argument both for the topological nature of the correlation functions of gauge-invariant operators and for the D-term relation \eqref{Relation1d}.   This argument relies on interpreting \eqref{ZS3} with $m = \zeta = 0$ as the partition function of a gauge-fixed  topological gauged quantum mechanics, albeit with an  unusual choice of integration cycle in the path integral.  Let us construct this gauged quantum mechanics starting from \eqref{ZS3}.   The first step is to rewrite \eqref{ZS3} as an integral over the whole Lie algebra, not just the Cartan.  Up to an overall constant, we have
\es{ZS3NotCartan}{
	Z = \int d\sigma\,  Z_\sigma\, \det{}^\prime_\text{adj} \frac{2 \sinh( \pi \sigma)}{\sigma} \,.
}
Indeed, in passing from \eqref{ZS3NotCartan} to \eqref{ZS3} one has to introduce a Vandermonde determinant equal to $\det{}^\prime_\text{adj} \sigma$ that cancels the denominator of the last factor in \eqref{ZS3NotCartan}, as well as divide by $|\cW|$, the order of the residual discrete gauge symmetry.

We then notice that 
\es{ProductFormula}{
	\frac{2 \sinh (\pi a)}{a} = {\cal C} \prod_{n =1}^\infty (n^2 + a^2)  \,,
}
where the prefactor ${\cal C}$ amounts to a divergent $a$-independent normalization.  It follows that up to an overall normalization factor, we can write \eqref{ZS3NotCartan} as 
\es{ZGhosts}{
	Z = \int d \sigma\, Z_\sigma \int D'c\, D' \tc \, \exp \left[ -\int d\varphi\, \tc\, \partial_\varphi (\partial_\varphi + \sigma) c  \right] \,,
}
where $c$ and $\tc$ are Lie-algebra valued fermionic ghosts that are periodic on the circle, and where the primes in the ghost integration measure mean that we are not integrating over the ghost zero modes.  Indeed, to show that \eqref{ZGhosts} reduces to \eqref{ZS3NotCartan}, note that integrating out the pair of modes of $c$ and $\tc$ whose $\vphi$ dependence is proportional to $e^{i n \vphi}$ and $e^{-i n \vphi}$, with $n \geq 1$, results in a factor of 
\es{Factor}{
	n^2 (n^2 + \sigma^2)  \,.
}
Using \eqref{ProductFormula}, we see that the product of \eqref{Factor} over all integer $n \geq 1$ gives \eqref{ZS3NotCartan} up to an unimportant overall normalization factor.
	
Lastly, we interpret \eqref{ZGhosts} as the ghost action corresponding to the gauge-fixing condition $\partial_\vphi {\cal A}_\vphi = 0$ for a 1d gauge field ${\cal A} = {\cal A}_\vphi d \vphi$.  Thus, combining all the steps mentioned above, we find that the theory \eqref{ZS3} is a gauge-fixed version of the gauged quantum mechanics
\es{ZGaugedQM}{
	Z = \int D{\cal A} \, DQ\, D \tQ \exp \left[- \ell \int_{-\pi}^\pi d \vphi\, \tQ {\cal D}_\vphi Q \right] \,, \qquad
	{\cal D}_\vphi \equiv \partial_\vphi + {\cal A}_\vphi \,.
}
In order to match \eqref{ZGhosts}, the gauge fixing condition is solved by ${\cal A}_\vphi = \sigma$.  
	
So far we have been cavalier about the proper integration cycle for $Q$, $\tQ$, ${\cal A}$, and the ghosts that would make the formal manipulations above well-defined.  We leave a careful discussion for future work, but we would like to point out that this integration cycle is likely to be quite non-trivial. For instance, we know from Section~\ref{REALITY} that when ${\cal A}$ is restricted to constant values $\sigma$ in the Cartan of the gauge algebra, one should choose a slightly unusual integration cycle for $Q$ and $\tQ$.  When ${\cal A}$ is not constant, one should choose an integration cycle which is middle-dimensional in the space of complex $Q$, $\tQ$, and ${\cal A}$ and that makes the path integral \eqref{ZGaugedQM} well-defined.  Hopefully such a cycle exists and can be used to show that, despite the lack of the usual factor of $i$ multiplying the gauge connection in the covariant derivative, the action \eqref{ZGaugedQM} still enjoys gauge invariance.

Assuming that the issues raised in the previous paragraph can be resolved, the description \eqref{ZGaugedQM} makes more transparent the properties of \eqref{ZS3} mentioned earlier.  In particular, the gauged quantum mechanics \eqref{ZGaugedQM} is topological\footnote{For quantum mechanics, ``topological'' is synonymous to ``zero Hamiltonian.'' We use the Schwarz type notion of a ``topological'' quantum mechanics, which means that Hamiltonian is precisely zero, not up to $\cQ$-exact terms.} because it is invariant under reparameterizations of the circle, so all correlation functions of gauge invariant operators must be topological.   In addition, the operator relation \eqref{Relation1d} is imposed pointwise by the path integral over ${\cal A}_\vphi$:  the gauge field ${\cal A}_\vphi$ acts as a Lagrange multiplier. 
	
It is worth noting that the topological quantum mechanics on a circle is an interesting theory which has previously appeared in the literature in the context of deformation quantization (see \cite{Ded} and the footnote 3 of \cite{Catt}) and the study of quantum geometry of phase spaces \cite{Ded}, as well as the analytic continuation of quantum mechanics path integral in \cite{WittenQM}, where the zero Hamiltonian quantum mechanics was the most basic example studied. The fields there were taken to be periodic on the circle, in which case the theory had a zero mode corresponding to translations along the circle. The proper definition of the theory involved regularizing that zero mode, which can be done in a variety of ways (fixing zero mode, delta function insertion, adding a small Hamiltonian, considering compact or finite volume phase spaces) and which can introduce various subtleties. In our case, the fields $Q$ and $\tQ$ naturally appear as antiperiodic on the circle, which is yet another way of solving the zero mode problem.

\subsubsection{Non-vanishing mass and FI parameters}
\label{NONCONFORMALPROPERTIES}
	
When $m$ and $\zeta$ are not necessarily vanishing, the discussion above is modified as follows.  The first modification is that all operators in the theory \eqref{ZS3} obey $\partial_\vphi {\cal O} = - \left[ {\cal R}_{\cal O}(\sigma) +  {\cal R}_{\cal O}(m r) \right] {\cal O}$, where ${\cal R}_{\cal O}$ is the representation of ${\cal O}$ under the product of the gauge and flavor groups.  For gauge invariant operators ${\cal R}_{\cal O} (\sigma) = 0$, but ${\cal R}_{\cal O}(m r)$ may not vanish when $m \neq 0$. Consequently, the correlation functions of ${\cal O}$ may no longer be topological.  The position dependence  of these correlation functions is rather simple, however.  We have
\es{CorrNonTop}{
	\langle \cO_1(\vphi_1) \cdots \cO_n (\vphi_n) \rangle = e^{- \sum_{i=1}^n {\cal R}_{\cO_i}(m r) \vphi_i}
	\times (\text{topological correlation function}) \,.
}
For example, if ${\cal O}_i$ has charge $q_j$ associated with the flavor $U(1)$ symmetry with real mass parameter $m_j$, the position dependence on $\vphi_i$ is $e^{- r \sum_j q_j m_j \phi_i}$.  The non-trivial position dependence can also be understood from the fact that, as explained in Section~\ref{COHOMOLOGY2}, in the presence of real mass parameters, the twisted translation $\hat P_\varphi$ is cohomologous to $-r \hat m$, where $\hat m$ is the real mass operator.
	
The second modification of the discussion in the previous section is that the RHS of the D-term relation \eqref{Relation1d} receives contributions proportional to the FI parameters.  Note that non-vanishing FI parameters do not affect the topological nature of the correlation function.
	
Lastly, the rewriting of the 1d Gaussian theory coupled to a matrix model \eqref{ZS3} as a gauge-fixed gauged quantum mechanics \eqref{ZGaugedQM} is straightforward to perform in the presence of non-zero $m$ and $\zeta$.  The covariant derivative in \eqref{ZGaugedQM} simply gets modified to ${\cal D}_\vphi + m r$ and the path integral acquires an extra factor of $e^{i\ell \int_{-\pi}^\pi \tr_\zeta {\cal A}}$. This factor is a 1d analog of an analytically continued Chern-Simons term.  A precise understanding of this term in the gauged quantum mechanics requires studying the middle-dimensional integration cycle in the space of complex fields $Q$, $\tQ$, and ${\cal A}$.  As already mentioned before, we postpone this task for future work.

\subsection{Correlators of twisted Higgs branch operators}

Having understood the main properties of the theory \eqref{ZS3}, let us move on to applications.  The main application of the theory \eqref{ZS3} (or \eqref{ZGaugedQM}) is to calculate correlation functions of twisted Higgs branch operators.  We again first focus on the conformal case, and then extend this discussion to the non-conformal deformations.
	
\subsubsection{The conformal case}

The most basic correlation functions are the 2- and 3-point correlators.  When $m = \zeta = 0$, we have 
\es{2pt3pt}{
	\langle {\cal O}_i(\varphi_1) {\cal O}_j(\varphi_2) \rangle = b_{ij} \,, \qquad
	\langle {\cal O}_i(\varphi_1) {\cal O}_j(\varphi_2) {\cal O}_k(\varphi_3)\rangle = c_{ijk} \,,\qquad 
	\varphi_1 < \varphi_2 < \varphi_3 \,,
}   
where $b_{ij}$ and $c_{ijk}$ are constants independent of the insertion points in the range where $\vphi_1 < \vphi_2 < \vphi_3$.  All higher-point correlation functions are determined by \eqref{2pt3pt} through a successive use of the OPE
\es{1dOPE}{
	{\cal O}_i (\varphi) {\cal O}_j (\varphi') = \sum_k c_{ij}{}^k  {\cal O}_k(\varphi') \,,
	\qquad \varphi < \varphi' \,,
}
This OPE should be interpreted as an identity when used inside correlation functions of adjacent insertions of twisted Higgs branch operators.  The OPE coefficients $c_{ij}{}^k$ can be extracted from the 2-point and 3-point functions \eqref{2pt3pt}:
\es{Gotc}{
	c_{ij}{}^k = b^{kl} c_{ijl} \,,
}   
where the matrix $b^{ij}$ is the inverse of $b_{ij}$ obeying $b^{ij} b_{jk} = \delta^i_k$.  In practice, it is convenient to work with a basis of operators that diagonalizes the 2-point function, such that $b_{ij} = b_i \delta_{ij}$ and $b^{ij} = (1/b_i) \delta^{ij}$.  The lack of an explicit position dependence in the coefficients $c_{ij}{}^k$ is a reflection of the topological nature of the correlation functions of gauge-invariant twisted Higgs branch operators.  
	
The quantities of main interest in any given application of the 1d theory \eqref{ZS3} are the OPE coefficients $c_{ij}{}^k$, because they contain all the information necessary to construct correlation functions of twisted Higgs branch operators.  In particular, in the special case where $\cO_k$ is the identity operator, $\cO_k = 1$, we have $c_{ij}{}^k = b_{ij}$, and thus from the $c_{ij}{}^k$ one can immediately extract the coefficients $b_{ij}$ and $c_{ijk}$ defined above.  The OPE coefficients $c_{ij}{}^k$ are thus quantities we will calculate in explicit examples in Section~\ref{CFTAPPLICATIONS}.  
	
\subsubsection{Non-vanishing mass and FI parameters}
	
When $m=0$ but $\zeta \neq 0$, the correlation functions of the twisted Higgs branch operators are still topological and take the form \eqref{2pt3pt}.  The discussion above holds in this case too.
	
When $m \neq 0$, the correlation functions acquire position dependence, as explained in Section~\ref{NONCONFORMALPROPERTIES}.  For the given ordering of the insertion points $\vphi_1 < \vphi_2 < \ldots$, we have
\es{2pt3ptMassive}{
	\langle {\cal O}_i(\varphi_1) {\cal O}_j(\varphi_2) \rangle &= b_{ij}  e^{- {\cal F}_{\cO_i}(m r) \vphi_1- {\cal F}_{\cO_j}(m r) \vphi_2}  \,, \qquad  \vphi_1 < \vphi_2 \\
	\langle {\cal O}_i(\varphi_1) {\cal O}_j(\varphi_2) {\cal O}_k(\varphi_3)\rangle 
	&= c_{ijk} e^{- {\cal F}_{\cO_i}(m r) \vphi_1- {\cal F}_{\cO_j}(m r) \vphi_2 - {\cal F}_{\cO_k}(m r) \vphi_3}  \,,\qquad 
	\varphi_1 < \varphi_2 < \varphi_3 \,,
}  
with $b_{ij}$ and $c_{ijk}$ being constants, and ${\cal F}$ being the map from the flavor symmetry algebra into complex $\dim {\cal R} \times \dim {\cal R}$ matrices defined in Footnote~\ref{FlavorFootnote}.  The 1d OPE \eqref{1dOPE} is replaced by
\es{1dOPEMassive}{
	{\cal O}_i (\varphi) {\cal O}_j (\varphi') = \sum_k c_{ij}{}^k  {\cal O}_k(\varphi') 
	e^{- {\cal F}_{\cO_i}(m r) \vphi- {\cal F}_{\cO_j}(m r) \vphi' + {\cal F}_{\cO_k}(m r) \vphi'} \,,
	\qquad \varphi < \varphi' \,,
}
with constant $c_{ij}{}^k$, which can now be used in any higher-point functions of twisted Higgs branch operators between adjacent insertions.  The relation between $c_{ij}{}^k$ and $b_{ij}$ and $c_{ijk}$ is then as in the conformal case.  
	
While in a general non-conformal QFT it is not always possible to find a basis of operators whose matrix of two-point functions is diagonal, in the 1d theory corresponding to the twisted Higgs branch operators studied here one can diagonalize $b_{ij}$ through the Gram-Schmidt procedure.

\subsection{2- and 3-point correlators of Higgs branch operators of the SCFT}
	
As advertised in the Introduction, when $m = \zeta = 0$, we can also infer the 2- and 3-point functions of the {\em untwisted} Higgs branch operators.  Concretely, let us make contact between \eqref{2pt3pt} and the 2- and 3-point functions of untwisted Higgs branch operators in flat space.  For a Higgs branch operator $\cO_{a_1 \cdots a_n}(\vec{x})$ of scaling dimension $\Delta_{\cal O} = n/2$ let us define the index free operator
\es{OHiggsUntwisted}{
	\cO^{(n)}(\vec{x}, U) = U^{a_1} \cdots U^{a_n} \cO_{a_1 \cdots a_n}(\vec{x})
}
by contracting the $\mathfrak{su}(2)_H$ indices with polarizations $U$.  Conformal invariance and $\mathfrak{su}(2)_H$ symmetry imply that we can choose a basis of operators where the only non-vanishing 2-point functions are
\es{2ptUntwisted}{
	\langle \cO_i^{(n)} (\vec{x}_1, U_1 ) \cO_j^{(n)} (\vec{x}_2, U_2)  \rangle = B_{ij} \frac{\langle U_1, U_2 \rangle^n}{\abs{\vec{x}_1 - \vec{x}_2}^n} \,,
} 
where $B_{ij}$ are constants and where $\langle U_1 , U_2 \rangle \equiv U_1^a U_2^b \varepsilon_{ab}$.\footnote{Recall that $\varepsilon_{12} = -\varepsilon_{21} = -1$.}  The 3-point functions of Higgs branch operators are also constrained by conformal and $\mathfrak{su}(2)_H$ symmetry to take the form
\es{3ptUntwisted}{
	\langle \cO_i^{(n_i)} (\vec{x}_1, U_1 ) \cO_j^{(n_j)} (\vec{x}_2, U_2) \cO_k^{(n_k)} (\vec{x}_3, U_3)  \rangle 
	= C_{ijk} \frac{\langle U_1, U_2 \rangle^{\frac{n_i + n_j - n_k }{2}} \langle U_1, U_3 \rangle^{\frac{n_i + n_k - n_j }{2}} \langle U_2, U_3 \rangle^{\frac{n_j + n_k - n_i }{2}} }
	{\abs{\vec{x}_1 - \vec{x}_2}^{\frac{n_i + n_j - n_k }{2}} \abs{\vec{x}_1 - \vec{x}_3}^{\frac{n_i + n_k - n_j }{2}} \abs{\vec{x}_2 - \vec{x}_3}^{\frac{n_j + n_k - n_i }{2}} }
} 
where $C_{ijk}$ are constants.
	
The untwisted operators are related to the twisted ones by setting $U^a = u^a_{\R^3}$ as in \eqref{TwistedFlat}.   A simple calculation shows that
\es{RelationsBC}{
	B_{ij} = \left( \frac{\ell}{2 \pi} \right)^n b_{ij} \,, \qquad
	C_{ijk} = \left( \frac{\ell}{2 \pi} \right)^{\frac{n_i + n_j + n_k}2} c_{ijk} \,.
}
If we calculate $b_{ij}$ and $c_{ijk}$ using the model \eqref{ZS3}, we can easily extract the coefficients appearing in the untwisted correlators \eqref{2ptUntwisted}--\eqref{3ptUntwisted} using \eqref{RelationsBC}.

\subsection{Star product, Higgs branch chiral ring, and deformation quantization}
\label{STAR}
	
More abstractly, in the conformal case $m = \zeta = 0$ we can represent the OPE \eqref{1dOPE} as the star product operation
\es{StarDef}{
	{\cal O}_i \star {\cal O}_j = \sum_k c_{ij}{}^k {\cal O}_k \,,
}
which can be thought of as a non-commutative multiplication operation on the algebra of twisted Higgs branch operators.   What is slightly less obvious is that the star product \eqref{StarDef} is also a non-commutative multiplication on the Higgs branch chiral ring.   Indeed, it is not hard to see that the twisted Higgs branch operators are in 1-to-1 correspondence with Higgs branch chiral ring operators---at any given $\varphi$, the twisted Higgs branch operators agree precisely with the Higgs branch chiral ring operators with respect to an ${\cal N} = 2$ subalgebra, which is to say that they can be represented by holomorphic functions on the Higgs branch with respect to an appropriate choice of complex structure.  To be concrete, let us take $\vphi = 0$ and  denote the holomorphic function associated with ${\cal O}(0)$ by $f_{\cal O}$.  The star product \eqref{StarDef} can therefore also be thought of as acting on the Higgs branch chiral ring:
\es{starChiral}{
	f_{\cO_i} \star f_{\cO_j} = \sum_k c_{ij}{}^k f_{\cO_k} \,.
}
As explained in \cite{Beem:2016cbd}, \eqref{starChiral} represents a deformation quantization of the usual commutative product $ f_{\cO_i \cO_j}  = f_{\cO_i} f_{\cO_j}$ given by the multiplication of the corresponding holomorphic functions, with the deformation parameter equal to $\ell^{-1}$.  Indeed, the star product \eqref{starChiral} reduces to this commutative product in the $\ell \to \infty$ limit.

It was explained in \cite{Beem:2016cbd} that the descent of the twisted Higgs branch operators from the untwisted ones of the 3d ${\cal N} = 4$ SCFT, which have scaling dimension equal to the $\mathfrak{su}(2)_H$ spin, $\Delta = n/2$, yields various very special properties of the 1d operator algebra, and consequently of the OPE \eqref{1dOPE} and of the star product \eqref{StarDef}.  In some cases, it was shown in \cite{Beem:2016cbd} that these properties determine the star product uniquely up to a finite number of parameters.  Since we will not use these properties directly, we refer the reader to the discussion in \cite{Beem:2016cbd}.  In the rest of this paper, we use the 1d theory of the previous section to compute the star product explicitly in a few examples and compare with the results of \cite{Beem:2016cbd}.  In particular, we determine the parameters left undetermined in \cite{Beem:2016cbd}.
	
In the non-conformal case when $m$ or $\zeta$ are non-zero, one can still define a star product operation as in \eqref{StarDef} using the coefficients $c_{ij}{}^k$ appearing in \eqref{1dOPEMassive}.   The star product defined in this way obeys all the properties discussed in \cite{Beem:2016cbd} except for the property referred to as evenness in \cite{Beem:2016cbd}.  There is thus more freedom in the star product of a non-conformal theory relative to the conformal case.

\subsection{Operator mixing}
\label{MIXING}
	
As a last point in the discussion on how to use \eqref{ZS3} to calculate explicit correlation functions, we stress that the definition of the twisted Higgs branch operators in terms of the fields of the 1d theory may suffer from operator mixing ambiguities even in the conformal case\@.  These ambiguities are reflected in non-zero two-point functions between operators of different scaling dimensions, and can be removed upon performing a Gram-Schmidt diagonalization procedure.   From the point of view of the parent ${\cal N} = 4$ SCFT, the freedom that is used for defining orthogonal operators is that of adding lower dimensional operators multiplied by powers of the curvature tensor.  Indeed, suppose we have identified a basis of orthogonal operators with scaling dimension strictly less than $\Delta$.  In defining an operator ${\cal O}_\Delta$ of dimension $\Delta$ that is orthogonal to all lower dimension operators, one can start with some choice $\widehat{\cal O}_\Delta$ that is not necessarily orthogonal to all lower dimension operators and consider the linear combination 
\es{CurvCouplings}{
	{\cal O}_\Delta = \widehat{\cal O}_\Delta + \sum_i \alpha_{1, i} R {\cal O}_{\Delta-2, i} + \sum_i \alpha_{2, i} R^2 {\cal O}_{\Delta - 4, i} + \cdots
}  
where $R^k$ denotes some contraction of the $k$th power of the Riemann tensor.  One should then adjust the coefficients $\alpha_{k, i}$ such that ${\cal O}_\Delta$ is orthogonal to all lower dimension operators.  The construction of orthogonal twisted Higgs branch operators  follows a similar recursive pattern, where we can simply replace $R^k$ by $1/\ell^{2k}$ after redefining the $\alpha_{k, i}$ by a dimensionless multiplicative constant.  Note that if there are any flavor symmetries present, then the lower dimension operators included in \eqref{CurvCouplings} should transform in the same representation of the flavor symmetry as $\widehat{\cal O}_\Delta$.  In practice, we remove the operator mixing by diagonalizing the matrix of 2-point coefficients $b_{ij}$ defined in \eqref{2pt3pt}.

\section{Applications to SCFTs}
\label{CFTAPPLICATIONS}
	
We now discuss specific examples in SCFTs, where $m = \zeta = 0$.
	
\subsection{SQED with $N$ charged hypermultiplet flavors}
\label{SQEDSECTION}

Our first example is ${\cal N} = 4$ SQED with $N$ charged hypermultiplet flavors.  Without adding any real masses or FI terms, this theory is believed to flow to an interacting SCFT in the IR\@.  The matter content of the 1d theory consists of fields $\tQ_I(\varphi)$ and $Q^I(\varphi)$, with $I = 1, \ldots, N$, and has the partition function
\es{PartSQED}{
	Z &= \int d\sigma\, Z_\sigma \,, \\
	Z_\sigma &=  \int D\tQ_I  D Q^I \exp\left[ - \ell \int d\varphi\, \left( \tQ_I \partial_\varphi Q^I + \sigma \tQ_I Q^I  \right) \right] = \frac{1}{\left[ 2 \cosh(\pi \sigma) \right]^N} \,.
}
	
In preparation for describing all the gauge-invariant operators of the 1d theory, let us observe that $\sigma$ acts as a Lagrange multiplier imposing the constraint $\tQ_I Q^I = 0$ as an operator relation. (See also the discussion in Section~\ref{TOPOLOGICALQM}.)  Indeed, taking a derivative of the integrand of \eqref{PartSQED} with respect to $\sigma$, one obtains the integrated identity
\es{Integrated}{
	\int d\varphi\, \langle 
	\tQ_I Q^I(\varphi)\, {\cal O}_1(\varphi_1)  \cdots {\cal O}_n(\varphi_n) \rangle = 0 
}
where ${\cal O}_i(\varphi_i)$ are any gauge-invariant insertions.  Let us assume without loss of generality that $\varphi_1 < \varphi_2 < \ldots \varphi_n$.   Because $\langle \tQ_I Q^I(\varphi)\, {\cal O}_1(\varphi_1)  \cdots {\cal O}_n(\varphi_n) \rangle$ is a correlation function of gauge-invariant operators, it is topological, or in other words
\es{TopCorr}{
	\langle \tQ_I Q^I(\varphi)\, {\cal O}_1(\varphi_1)  \cdots {\cal O}_n(\varphi_n) \rangle = \alpha_k \,, \qquad
	\text{if $\varphi \in (\varphi_{k}, \varphi_{k+1})$} \,,
}
for some constants $\alpha_k$, with $1 \leq k \leq n$.  Here, we identified $\varphi_{n+1} =  \varphi_1 + 2 \pi$. Then \eqref{Integrated} reduces to
\es{alphaSum}{
	\sum_{k=1}^n \alpha_k (\varphi_{k+1} - \varphi_k) = 0 \,.
}
This equation should hold for any $\varphi_k$ obeying $\varphi_1 < \varphi_2 < \ldots \varphi_n < \varphi_{n+1} = \varphi_1 + 2 \pi$, which implies that $\alpha_k = 0$ for all $k$.  Consequently $\tQ_I Q^I(\varphi) = 0$ in all correlation functions.  As explained in Section~\ref{TOPOLOGICALQM}, this relation also follows from the D-term relations of the 3d theory.
	
As a further corrolary, we have that, up to mixings with lower dimension operators as discussed in Section~\ref{MIXING}, composite operators $\tQ_I Q^I {\cal O}(\varphi)$ also vanish in correlation functions.  Here, ${\cal O}$ is any gauge-invariant operator.  Indeed, a convenient definition of $\tQ_I Q^I {\cal O}(\varphi)$ is through the limit
\es{limDef}{
	\lim_{\substack{\varphi' \to \varphi \\ \varphi' > \varphi}} \tQ_I Q^I(\varphi) {\cal O}(\varphi') \,.
}
Because $\tQ_I Q^I(\varphi) = 0$, as shown above, it follows that the operator defined in \eqref{limDef} also vanishes in correlation functions.  Any other definition of $\tQ_I Q^I {\cal O}(\varphi)$ differs from \eqref{limDef} by lower dimension operators, so one concludes that indeed, $\tQ_I Q^I {\cal O}(\varphi)$ equals a linear combination of lower dimension operators, with the precise linear combination depending on the precise definition of $\tQ_I Q^I {\cal O}(\varphi)$.  We call such an operator redundant.

Let us now describe the twisted Higgs branch operators of this theory.  They are in 1-to-1 correspondence with the Higgs branch chiral ring operators of the SCFT, or equivalently with the holomorphic functions on the Higgs branch, as described in Section~\ref{STAR}.\footnote{The Higgs branch of SQED with $N$ hypermultiplets is the hyperk\"ahler quotient $\HH^N /// U(1)$.}  The twisted Higgs branch operators can be constructed as gauge invariant words in $Q^I$ and $\tQ_I$ modulo the complex D-term relation $\tQ_I Q^I = 0$ proven above, with $Q^I$ and $\tQ_I$ having gauge charges $+1$ and $-1$, respectively.  The result is an algebra of operators generated by the traceless bilinears
\es{AdjOps}{
	{\cal J}_I{}^J = \tQ_I Q^J - \frac 1N \delta_I^J \tQ_K  Q^K \,,
}
obeying the nilpotency constraint ${\cal J}_I{}^J {\cal J}_J{}^K = 0$. This nilpotency constraint should be interpreted as the relation $f_{{\cal J}_I{}^J} f_{ {\cal J}_J{}^K} = f_{{\cal J}_I{}^J {\cal J}_J{}^K}= 0$ in the Higgs branch chiral ring, which means  that the operator ${\cal J}_I{}^J {\cal J}_J{}^K$ is redundant.  This relation can also be recovered from the $\ell \to \infty$ limit of the star product, namely $ \lim_{\ell \to \infty} {\cal J}_I{}^J \star {\cal J}_J{}^K = 0$, as we will see below.

Note that given that $\tQ_K  Q^K = 0$, one need not write the second term in \eqref{AdjOps} that makes manifest the tracelessness of ${\cal J}_I{}^J$.  However, we find it convenient to explicitly remove the $SU(N)$ traces in the definitions of the various operators considered here, as in \eqref{AdjOps}, because with such a traceless definition group theory arguments guarantee that there is no mixing with lower dimension operators.
	
The other linearly independent twisted Higgs branch operators can be taken to be products of ${\cal J}_I{}^J$ symmetrized in their upper and lower indices separately and with no indices contracted and the traces removed, namely
\es{Jp}{
	{\cal J}_{I_1 I_2 \ldots I_p}{}^{J_1 J_2 \ldots J_p} \equiv {\cal J}_{(I_1}^{(J_1} {\cal J}_{I_2}^{J_2} \cdots {\cal J}_{I_p)}^{J_p)} - \text{traces} \,.
}
For instance, we have
\es{J2}{
	{\cal J}_{I_1 I_2}{}^{J_1 J_2} &= \tQ_{I_1} \tQ_{I_2} Q^{J_1} Q^{J_2}
	- \frac{4 \tQ_K Q^K}{N+2}  \delta_{(I_1}^{(J_1} \tQ_{I_2)} Q^{J_2)}  + \frac {2 (\tQ_K Q^K)^2}{(N+1)(N+2)} \delta_{(I_1}^{(J_1} \delta_{I_2)}^{J_2)}  \,.
}
	
From a group theory perspective, the Higgs branch chiral ring admits an action of the $SU(N)$ flavor symmetry of the SQED under which $Q^I$ and $\tQ_I$ transform as a fundamental and as an anti-fundamental, respectively.\footnote{ Under the complexified action of $\mathfrak{sl}(N) \cong \mathfrak{su}(N)$, the Higgs branch, seen as a complex manifold, can be identified with the minimal nilpotent orbit of $\mathfrak{sl}(N)$.  The Higgs branch chiral ring can be identified with the ring of holomorphic functions on the Higgs branch.}   The algebra of operators decomposes under this action as
\es{Decomp}{
	\bigoplus_{p=0}^\infty [p, 0, 0, \ldots, 0, p] \,,
}
where the term $[p, 0, 0, \ldots, 0, p]$, appearing only once, is represented precisely by \eqref{Jp}.  
	
As reviewed in Section~\ref{STAR}, the correlation functions in the 1d theory can be deduced from the non-commutative star product defined in \eqref{StarDef}.  In the case of minimal nilpotent orbits of classical groups (other than $SU(2)$), it was shown in \cite{Beem:2016cbd} that the star product is uniquely determined (see for instance\cite{Joung:2014qya}) by its properties, and consequently all correlation functions of the operators \eqref{Jp} are also uniquely determined.\footnote{In the case of $SU(2)$ the star product is determined up to a free parameter.  There exist different SCFTs for which this parameter takes distinct values.}  Nevertheless, it is instructive to see how these correlation functions are computed within our formalism.
	
To avoid dealing with $SU(N)$ indices, it is convenient to contract them into polarization vectors $(y^I, \bar y_I)$ obeying $\bar y \cdot y \equiv \bar y_I y^I = 0$.  So let us define
\es{JpDef}{
	{\cal J}^{(p)}(\varphi, y, \bar y) = {\cal J}_{I_1 I_2 \ldots I_p}{}^{J_1 J_2 \ldots J_p}
	y^{I_1} \cdots y^{I_p} \bar y_{J_1} \cdots \bar y_{J_p} \,.
}
Using \eqref{GSummary} for each flavor, one can easily express the two-point functions of ${\cal J}^{(p)}$ as 
\es{TwoPointJp}{
	\langle {\cal J}^{(p)}(\varphi_1, y_1, \bar y_1) {\cal J}^{(p)}(\varphi_2, y_2, \bar y_2) \rangle
	= (\bar y_1 \cdot y_2)^p (y_1 \cdot \bar y_2)^p \frac{ \int d\sigma\, Z_\sigma \, \left[ G_\sigma(\varphi) G_\sigma(-\varphi) \right]^p }{\int d\sigma\, Z_\sigma} \,,
}
with the propagator $G_\sigma(\varphi)$ given in \eqref{GSummary}.  We have
\es{IntegralGG}{
	\int d\sigma\, Z_\sigma \, \left[ G_\sigma(\varphi) G_\sigma(-\varphi) \right]^p 
	= \int d\sigma \frac{(-1)^p  \sech^{N + 2p}(\pi \sigma)}{2^{N + 2p} \ell^{2p}}
	= \frac{(-1)^p \Gamma\left( \frac{N}{2} + p \right)}{2^{N+ 2p} \sqrt{\pi} \ell^{2p} \Gamma\left( \frac{N+1}{2} + p \right)}
}
which immediately implies
\es{JpJp}{
	\langle {\cal J}^{(p)}(\varphi_1, y_1, \bar y_1) {\cal J}^{(p)}(\varphi_2, y_2, \bar y_2) \rangle
	= \frac{(-1)^p}{(2 \ell)^{2p}} \frac{\Gamma\left(\frac{N+1}{2} \right) \Gamma\left(\frac{N}{2} + p \right)}
	{\Gamma\left(\frac{N}{2} \right) \Gamma\left(\frac{N+1}{2} + p \right)} (\bar y_1 \cdot  y_2 )^p (y_1 \cdot  \bar y_2 )^p  \,.
}
	
It is not hard to compare this relation with the general expectation of \cite{Joung:2014qya}.  To do so, let us group together the ${\cal J}^{(p)}$ into a single quantity
\es{JTotal}{
	{\cal J}(\varphi, y, \bar y) = \sum_{p=0}^\infty  \ell^p {\cal J}^{(p)} (\varphi, y, \bar y)
}
from which ${\cal J}^{(p)} (\varphi, y, \bar y)$ can be identified with the term of total degree $2p$ in $(y, \bar y)$, and the factor of $\ell^p$ was inserted such that all terms in the sum have the same scaling dimension.  Then \eqref{JpJp} implies
\es{JTotJTot}{
	\langle {\cal J}(\varphi_1, y_1, \bar y_1) {\cal J}(\varphi_2, y_2, \bar y_2) \rangle
	= {}_3 F_2 \left( \frac{N}{2}, \frac{N}{2}, 1; \frac{N}{2} \frac{N+1}{2}; -\frac{(\bar y_1 \cdot  y_2 ) (y_1 \cdot  \bar y_2 )}4  \right) \,.
}
Comparing with Eq.~(1.3) of \cite{Joung:2014qya}, we see that these expressions agree precisely with the bilinear form on the generalized higher spin algebra $\mathfrak{hs}_\lambda (\mathfrak{sl}(N))$, with the parameter $\lambda$ taking the value $\lambda = 0$!\footnote{In making this comparison, one has to convert the matrix polarization $V$ used in that reference to the vector polarizations used here.  The relation is $V_I{}^J = \bar y_I y^J$.  Then $(\bar y_1 \cdot  y_2 ) (y_1 \cdot  \bar y_2 ) = \tr (V_1 V_2)$. \label{VFootnote}}
	
One can also calculate three-point functions.  For instance, for $\varphi_1 < \varphi_2 < \varphi_3$ we have
\es{J1J1J1}{
	&\langle {\cal J}^{(1)}(\varphi_1, y_1, \bar y_1) {\cal J}^{(1)}(\varphi_2, y_2, \bar y_2) {\cal J}^{(1)}(\varphi_3, y_3, \bar y_3) \rangle \\
	&\qquad= -\frac{N}{8 \ell^3 (N+1)} \left[ (\bar y_1 \cdot  y_3 ) (\bar y_3 \cdot  y_2 )(\bar y_2 \cdot  y_1 )
	-  (\bar y_1 \cdot  y_2 ) (\bar y_2 \cdot  y_3 )(\bar y_3 \cdot  y_1 ) \right] \,,
}    
which also agrees with the results of \cite{Joung:2014qya}.\footnote{To compare, in the convention of Footnote~\ref{VFootnote}, we can write $(\bar y_1 \cdot  y_3 ) (\bar y_3 \cdot  y_2 )(\bar y_2 \cdot  y_1 ) = \tr (V_1 V_2 V_3)$ and $ (\bar y_1 \cdot  y_2 ) (\bar y_2 \cdot  y_3 )(\bar y_3 \cdot  y_1 ) = \tr (V_3 V_2 V_1)$.  Then \eqref{J1J1J1}, with $\ell=1$, matches the terms cubic in $V$ in the expansion of Eq.~(1.4) of \cite{Joung:2014qya}.\label{V3Footnote}}   From this expression as well as from \eqref{JpJp} and \eqref{2pt3pt}--\eqref{StarDef}, we can extract the star product of the generator ${\cal J}_I{}^J$:
\es{starSQED}{
	{\cal J}_I{}^J \star  {\cal J}_K{}^L
	= {\cal J}_{IK}{}^{JL} - \frac{1}{2 \ell} \left( \delta_K^J {\cal J}_I{}^L
	- \delta_I^L {\cal J}_K{}^J \right) 
	-\frac{N}{ 4 \ell^2 (N+1)}  \left( \delta_I^L \delta_K^J - \frac{1}{N} \delta_I^J \delta_K^L \right) \,.
}     
In the $\ell \to \infty$ limit, this expression reduces to the commutative product on the Higgs branch chiral ring, ${\cal J}_I{}^J  {\cal J}_K{}^L = {\cal J}_{IK}{}^{JL}$, or more precisely to the relation $f_{{\cal J}_I{}^J}  f_{{\cal J}_K{}^L} = f_{{\cal J}_{IK}{}^{JL}}$ written in terms of the holomorphic functions on the Higgs branch.  This commutative product follows from \eqref{AdjOps} and \eqref{J2} as well as from the condition that in the Higgs branch chiral ring we have the relation $\tQ_K Q^K = 0$.

\subsection{$N$-node quiver}
\label{NNODESECTION}
	
The next example we study is that of an $N$-node Abelian quiver gauge theory with $N$ hypermultiplets with charges $(1, -1, 0, 0, \ldots)$, $(0, 1, -1, 0, \ldots)$, and so on.  One has to mod out by the overall $U(1)$ since no matter fields are charged under it---the gauge group is $U(1)^N / U(1)$.  This theory is the mirror dual of SQED with $N$ charged hypermultiplets, as explained in \cite{Intriligator:1996ex}.  
	
The 1d theory describing the twisted Higgs branch operators is
\es{3node}{
	Z &= \int \left( \prod_{j=1}^N d\sigma_j \right)\,  \delta\left( \frac 1N \sum_{j=1}^N \sigma_j \right) Z_\sigma \,, \\
	Z_\sigma &= \int \left( \prod_{j=1}^N D\tQ_j DQ_j \right) \exp 
	\left[-\ell \int d\varphi \sum_{j=1}^N \left( \tQ_j \partial_\varphi Q_j + (\sigma_j - \sigma_{j+1}) \tQ_j Q_j \right) \right]
}
where for the purpose of writing a succinct formula we have defined $\sigma_{N+1} = \sigma_1$.  Integrating out $Q_j$ and $\tQ_j$ gives
\es{calZ3node}{
	Z_\sigma = \prod_{j=1}^N \frac{1}{2 \cosh( \pi (\sigma_j - \sigma_{i+1})) } \,.
}
The $S^3$ partition function itself can be calculated using the trick of writing each factor in \eqref{calZ3node} as a Fourier transform:
\es{identityBasic}{
	\frac{1}{2 \cosh (\pi (\sigma_j - \sigma_{j+1})) } = \int d \tau_j \frac{e^{2 \pi i (\sigma_j - \sigma_{j+1}) \tau_j}}{2 \cosh (\pi \tau_j)} \,.
}
Further performing the integral over $\sigma_j$ and $\tau_j$, one finds
\es{NNodeDual}{
	Z = \int d\tau \frac{1}{\left[2 \cosh(\pi \tau) \right]^N} = 
	\frac{\Gamma\left( \frac{N}{2} \right)}{2^N \sqrt{\pi}  \Gamma\left( \frac{N+1}{2} \right)} \,.
}
This expression agrees precisely with the partition function of SQED with $N$ charged hypermultiplets, as should be the case since the two theories are each other's mirror duals.  The Fourier transform in \eqref{identityBasic} effectively implements the mirror symmetry duality.  See also~\cite{Kapustin:2010xq,Gulotta:2011si} and \cite{Kapustin:1999ha}.

Let us now discuss the twisted Higgs branch operators of this theory and their correlation functions. The Higgs branch is the hyperk\"ahler cone $\C^2 / \Z^N$, and the Higgs branch chiral ring, whose operators are in 1-to-1 correspondence with the twisted Higgs branch operators, is generated by three operators 
\es{Ops3Node}{
	{\cal X} = Q_1 Q_2 \cdots Q_N \,, \qquad
	{\cal Y} = \tQ_1 \tQ_2 \cdots \tQ_N \,, \qquad
	{\cal Z} = \tQ_1 Q_1 =  \tQ_2 Q_2  = \ldots =  \tQ_N Q_N
}
modulo the relation ${\cal X}{\cal Y} = {\cal Z}^N$.  This relation should be interpreted as the $\ell \to \infty$ limit of the star product ${\cal X} \star {\cal Y}$ or as the commutative product on the Higgs branch chiral ring $f_{\cal X} f_{\cal Y} = f_{{\cal Z}^N} = (f_{\cal Z})^N$.   The relations in the last equation of \eqref{Ops3Node} can be seen from the 1d theory \eqref{3node} precisely in the same way as the relation $\tQ_I Q^I = 0$ was derived in SQED around equations \eqref{Integrated}--\eqref{limDef}.  They are imposed by the integration variables $\sigma_j$ which act as Lagrange multipliers.
	
The 1d topological algebra of this theory was also studied in \cite{Beem:2016cbd} quite expliclity in the cases $N=3, 4$.   When $N=3$, for instance, Ref.~\cite{Beem:2016cbd} found that some of the abstract properties on the star product determine it up to two parameters that are denoted by $\alpha_3$ and $\kappa_3$:
\es{3nodeStar}{
	N=3:\qquad  \qquad {\cal Z} \star {\cal Z} &= {\cal Z}^2 - \frac{\alpha_3}{\ell^2} \,, \\
	{\cal Z} \star {\cal X} &= {\cal Z} {\cal X} + \frac 1{2\ell} {\cal X} \,, \\
	{\cal Z} \star {\cal Y} &= {\cal Z}{\cal Y} - \frac 1{2\ell}  {\cal Y} \,, \\
	{\cal X} \star {\cal Y} &= {\cal Z}^3 - \frac 3{2\ell} {\cal Z}^2 - \frac{3 \alpha_3 + \kappa_3}{4 \alpha_3 \ell^2}  {\cal Z} + \frac{3 \alpha_3 + \kappa_3}{2 \ell^3}  \,,
}
etc.  The parameter $\alpha_3$ can be calculated using the supersymmetric localization results of \cite{Jafferis:2010un} combined with the prescription in \cite{Closset:2012vg}.  It is found that \cite{Beem:2016cbd}
\es{GotAlpha}{
	\alpha_3 = \frac{\pi^2 - 8}{4\pi^2} \,.
}  
Lastly, the relation between $\alpha_3$ and $\kappa_3$ was determined in \cite{Beem:2016cbd} using the remaining properties of the star product.  The simple form of the result, namely $\kappa_3 = -1/4$  is suggestive of the existence of an analytical derivation of it.  The computation we are about to perform represents such a derivation.  
	
A similar analysis was performed in~\cite{Beem:2016cbd} in the case $N=4$, where it was found that the first few star product relations are
\es{4nodeStar}{
	N=4:  \qquad {\cal Z} \star {\cal Z} &= {\cal Z}^2 - \frac{\alpha_4}{\ell^2} \,, \\
	{\cal Z} \star {\cal X} &= {\cal Z} {\cal X} + \frac 1{2\ell} {\cal X} \,, \\
	{\cal Z} \star {\cal Y} &= {\cal Z}{\cal Y} - \frac 1{2\ell}  {\cal Y} \,, \\
	{\cal X} \star {\cal Y} &= {\cal Z}^4 - \frac 2{\ell} {\cal Z}^3
	- \frac{- 2 \kappa_4 (\lambda_4 - 4 \alpha_4) + (1 + \lambda_4) (-5 + 6 \lambda_4 - 14 \alpha_4) \alpha_4}
	{7 (\kappa_4 + \alpha_4 (2 - 3 \lambda_4 + 5 \alpha_4)) \ell^2} {\cal Z}^2 \\
	&{}+ \frac{-2 \kappa_4 + \alpha_4 + \lambda_4 \alpha_4}{5 \alpha_4 \ell^3} {\cal Z} 
	- 2 \frac{-2 \kappa_4 + \alpha_4 + \lambda_4 \alpha_4}{5 \ell^4}  \,,  
}
etc., where $\alpha_4$, $\lambda_4$, and $\kappa_4$ are constants.  Using existing supersymmetric localization computations, one can determine $\alpha_4$, while imposing the other properties of the topological operator algebra restricts the space of allowed values of $\kappa_4$ and $\lambda_4$ to a curve.  Which point on this curve corresponds to the $N=4$ quiver theory was not determined.  Our computation below determines it.

Let us now compute various correlation functions from which we can reproduce the algebras \eqref{3nodeStar} and \eqref{4nodeStar} as well as generalizations thereof.   To simplify the following formulas, let us define the integration measure
\es{Measure3Node}{
	d\mu(\sigma_j) \equiv \left( \prod_{i=1}^N d\sigma_j \right)\,  \delta\left( \frac 1N \sum_{j=1}^N \sigma_j \right) Z_\sigma
}
and define $\sigma_{jk} = \sigma_j - \sigma_k$ and $\varphi_{jk} = \varphi_j - \varphi_k$.  Let us start with the 2-point functions 
\es{XYTwo}{
	\langle {\cal Z}(\varphi_1) {\cal Z}(\varphi_2) \rangle &=
	\frac{\int d\mu(\sigma_j) \left[ G_{\sigma_{12}}(\varphi_{12}) G_{\sigma_{12}}(-\varphi_{12})
	+ G_{\sigma_{12}}(0)^2 \right] }{Z} \,, \\
	\langle {\cal X}(\varphi_1) {\cal Y}(\varphi_2) \rangle &= \frac{\int d\mu(\sigma_j)  \prod_{j=1}^N G_{\sigma_{j(j+1)}}(\varphi_{12})
	}{Z}  \,,
}
and the 3-point function
\es{XYZThree}{
	\langle {\cal Z}(\varphi_1) {\cal X}(\varphi_2) {\cal Y}(\varphi_3) \rangle &= \frac{\int d\mu(\sigma_i)  
	\prod_{j=2}^N G_{\sigma_{j(j+1)}}(\varphi_{23}) 
	\left[ G_{\sigma_{12}}(\varphi_{23}) G_{\sigma_{12}}(0)
	+ G_{\sigma_{12}}(\varphi_{13}) G_{\sigma_{12}}(\varphi_{21}) \right] }{Z} \,.
}
	
In these correlation functions, one can again pass to the mirror dual integration variable $\tau$ by performing a Fourier transform.  The result is
\es{XYDual}{
	\langle {\cal Z}(\varphi_1) {\cal Z}(\varphi_2) \rangle &= 
	\frac 1{\ell^2} \frac 1Z \int d\tau \frac{1}{\left[2 \cosh(\pi \tau) \right]^N} \left(i \tau \right)^2 \,, \\
	\langle {\cal X}(\varphi_1) {\cal Y}(\varphi_2) \rangle &= \frac{1}{\ell^N} \frac 1Z \int d\tau \frac{1}{\left[2 \cosh(\pi \tau) \right]^N} 
	\left( i \tau - \frac 12 \right)^N \,, \qquad
	\text{if $\varphi_1 < \varphi_2$} \\
	\langle {\cal Z}(\varphi_1) {\cal X}(\varphi_2) {\cal Y}(\varphi_3) \rangle &= \frac{1}{\ell^{N+1}} \frac 1Z \int d\tau \frac{1}{\left[2 \cosh(\pi \tau) \right]^N} 
	\left( i \tau \left(  i \tau - \frac 12 \right)^N  \right) \,, \qquad
	\text{if $\varphi_1 < \varphi_2 < \varphi_3$}
}

These calculations are sufficient to determine the dimensionless parameters $\alpha_3$ and $\kappa_3$ entering the algebra \eqref{3nodeStar} in the $N=3$ case.  Before we do so, however, let us compute 2- and 3-point functions of some of the composite operators as well.  While so far, the operator mixing discussed in Section~\ref{MIXING} has not been important, it does become important for composite operators.  Our strategy is to first calculate the matrix of two-point functions in some conveniently chosen basis of operators, and afterwards perform a change of basis to an orthogonal set of operators.  
	
Let us focus on the operators ${\cal Z}^p$ with $p \leq N$.  A rather convenient basis is
\es{ZpConvenient}{
	\widehat{{\cal Z}^p} = \prod_{j=1}^p \tQ_j Q_j
}
where the hat on $\widehat{{\cal Z}^p}$ signifies that we do not expect these operators to be orthogonal.  We reserve the notation ${\cal Z}^p$ for the orthogonal operators.  Following the same path as above, one can calculate
\es{ZpZqCorr}{
	\langle  \widehat{{\cal Z}^p} (\varphi_1)  \widehat{{\cal Z}^q} (\varphi_2) \rangle 
	&= \frac 1{\ell^{p+q}} \frac 1Z \int d\tau \frac{1}{\left[2 \cosh(\pi \tau) \right]^N} \left(i \tau \right)^{p+q} \,, \\
	\langle \widehat{{\cal Z}^p}(\varphi_1) {\cal X}(\varphi_2) {\cal Y}(\varphi_3) \rangle &= \frac{1}{\ell^{N+p}} 
	\frac 1Z \int d\tau \frac{1}{\left[2 \cosh(\pi \tau) \right]^N} 
	\left( (i \tau)^p \left(  i \tau - \frac 12 \right)^N  \right) \,, 
}
where in the second equation we assumed $\varphi_1 < \varphi_2 < \varphi_3$.  A closed analytical expression for the integrals appearing in \eqref{ZpZqCorr} does not seem to be available for a generic value of $N$, $p$, and $q$, but these integrals can be performed analytically on a case-by-case basis.  Once these integrals have been performed, we can construct the orthogonal operators ${\cal Z}^p$ recursively via the Gram-Schmidt procedure:
\es{calZp}{
	{\cal Z}^p = \widehat{{\cal Z}^p} - \sum_{q=0}^{p-1} \frac{ \langle  \widehat{{\cal Z}^p} (\varphi_1)  {\cal Z}^q (\varphi_2) \rangle }{ \langle  {\cal Z}^q (\varphi_1)  {\cal Z}^q (\varphi_2) \rangle } {\cal Z}^q \,.
}
	
In the case $N=3$, we have
\es{N3Z}{
	{\cal Z} = \widehat{{\cal Z}} \,, \qquad
	{\cal Z}^2 = \widehat{{\cal Z}^2} - \frac{8 - \pi^2}{4 \pi^2 \ell^2} \,, 
	\qquad {\cal Z}^3 = \widehat{{\cal Z}^3} - \frac{48 - 5 \pi^2}{4 (\pi^2 - 8)} \widehat{{\cal Z}} \,, \qquad \text{etc.}
}
Assuming $\varphi_1 < \varphi_2 < \varphi_3$, we have
\es{XYZ3Node}{
	\langle {\cal Z}(\varphi_1) {\cal Z}(\varphi_2) \rangle &=
	-\frac{\pi^2-8}{4 \pi^2 \ell^2} \,, \\
	\langle {\cal Z}^2(\varphi_1) {\cal Z}^2(\varphi_2) \rangle &=
	\frac{\pi^4 - 8 \pi^2 - 16}{4 \pi^4 \ell^4} \,, \\   
	\langle {\cal Z}^3(\varphi_1) {\cal Z}^3(\varphi_2) \rangle &=
	-\frac{9 \pi^4 - 152 \pi^2 +624}{16 \pi^2 (\pi^2 - 8) \ell^6} \,, \\      
	\langle {\cal X}(\varphi_1) {\cal Y}(\varphi_2) \rangle &=  -\frac{12 - \pi^2}{4 \pi^2 \ell^3} \,, \\
	\langle {\cal Z}(\varphi_1) {\cal X}(\varphi_2) {\cal Y}(\varphi_3) \rangle &= 
	- \frac{12 - \pi^2}{8 \pi^2 \ell^4}  \,, \\
	\langle {\cal Z}^2(\varphi_1) {\cal X}(\varphi_2) {\cal Y}(\varphi_3) \rangle &= 
	- \frac{3(16 + 8 \pi^2 - \pi^4)}{8 \pi^4 \ell^5}  \,, \\
	\langle {\cal Z}^3(\varphi_1) {\cal X}(\varphi_2) {\cal Y}(\varphi_3) \rangle &= 
	- \frac{9 \pi^4 - 152 \pi^2 +624}{16 \pi^2 (\pi^2 - 8) \ell^6}  \,.
}
Combining these expressions with \eqref{2pt3pt}--\eqref{StarDef}, one can derive the star product rules \eqref{3nodeStar} with 
\es{alphaRel}{
	\alpha_3  = \frac{\pi^2-8}{4 \pi^2} \,, \qquad
	\kappa_3 =  -\frac 14 \,.
}
We have thus provided a direct derivation of the result $\kappa_3 = -1/4$ that was found numerically in \cite{Beem:2016cbd}.  A similar exercise for $N=4$ gives
\es{ZZCorr}{
	\langle {\cal Z} (\varphi_1) {\cal Z}(\varphi_2) \rangle &= 
	\frac{6 - \pi^2}{12 \pi^2 \ell^2} \,, \\
	\langle {\cal Z}^2 (\varphi_1) {\cal Z}^2(\varphi_2) \rangle &= 
	\frac{-45 - 30 \pi^2 + 4 \pi^4}{180 \pi^4 \ell^4} \,, \\
	\langle {\cal Z}^3 (\varphi_1) {\cal Z}^3(\varphi_2) \rangle &= 
	-\frac{3 (525 - 170 \pi^2 + 12 \pi^4)}{2880 \pi^2 (-6 + \pi^2) \ell^6} \,, \\
	\langle {\cal Z}^4 (\varphi_1) {\cal Z}^4(\varphi_2) \rangle &= 
	\frac{55125 + 17850 \pi^2 - 6160 \pi^4 + 384 \pi^6}{7350 \pi^2 (-45 - 30 \pi^2 + 4 \pi^4) \ell^8} \,, \\
	\langle {\cal X}(\varphi_1) {\cal Y}(\varphi_2) \rangle &=  \frac{15 - \pi^2}{30 \pi^2 \ell^4} \,, \\
	\langle {\cal Z}(\varphi_1) {\cal X}(\varphi_2) {\cal Y}(\varphi_3) \rangle &= 
	\frac{15 - \pi^2}{60 \pi^2  \ell^5}  \,, \\
	\langle {\cal Z}^2(\varphi_1) {\cal X}(\varphi_2) {\cal Y}(\varphi_3) \rangle &= 
	\frac{16 \pi^4 - 84 \pi^2 - 315}{1260 \pi^4 \ell^6}  \,, \\
	\langle {\cal Z}^3(\varphi_1) {\cal X}(\varphi_2) {\cal Y}(\varphi_3) \rangle &= 
	- \frac{3(525 - 170 \pi^2 + 12 \pi^4)}{1400 \pi^2 (\pi^2 - 6) \ell^7}  \,,   \\
	\langle {\cal Z}^4(\varphi_1) {\cal X}(\varphi_2) {\cal Y}(\varphi_3) \rangle &= 
	\frac{55125 + 17850 \pi^2 - 6160 \pi^4 + 384 \pi^6}{7350 \pi^2 (-45 - 30 \pi^2 + 4 \pi^4) \ell^8} \,.   
}
From these correlation functions one can reproduce the algebra \eqref{4nodeStar} with
\es{al4}{
	\alpha_4 = \frac{\pi^2 - 6}{12 \pi^2} \,, \qquad
	\kappa_4 = \frac{1}{16} \,, \qquad
	\lambda_4 = \frac 32 \,.
}
One can see from Figure~5 of \cite{Beem:2016cbd} that these values of the parameters lie in the region allowed by the numerical bounds.  Extending the analysis above to $N>4$ is then straightforward, but we will not perform it explicitly here.

\subsection{$U(2)$ with adjoint hypermultiplet and fundamental hypermultiplet}
	
We can also study the much more intricate example of an ${\cal N} = 8$ SCFT and make a comparison with the results of \cite{Chester:2014mea}.  The ${\cal N} = 8$ SCFT we consider is the infrared limit of $U(2)$ gauge theory with an adjoint hypermultiplet and a fundamental hypermultiplet.  Let the twisted fields corresponding to the adjoint hypermultiplet be denoted by $X^i{}_j$ and $\tX_i{}^j$ and those corresponding to the fundamental hypermultiplet by $Q^i$ and $\tQ_i$, where $i, j = 1, 2$ are gauge indices. The 1d twisted Higgs branch theory is 
\es{ZtopU2}{
	Z = \frac 12 \int d \sigma_1 d\sigma_2\, 4 \sinh^2 (\pi(\sigma_1 - \sigma_2)) \int DQ^i D\tQ_i DX^i{}_j D\tX_i{}^j e^{-S}
}
with
\es{SU2}{
	S = \ell \int d\varphi \, \left[\tQ_i \dot{Q}^i + \tX_i{}^j \dot{X}^i{}_j + \sigma_1 \tQ_1 Q^1 + \sigma_2 \tQ_2 Q^2 + (\sigma_1 - \sigma_2) (\tX_1{}^2 X^1{}_2 - \tX_2{}^1 X^2{}_1)  \right]
}
Integrating out the $Q$'s and $X$'s we get the matrix model \cite{Kapustin:2009kz}
\es{ZTopU2KWY}{
	Z = \frac 12 \int d \sigma_1 d\sigma_2\, \frac{\sinh^2 (\pi(\sigma_1 - \sigma_2)) }{16 \cosh^2 (\pi (\sigma_1 - \sigma_2)) \cosh(\pi \sigma_1) \cosh(\pi \sigma_2) } = \frac{1}{16 \pi} \,. 
}
	
The $U(2)$ gauge theory with a fundamental and an adjoint hypermultiplet is believed to flow to the same IR fixed point as the ${\cal N} = 8$ $U(2)$ Yang-Mills theory.  The IR fixed point SCFT has two ${\cal N} = 8$ stress tensor multiplets, one of which corresponds to a free sector and one to an interacting sector.  Intuitively, the free sector corresponds to the IR limit of the diagonal $U(1)$ in the Yang-Mills description, while the interacting sector corresponds to the IR limit of  $SU(2)$ Yang-Mills theory, as will be made more precise shortly.
	
It was shown in \cite{Chester:2014mea} that upon decomposition to ${\cal N} = 4$ SCFT notation, the 1d Higgs branch theory has a flavor $\mathfrak{su}(2)_F$ symmetry that is a subgroup of the $\mathfrak{so}(8)$ R-symmetry.  Under $\mathfrak{su}(2)_F$, $(\tX, X^T)$ form a doublet.\footnote{Because of the D-term relations, we may construct operators only from $X$ and $\tX$.  Indeed, the equations of motion for the auxiliary field $D_{ab}$ imply $\tQ_j Q^i + \tX_j{}^k X^i{}_k - \tX_k{}^i X^k{}_j = 0$, so every pair $\tQ_j Q^i$ can be replaced by $ -\tX_j{}^k X^i{}_k + \tX_k{}^i X^k{}_j$.  Since gauge-invariant operators can only contain an equal number of $Q$'s and $\tQ$'s such replacements yield expressions depending only on $X$ and $\tX$.} In order to match the notation in \cite{Chester:2014mea}, let us introduce polarization variables $\bar y^{\bar a}$, $\bar a = 1, 2$, and denote the operators in the 1d theory by 
\es{OFlaforDef}{
	{\cal O}_{2 j_F}(\varphi, \bar y) = {\cal O}_{\bar a_1 \ldots \bar a_{2 j_F}} \bar y^{\bar a_1} \cdots \bar y^{\bar a_{j_F}} \,, 
}
where $j_F$ is the spin of the $\mathfrak{su}(2)_F$ representation.
	
We will identify 3 operators in the 1d theory and compute their correlation functions:
\begin{itemize}
	\item The twisted Higgs branch representative of the ${\cal N} = 8$ free field multiplet.  The ${\cal N} = 8$ free field multiplet consists of 8 scalar operators of scaling dimension $1/2$ and 8 spin-$1/2$ operators of scaling dimension $1$.  Under the decomposition to ${\cal N} = 4$ supersymmetry, $4$ of the scalar operators are interpreted as Higgs branch operators (transforming under $\mathfrak{su}(2)_H \oplus \mathfrak{su}(2)_F$ as $({\bf 2}, {\bf 2})$, while the other $4$ are Coulomb branch operators.   From the $4$ Higgs branch operators one can construct the twisted Higgs branch operator ${\cal O}_{1, \text{free}}(\varphi, \bar y)$.
	\item The twisted Higgs branch representatives of the free and of the interacting ${\cal N} = 8$ stress tensor multiplets.   Any ${\cal N } = 8$ stress tensor multiplet contains $35$ scalar operators of scaling dimension $1$, $9$ of which being Higgs branch operators from an ${\cal N} = 4$ point of view.  From them, one can construct twisted Higgs branch operators ${\cal O}_2(\varphi, \bar y)$.  We will denote the operator corresponding to the free stress tensor multiplet by ${\cal O}_{2, \text{free}}(\varphi, \bar y)$ and the one corresponding to the interacting stress tensor multiplet by ${\cal O}_{2, \text{int}}(\varphi, \bar y)$.

	\end{itemize}

\subsubsection{Free ${\cal N} = 8$ multiplet}
	
The free multiplet operator ${\cal O}_{1, \text{free}} (\varphi, \bar y)$ is 
\es{O1hat}{
	{\cal O}_{1, \text{free}}(\varphi, \bar y) =  \bar y^1 \tr \tX(\varphi) + \bar y^2 \tr X(\varphi) \,.
}
From \eqref{ZtopU2}, we see that $\tr \tX$ and $\tr X$ only appear in the kinetic term, so computing correlation functions of these operators can be performed using the propagator
\es{XtX}{
	\langle \tr  X(\varphi_1) \tr \tX(\varphi_2) \rangle = \frac{\sgn (\varphi_1 - \varphi_2)}{2 \ell}  \,.
}
(No integrals over $\sigma$ are necessary to establish \eqref{XtX}.)  Using this expression and \eqref{O1hat}, one obtains
\es{O1TwoPoint}{
	\langle {\cal O}_{1, \text{free}}(\varphi_1, \bar y_1)  {\cal O}_{1, \text{free}}(\varphi_2, \bar y_2)\rangle = \frac{1}{\ell} \langle \bar y_1, \bar y_2 \rangle \sgn (\varphi_1 - \varphi_2) \,.
}
where the angle bracket notation is defined by
\es{bracketDef}{
	\langle \bar y_i ,\bar y_j \rangle \equiv \bar y_i^{\bar a} \varepsilon_{\bar a \bar b} \bar y_j^{\bar b} \,, \qquad \varepsilon_{21} =-\varepsilon_{12} = 1 \,.
}
Higher point functions of ${\cal O}_{1, \text{free}}(\varphi, \bar y)$ can be computed using Wick contractions using \eqref{O1TwoPoint}.
	
\subsubsection{Free ${\cal N} = 8$ stress tensor multiplet}
	
There are two $\mathfrak{su}(2)_F$ triplets of linearly independent operators that are quadratic in $X$ corresponding to the two stress tensor multiplets of the theory.  It is easy to identify the one corresponding to the free ${\cal N} = 8$ multiplet because this is the only one appearing in the OPE of ${\cal O}_{1, \text{free}} \times {\cal O}_{1, \text{free}}$:  it is simply the square of the free ${\cal N} = 8$ operator ${\cal O}_{1, \text{free}} (\varphi, \bar y)$, 
\es{O2Def}{
	{\cal O}_{2, \text{free}} (\varphi, \bar y) =(\bar y^1)^2 (\tr \tX)^2 + 2 \bar y^1 \bar y^2 (\tr \tX) ( \tr X )+ (\bar y^2)^2 (\tr X)^2 \,.
}
Again using \eqref{XtX} gives
\es{O2freeTwo}{
	\langle {\cal O}_{2, \text{free}}(\varphi_1, \bar y_1) {\cal O}_{2, \text{free}}(\varphi_2, \bar y_2)  \rangle &=  
	\frac{2}{\ell^2} \langle \bar y_1, \bar y_2 \rangle^2  \,.
}
	
\subsubsection{Interacting ${\cal N} = 8$ stress tensor multiplet}
	
The interacting stress tensor multiplet must be orthogonal to the free one.  To obtain ${\cal O}_{2, \text{int}}$, we first compute the matrix of 2-point functions
\es{TwoPointStress}{
	\begin{pmatrix}
		\langle (\tr  X)^2(\varphi)\, (\tr \tX)^2(0 ) \rangle & \langle (\tr  X)^2(\varphi)\, (\tr \tX^2)(0 ) \rangle \\
		\langle (\tr X^2)(\varphi)\, (\tr \tX)^2(0 ) \rangle & \langle (\tr X^2)(\varphi)\, (\tr \tX^2)(0 ) \rangle
	\end{pmatrix}
	= \begin{pmatrix}
		\frac{1}{2\pi^2} & \frac{1}{4\pi^2} \\
		\frac{1}{4\pi^2}  & \frac{7}{24 \pi^2}
	\end{pmatrix} \,.
} 
We can then easily see that 
\es{Orthog}{
	\left \langle  \left[ (\tr \tX^2)(\varphi) - \frac 12 (\tr \tX)^2(\varphi) \right] (\tr X)^2 (0) \right\rangle = 0  \,,
}
which implies that the $(\bar y^1)^2$ component of ${\cal O}_{2, \text{int}}(\varphi, \bar y)$ is $(\tr \tX^2)(\varphi) - \frac 12 (\tr \tX)^2(\varphi)$ up to an overall normalization factor of our choice.   The $\mathfrak{su}(2)_F$ symmetry then implies
\es{O2Defs}{
	{\cal O}_{2, \text{int}} (\varphi, \bar y) &=   (\bar y^1)^2 \left( (\tr \tX^2) - \frac 12 (\tr \tX)^2 \right) +2 \bar y^1 \bar y^2 \left( (\tr X \tX^T) - \frac 12 (\tr X) (\tr \tX) \right)\\
	&{}+ (\bar y^2)^2 \left( (\tr X^2) - \frac 12 (\tr X)^2 \right)   \,.
}
Computing the two-point function of ${\cal O}_{2, \text{int}}$ is more challenging, as one now has to use the non-trivial propagators coming from \eqref{SU2}.  A careful calculation shows that the two-point function is
\es{O2Norm}{
	\langle {\cal O}_{2, \text{int}}(\varphi_1, \bar y_1) {\cal O}_{2, \text{int}}(\varphi_2, \bar y_2)  \rangle &=
	\frac{\langle \bar y_1, \bar y_2 \rangle^2}{4 \ell^2} \frac{1}{Z} \int d \sigma_1 d\sigma_2\, \frac{\sinh^2 (\pi \sigma_{12})  \left[ 5 + \cosh (2 \pi \sigma_{12} ) \right]}{16 \cosh^4 (\pi \sigma_{12}) \cosh(\pi \sigma_1) \cosh(\pi \sigma_2) }  \,    \\
	&= \frac{2}{3 \ell^2} \langle \bar y_1, \bar y_2 \rangle^2 \,,
}
where $\sigma_{12} \equiv \sigma_1 - \sigma_2$.
	
\subsubsection{Four-point functions}
	
We can use the formalism we have developed to calculate the 4-point functions of ${\cal O}_{2, \text{free}} (\varphi, \bar y)$ and ${\cal O}_{2, \text{int}} (\varphi, \bar y)$ and compare with \cite{Chester:2014mea}.   In \cite{Chester:2014mea} it was found that the 4-point function of an operator ${\cal O}_2(\varphi, \bar y)$ corresponding to an ${\cal N} = 8$ stress tensor multiplet (which could be any linear combination of ${\cal O}_{2, \text{free}} (\varphi, \bar y)$ and ${\cal O}_{2, \text{int}} (\varphi, \bar y)$) is
\es{FourPointExpectation}{
	\langle {\cal O}_2(\varphi_1, \bar y_1) {\cal O}_2(\varphi_2, \bar y_2) 
	& {\cal O}_2(\varphi_3, \bar y_3) {\cal O}_2(\varphi_4, \bar y_4)  \rangle
	= C^2 \langle \bar y_1, \bar y_2 \rangle^2 \langle \bar y_3, \bar y_4 \rangle^2 \\
	&\times
	\biggl[1 + \frac{1}{16} \lambda_{(B, 2)}^2 + \frac 14 \lambda_\text{stress}^2 \frac{2 - \bar w}{\bar w} + \frac{1}{16} \lambda_{(B, +)}^2 \frac{6 - 6 \bar w + \bar w^2}{\bar w^2}  \biggr] \,.
}
Here, $\bar w$ is defined as
\es{wDef}{
	\bar w \equiv \frac{\langle \bar y_1, \bar y_2 \rangle \langle \bar y_3, \bar y_4 \rangle}{\langle \bar y_1, \bar y_3 \rangle \langle \bar y_2, \bar y_4 \rangle} \,,
} 
the constant $C$ is given by the normalization of the operator, 
\es{TwoStress}{
	\langle {\cal O}_2(\varphi_1, \bar y_1) {\cal O}_2(\varphi_2, \bar y_2)    \rangle = C \langle \bar y_1, \bar y_2 \rangle^2 \,,
}
and $\lambda_{(B, 2)}^2$, $\lambda_{(B, +)}^2$, and $\lambda_\text{stress}^2$ are the squares of the various OPE coefficients of ${\cal N} = 8$ superconformal multiplets appearing in the OPE of the ${\cal N} = 8$ stress tensor multiplet with itself.
	
The four-point function of ${\cal O}_{2, \text{free}} (\varphi, \bar y)$ does not require any integrals, as it again only uses \eqref{XtX}.  When $\varphi_1 < \varphi_2 < \varphi_3 < \varphi_4$, we obtain
\es{O21Four}{
	\langle {\cal O}_{2, \text{free}}(\varphi_1, \bar y_1) {\cal O}_{2, \text{free}}(\varphi_2, \bar y_2) 
	{\cal O}_{2, \text{free}}(\varphi_3, \bar y_3) {\cal O}_{2, \text{free}}(\varphi_4, \bar y_4)  \rangle
	= \frac{4}{\ell^4} \langle \bar y_1, \bar y_2 \rangle^2 \langle \bar y_3, \bar y_4 \rangle^2 \frac{6 + 2 \bar w - 2 \bar w^2}{\bar w^2}  \,,
}
Obtaining ${\cal O}_{2, \text{int}}(\varphi, \bar y)$ is slightly more complicated.  The final result is
\es{O22Four}{
	\langle {\cal O}_{2, \text{int}}(\varphi_1, \bar y_1) {\cal O}_{2, \text{int}}(\varphi_2, \bar y_2) 
	{\cal O}_{2, \text{int}}(\varphi_3, \bar y_3) {\cal O}_{2, \text{int}}(\varphi_4, \bar y_4)  \rangle
	= \frac{8}{15 \ell^4} \langle \bar y_1, \bar y_2 \rangle^2 \langle \bar y_3, \bar y_4 \rangle^2 \frac{4 + \bar w - \bar w^2}{\bar w^2}  \,.
}
	
Comparing these expression with \eqref{FourPointExpectation}, we find 
\es{lamFree}{
	\text{Free stress tensor:}\qquad  \lambda_\text{stress}^2 &= 16 \,, \qquad
	\lambda_{(B, +)}^2 = 16 \,, \qquad \lambda_{(B, 2)}^2 = 0 \,, \\
	\text{Interacting stress tensor:}\qquad  \lambda_\text{stress}^2 &= 12 \,, \qquad
	\lambda_{(B, +)}^2 = \frac{64}{5} \,, \qquad \lambda_{(B, 2)}^2 = 0 \,.  
}
The expressions for the OPE coefficients of the free ${\cal N} = 8$ stress tensor multiplet match the result of \cite{Chester:2014mea} in the free ${\cal N} = 8$ theory (of 8 free massless scalars and 8 free massless Majorana fermions), while the corresponding expressions obtained for the interacting stress tensor match those obtained in \cite{Chester:2014mea} for the $U(2)_2 \times U(1)_{-2}$ ABJ theory.  The former theory is the infrared limit of ${\cal N} = 8$ super Yang-Mills theory with gauge group $U(1)$, while the latter theory is the infrared limit of ${\cal N} = 8$ Yang-Mills theory with gauge group $SU(2)$.    These results show quite explicitly how, at the level of the ${\cal N} = 4$ Higgs branch theory, the IR limit of ${\cal N} = 8$ $U(2)$ Yang-Mills theory (or the $U(2)$ gauge theory with one fundamental and one adjoint hypermultiplet) is a product between a free theory and the IR limit of $SU(2)$ Yang-Mills theory.

\section{Applications to ${\cal N} = 4$ QFTs on $S^3$ with non-vanishing mass and FI parameters}
\label{NONCONFORMALAPPLICATIONS}
	
Let us now present a few examples of correlation functions in non-conformal theories with either $m$ or $\zeta$ non-vanishing.
	
\subsection{Deformation by FI parameters}

\subsubsection{SQED with non-zero FI parameter}
	
Turning on a non-zero FI parameter in SQED is easily implemented by replacing $Z_\sigma$ in \eqref{PartSQED} and subsequent formulas in Section~\ref{SQEDSECTION} by
\es{ZNewZeta}{
	Z_\sigma = \frac{e^{2 \pi i \zeta \ell \sigma}}{\left[ 2 \cosh(\pi \sigma) \right]^N} \,.
}
The $S^3$ partition function is
\es{ZZeta}{
	Z = \frac{\Gamma\left( \frac{N}{2} - i \zeta \ell \right)\Gamma\left( \frac{N}{2} + i \zeta \ell  \right)}{2\pi (N-1)!} \,.
} 
The two-point function of ${\cal J}^{(p)}$ is then still given by \eqref{TwoPointJp}.  The integrals evaluate to
\es{JpJpFI}{
	\langle {\cal J}^{(p)}(\varphi_1, y_1, \bar y_1) {\cal J}^{(p)}(\varphi_2, y_2, \bar y_2) \rangle
	= \frac{(-1)^p}{\ell^{2p}} \frac{\Gamma(N) \Gamma\left(\frac{N}{2} - i \zeta \ell +p\right) \Gamma\left(\frac{N}{2} + i \zeta \ell + p \right)}
	{\Gamma(N+2p) \Gamma\left(\frac{N}{2} - i \zeta \ell \right) \Gamma\left(\frac{N}{2} + i \zeta \ell \right)} (\bar y_1 \cdot  y_2 )^p (y_1 \cdot  \bar y_2 )^p  \,.
}
These two-point functions can be combined into a single formula upon using the definition \eqref{JTotal}. We have
\es{JTotJTotFI}{
	\langle {\cal J}(\varphi_1, y_1, \bar y_1) {\cal J}(\varphi_2, y_2, \bar y_2) \rangle
	= {}_3 F_2 \left( \frac{N}{2} - i \zeta \ell, \frac{N}{2} + i \zeta \ell , 1; \frac{N}{2} \frac{N+1}{2}; -\frac{(\bar y_1 \cdot  y_2 ) (y_1 \cdot  \bar y_2 )}4  \right) \,.
}
Comparing with Eq.~(1.4) of \cite{Joung:2014qya} we see that \eqref{JTotJTotFI} agrees with the bilinear form of the generalized higher spin algebra $\mathfrak{hs}_\lambda(\mathfrak{sl}(N))$ with parameter $\lambda = \pm 2 i \zeta \ell / N$.
	
One can also compute 3-point functions.  We have, for example,
\es{J1J1J1FI}{
	&\langle {\cal J}^{(1)}(\varphi_1, y_1, \bar y_1) {\cal J}^{(1)}(\varphi_2, y_2, \bar y_2) {\cal J}^{(1)}(\varphi_3, y_3, \bar y_3) \rangle \\
	&\qquad\qquad= -\frac{N \left(1 + \frac{4 \zeta^2 \ell^2}{N^2} \right)}{8 \ell^3 (N+1)} \biggl[ 
	\left(1 - \frac{2i \zeta \ell}{N+2} \right) (\bar y_1 \cdot  y_3 ) (\bar y_3 \cdot  y_2 )(\bar y_2 \cdot  y_1 ) \\
	&\qquad\qquad{}-  \left(1 + \frac{2i \zeta \ell}{N+2} \right) (\bar y_1 \cdot  y_2 ) (\bar y_2 \cdot  y_3 )(\bar y_3 \cdot  y_1 ) \biggr]    \,,
}    
which matches Eq.~(1.4) of~\cite{Joung:2014qya} upon making the identification $\lambda = -2i \zeta \ell /N$.  (See Footnotes~\ref{VFootnote} and \ref{V3Footnote}.)  The star product of the generators of the chiral ring becomes
\es{starFI}{
	{\cal J}_I{}^J \star  {\cal J}_K{}^L
	&= {\cal J}_{IK}{}^{JL} 
	+ \frac{i \zeta}{N+2} \biggl(\delta_K^J {\cal J}_I{}^L +  \delta_I^L {\cal J}_K{}^J
	- \frac{2}{N} (\delta_I^J {\cal J}_K{}^L + \delta_K^L {\cal J}_I{}^J)  \biggr) \\
	&{}-\frac{\zeta^2}{ N (N+1)} \left( \delta_I^L \delta_K^J - \frac{1}{N} \delta_I^J \delta_K^L \right) \\
	&{}- \frac{1}{2 \ell}  \delta_K^J {\cal J}_I{}^L
	+  \frac{1}{2 \ell}  \delta_I^L {\cal J}_K{}^J  
	-\frac{N}{ 4 \ell^2 (N+1)}  \left( \delta_I^L \delta_K^J - \frac{1}{N} \delta_I^J \delta_K^L \right) \,.
}
	
In the limit $\ell \to \infty$, \eqref{starFI} reduces to the commutative product on the deformed Higgs branch chiral ring.  Indeed, using \eqref{AdjOps} and \eqref{J2}, one can check that the multiplication of ${\cal J}_I{}^J {\cal J}_K{}^L$ yields the $\ell \to \infty$ limit of \eqref{starFI} provided that the relation $\tQ_I Q^I = i \zeta$ is satisfied, as appropriate for the deformed Higgs branch chiral ring.

\subsubsection{$N$-node quiver with non-zero FI parameters}
	
The $N$ node quiver has gauge group $U(1)^N / U(1)$ containing $N-1$ Abelian factors.  Consequently, there are $N-1$ linearly independent FI parameters that can be introduced.  Let us introduce an FI parameter $\zeta_j$ for each one of the $N$ gauge group factors with the constraint 
\es{zetaSum}{
	\sum_{j=1}^N \zeta_j = 0 \,.
} 
The deformation to non-zero $\zeta_j$'s is realized by modifying the expression of $Z_\sigma$ in \eqref{calZ3node} to
\es{bfZNNodeFI}{
	Z_\sigma = \prod_{j=1}^N \frac{e^{2 \pi i \ell \zeta_j  \sigma_j } }{2 \cosh( \pi (\sigma_j - \sigma_{i+1})) } \,.
} 
Because the $\zeta_j$ sum to zero, it is possible to write them as
\es{zetaomega}{
	\zeta_j = \omega_{j-1} - \omega_{j} \,, 
} 
for some $\omega_j$, and then using summation by parts one can write
\es{sumRewriting}{
	\sum_{j=1}^N \zeta_j \sigma_j = \sum_{j=1}^N \sigma_j (\omega_{j-1} - \omega_{j}) 
	= -\sum_{j=1}^N \omega_j (\sigma_j - \sigma_{j+1}) \,.
}
This expression can be substituted into \eqref{bfZNNodeFI}.
Upon performing the Fourier transform to the $\tau_j$ coordinates using now 
\es{identity}{
	\frac{e^{-2 \pi i \ell \omega_j(\sigma_j - \sigma_{j+1})}}{2 \cosh (\pi (\sigma_j - \sigma_{j+1})) } = \int d \tau_j \frac{e^{2 \pi i \ell (\sigma_j - \sigma_{j+1}) \tau_j}}{2 \cosh (\pi (\tau_j + \ell \omega_j))}  \,,
}
we obtain
\es{ZNNodeFI}{
	Z = \int d\tau \frac{1}{\prod_{j=1}^N \left[2 \cosh(\pi (\tau + \ell \omega_j)) \right]} \,.
}
The $S^3$ partition function \eqref{ZNNodeFI} agrees with that of SQED with $N$ hypermultiplets with real masses $\omega_j$, as required by mirror symmetry. (See~\cite{Kapustin:2010xq} where this equivalence was first shown at the level of the $S^3$ partition function.)  Note that an overall shift in $\omega_j$ can be ``gauged away'' by shifting the integration variable $\tau$.  We will thus impose a gauge fixing condition 
\es{sumom}{
	\sum_{j=1}^N \omega_j = 0 \,.
}

In the presence of the FI terms, we can use a modified definition of the operators \eqref{Ops3Node}:
\es{XYZFI}{
	{\cal X} &= Q_1 Q_2 \cdots Q_N \,, \qquad
	{\cal Y} = \tQ_1 \tQ_2 \cdots \tQ_N \,,\\
	{\cal Z} &= \tQ_1 Q_1 - i \omega_1 
	=  \tQ_2 Q_2 - i \omega_2 = \ldots 
	=  \tQ_N Q_N - i \omega_N \,.
}
They obey the classical relation ${\cal X} {\cal Y} = ({\cal Z} + i \omega_1) ({\cal Z} + i \omega_2) \cdots ({\cal Z} + i \omega_N)$, corresponding to the deformation of the Kleinian singularity ${\cal X} {\cal Y} = {\cal Z}^N$ with parameters $\omega_j$.
	
With the definition in \eqref{XYZFI}, we have
\es{ZZ}{
	\langle {\cal Z}(\varphi_1) {\cal Z}(\varphi_2) \rangle = 
	\frac{1}{\ell^2} \frac{1}{Z} \int d \tau \, \frac{(i \tau)^2}{\prod_{j=1}^N \left[ 2 \cosh (\pi (\tau + \ell \omega_j)) \right]}  \,.
} 
Then the second equation in~\eqref{XYTwo} still holds, and we have
\es{XYFI}{
	\langle {\cal X}(\varphi_1) {\cal Y}(\varphi_2) \rangle 
	= \frac{1}{\ell^N} \frac{1}{Z} \int d \tau \, \frac{\prod_{j=1}^N \left( i (\tau + \ell \omega_j) - \frac 12 \right)}{\prod_{j=1}^N \left[ 2 \cosh (\pi (\tau + \ell \omega_j)) \right]}  \,, \qquad
	\text{for $\varphi_1< \varphi_2$}\,.
}
	
More generally, defining
\es{ZpConvenientZeta}{
	\widehat{{\cal Z}^p} = \prod_{j=1}^p (\tQ_j Q_j - i \omega_j)
}
we find that for $\varphi_1 < \varphi_2 < \varphi_3$ we have
\es{ZpZqCorrZeta}{
	\langle  \widehat{{\cal Z}^p} (\varphi_1)  \widehat{{\cal Z}^q} (\varphi_2) \rangle 
	&= \frac 1{\ell^{p+q}} \frac 1Z \int d\tau \frac{\left(i \tau \right)^{p+q}}{\prod_{j=1}^N \left[ 2 \cosh (\pi (\tau + \ell \omega_j)) \right]}  \,, \\
	\langle \widehat{{\cal Z}^p}(\varphi_1) {\cal X}(\varphi_2) {\cal Y}(\varphi_3) \rangle &= \frac{1}{\ell^{N+p}} 
	\frac 1Z \int d\tau \frac{(i \tau)^p \prod_{j=1}^N \left( i (\tau + \ell \omega_j) - \frac 12 \right)}{\prod_{j=1}^N \left[ 2 \cosh (\pi (\tau + \ell \omega_j)) \right]} 
	\,.
}
From these expressions, it is straightforward to extract the corresponding star product deformed by the parameters $\omega_j$.

\subsection{Introducing mass parameters}

\subsubsection{Mass-deformed $N$-node quiver}
	
The $N$-node quiver has a $U(1)$ flavor symmetry under which the $Q_i$ carry charge $+1/N$ while $\tQ_i$ carry charge $-1/N$.  This normalization of the $U(1)$ charge is such that the operators ${\cal X}$ and ${\cal Y}$ carry charges $+1$ and $-1$, respectively.
	
We can introduce a real mass term associated with this flavor symmetry by adding 
\es{MassTermNNode}{
	-\ell \int d\varphi\,  \frac{mr}{N} \tQ_i Q_i
}
to the exponent of \eqref{3node}.    This amounts to replacing $Z_\sigma$ in \eqref{calZ3node} by
\es{ZMassNNode}{
	Z_\sigma = \prod_{j=1}^N \frac{1}{2 \cosh \left( \pi (\sigma_j - \sigma_{j+1} + mr/N) \right) } \,.
}
The partition function is given by the equation
\es{PartFnNNodeMass}{
	Z = \frac{\Gamma\left( \frac{N}{2} - i mr \right)\Gamma\left( \frac{N}{2} + i mr \right)}{2\pi (N-1)!} \,,
}  
which, upon the replacement $mr \to \zeta \ell$, can be seen to agree with Eq.~\eqref{ZZeta} of the partition function of SQED with $N$ charged hypers and FI parameter $\zeta$.  Indeed, under mirror symmetry the real masses and FI parameters are interchanged.
	
Eqs.~\eqref{XYTwo} and \eqref{XYZThree} still hold, with the only change that $\sigma_{j(j+1)}$ is replaced by $\sigma_{j(j+1)} + mr/N$.  We obtain, for instance, that
\es{ZZMass}{
	\langle {\cal Z} (\varphi_1) {\cal Z}(\varphi_2) \rangle &= 
	\frac 1{\ell^2} \frac 1Z \int d\tau \frac{e^{2 \pi i mr \tau}}{\left[2 \cosh(\pi \tau) \right]^N} \left(i \tau \right)^2 \,.
}
More generally, we can define the operators $\widehat{{\cal Z}^p}$ with $p \leq N$, whose matrix of two point functions is given by
\es{ZpZqMass}{
	\langle \widehat{{\cal Z}^p}(\varphi_1) \widehat{{\cal Z}^q}(\varphi_2) \rangle &= 
	\frac 1{\ell^{p+q}} \frac 1Z \int d\tau \frac{e^{2 \pi i mr \tau}}{\left[2 \cosh(\pi \tau) \right]^N} \left(i \tau \right)^{p+q} \,.
}
	
The mixing of these operators can be removed by performing a Gram-Schmidt procedure as was the case for SCFTs.  For example, we can remove the mixing with the identity operator by subtracting the expectation values of the operators.  Explicitly, \eqref{ZZMass}--\eqref{ZpZqMass} imply that the connected correlation function of ${\cal Z}$ is
\es{ZZConn}{
	\langle {\cal Z}(\varphi_1) {\cal Z}(\varphi_2) \rangle - \langle {\cal Z}(\varphi_1) \rangle \langle {\cal Z}(\varphi_2) \rangle = -\frac{\psi^{(1)}\left( \frac N2 - imr \right) + \psi^{(1)}\left( \frac N2 + imr \right)}{4 \pi^2 \ell^2} \,,
}
where $\psi^{(n)}(z)$ is the polygamma function.  One can see that this function vanishes as $m \to \infty.$

\subsubsection{Mass-deformed SQED}
	
The SQED theory with $N$ charged hypermultiplets has an $SU(N)$ flavor symmetry.  One can introduce $N-1$ real mass parameters corresponding to the $U(1)^{N-1}$ Cartan of $SU(N)$ by adding 
\es{MassTermSQED}{
	-\ell \int d\varphi\,  \sum_{I=1}^N m_I r\, \tQ_I Q^I \,, \qquad \sum_{I=1}^N m_I = 0 
}
to the exponent of the second equation in \eqref{PartSQED}.  The condition $\sum_{I=1}^N m_I = 0 $ ensures that the $m_I$ are real masses for the Cartan of $SU(N)$.  The expression for the $S^3$ partition function in \eqref{PartSQED} gets replaced by
\es{ZMassSQED}{
	Z = \int d \sigma\, Z_\sigma \,, \qquad
	Z_\sigma = \frac{1}{\prod_{I=1}^N 2 \cosh (\pi (\sigma + m_I r) )} \,.
}
The $S^3$ partition function agrees with that of the $N$-node quiver \eqref{ZNNodeFI} upon the replacement $m_I r \to \ell \omega_I$, in agreement with mirror symmetry.
	
While it is possible to perform computations for arbitrary $N$, for simplicity let us give an example in the case $N=2$ where we take $m_1 = -m_2 = m$.  The partition function in \eqref{ZMassSQED} evaluates to 
\es{PartN2}{
	Z = mr  \csch (2 \pi mr)
}
in this case.  Let us define the quadratic operators 
\es{J123Defs}{
	{\cal J}_3 = \frac 12 \left( \tQ_1 Q^1 - \tQ_2 Q^2 \right) \,, \qquad
	{\cal J}_+ = \tQ_1 Q^2 \,, \qquad
	{\cal J}_- = \tQ_2 Q^1 \,.
}
The operator ${\cal J}_3$ is neutral under the $U(1)$ Cartan of flavor $SU(2)$ symmetry, so it's correlation functions are independent of position.  We obtain, for instance, 
\es{J3Corr}{
	\langle {\cal J}_3(\varphi) \rangle &= -\frac{1-2 \pi mr \coth(2 \pi mr)}{4 \pi m r \ell} \,, \\
	\langle {\cal J}_3(\varphi_1) {\cal J}_3(\varphi_2) \rangle - \langle {\cal J}_3(\varphi_1) \rangle  
	\langle {\cal J}_3(\varphi_2) \rangle
	&= \frac{\left[1 + 8 \pi^2 m^2 r^2 - \cosh(4 \pi mr) \right] \csch^2(2 \pi mr)}
	{32 \pi^2 m^2 r^2 \ell^2} \,.
} 
On the other hand, the operators ${\cal J}_{\pm}$ carry charges $\pm 2$ under the Cartan of the flavor $SU(2)$.  Their expectation values must vanish because they cannot mix with the identity operator.  Their correlation functions, however, do depend on position as in \eqref{2pt3ptMassive} with ${\cal F}_{{\cal J}_\pm} (mr)= \pm 2 mr$.  We obtain
\es{JpmCorr}{
	\langle {\cal J}_+(\varphi_1) {\cal J}_-(\varphi_2) \rangle 
	= \frac{e^{2 mr (\varphi_1 - \varphi_2) } \left[1-2 mr \pi \coth(2 mr \pi) \right] \left[\coth(2 mr \pi) - \sgn(\varphi_1 - \varphi_2) \right]}{16 \pi \ell^2 mr} \,.
}
One can see that both \eqref{J3Corr} and \eqref{JpmCorr} interpolate between a non-trivial topological expression at $mr = 0$ and they both tend to zero as $mr \to \infty$.  Indeed, if we interpret $mr$ as the RG scale, then at small $mr$ we are probing the UV SCFT, while at large $mr$ we are probing the infrared.

\section{Discussion}
\label{DISCUSSION}
	
In this paper we used supersymmetric localization to derive a 1d theory coupled to a matrix model, given in \eqref{ZS3}, that can be used to calculate correlation functions of twisted Higgs branch operators of ${\cal N}  = 4$ QFTs on $S^4$.  In the case of ${\cal N} = 4$ SCFTs, this theory provides a Lagrangian realization of the protected Higgs branch topological sector discussed in \cite{Chester:2014mea, Beem:2016cbd}.  The immediate practical application of \eqref{ZS3} is to the computation of 2- and 3-point functions of Higgs branch operators.
	
Our results can be used to perform more detailed tests of mirror symmetry.  We have seen, for instance, that in the $N$-node necklace quiver, the twisted operator ${\cal Z}$ has the 2-point function (see \eqref{XYDual})
\es{ZZNNode}{
	\langle {\cal Z}(\vphi_1) {\cal Z}(\vphi_2) \rangle = \frac{1}{\ell^2} \frac{1}{Z} \int d\tau \frac{1}{\left[ 2 \cosh (\pi \tau) \right]^N} (i \tau)^2 \,.
}
This theory is mirror dual to SQED with $N$ flavors.  One expects the twisted Higgs branch operator ${\cal Z}$ in the $N$-node quiver to be mirror dual to the twisted Coulomb branch operator $\Phi$ constructed from the vectormultiplet scalars in SQED\@.  In Section~\ref{LOCQC} we explained that the 2-point function of $\Phi$ can be computed by replacing each insertion of $\Phi$ by $2 \sigma / r$ in the KWY matrix model, thus obtaining
\es{PhiPhiSQED}{
	\langle \Phi(\vphi_1) \Phi(\vphi_2) \rangle = \frac{64 \pi^2}{\ell^2}
	\frac{1}{Z} \int d\sigma \frac{1}{\left[ 2 \cosh (\pi \sigma) \right]^N} \sigma^2 \,.
} 
Comparing \eqref{ZZNNode} and \eqref{PhiPhiSQED}, we can thus identify ${\cal Z}$ in the $N$-node quiver with $\pm i \Phi / (8 \pi)$ in SQED\@.  A similar exercise shows that, at least for $p \leq N$, ${\cal Z}^p$  in the $N$-node quiver can be identified with $\left[\pm i \Phi / (8 \pi) \right]^p$ in SQED\@. These are, of course, rather simple tests of mirror symmetry.  It should be possible to perform more non-trivial tests in non-Abelian gauge theories.
	
There are a few generalizations of our results that we have left for the future.   One such generalization is to ${\cal N} = 4$ gauge theories that include twisted vectormultiplets and twisted hypermultiplets, which would then open the possibility of including Chern-Simons interactions.  Another such generalization would be to complete the Coulomb branch localization computation by allowing for insertions of monopole operators.  Yet another such generalization would be to Higgs branch operators in theories with 8 supercharges defined in a different number of spacetime dimensions.  We hope to report on these questions in future publications.

\section*{Acknowledgments}
	
We thank Chris Beem, Cyril Closset, Thomas Dumitrescu, Jaume Gomis, Bruno Le Floch, Wolfger Peelaers, Herman Verlinde, and Edward Witten for useful discussions.  The work of SSP was supported in part by the US NSF under Grant No.~PHY-1418069, and that of RY by NSF Grant No.~PHY-1314198. Work of MD was supported in part by Walter Burke Institute for Theoretical Physics and the U.S. Department of Energy, Office of Science, Office of High Energy Physics, under Award Number DE-SC0011632, as well as Sherman Fairchild foundation.

\appendix
\section{Conventions}
\label{conventions}
	
Curved space vector indices are denoted by $\mu,\nu,\ldots$, while frame indices are denoted by $i,j,\ldots=1,2,3$. We label the doublet (spinor) representation of the $SU(2)_{\text{rot.}}$ frame rotation group by $\alpha,\beta,\ldots=1,2$, of $SU(2)_H$ by $a,b,\ldots=1,2$, and of $SU(2)_C$ by $\dot{a},\dot{b},\ldots=1,2$. Spinor indices are raised and lowered from the left with the antisymmetric tensors $\varepsilon_{\alpha\beta}$ and $\varepsilon^{\alpha\beta}$, where $\varepsilon_{12}=-\varepsilon^{12} = -1$. The same conventions are used for raising and lowering  $SU(2)_C\times SU(2)_H$ indices (e.g., $\lambda^{a\dot{a}}\equiv \varepsilon^{a b}\varepsilon^{\dot{a}\dot{b}}\lambda_{b\dot{b}}$). When $SU(2)$ spinor indices are suppressed their contraction is defined with the convention:
\begin{align}
(\psi\chi) \equiv \psi^{\alpha}\chi_{\beta} = (\chi\psi) \ed
\end{align}
In particular, for any three spinors $x$, $y$ and $z$ (either commuting or anti-commuting) we have the Fierz identity:
\begin{align}
x_{\alpha}(yz)+(xy)z_{\alpha}+x_{\beta} y_{\alpha} z^{\beta} =0\ed
\end{align}
We will always take variation spinors $\xi$, as in \eqref{Avar}--\eqref{qtpsitvar}, to be commuting, while the $\delta_{\xi}$ symbol itself to be anti-commuting.
	
The flat space gamma matrices are the usual Pauli matrices,  $(\gamma^{i})_{\alpha}{}^{\beta} \equiv \sigma^{i}$, which satisfy
\begin{align}
\gamma_{i}\gamma_{j} &= \delta_{ij} + i \varepsilon_{ijk}\gamma^{k} \ecq (\varepsilon_{123} = 1) \ec  \\
(\gamma^{i})_{\alpha}{}^{\beta} (\gamma_{i})_{\gamma}{}^{\delta} &= 2 \delta_{\alpha}{}^{\delta}\delta_{\gamma}{}^{\beta} - \delta_{\alpha}{}^{\beta}\delta_{\gamma}{}^{\delta} \ed
\end{align}
	
Given a Euclidean metric $g_{\mu\nu}$ an orthonormal frame is defined by
\begin{align}
g_{\mu\nu} = e^{i}{}_{\mu}e^{j}{}_{\nu}\delta_{ij} \ecq \delta^{ij} = g^{\mu\nu}e^{i}{}_{\mu}e^{j}{}_{\nu} \ed
\end{align}
A spin connection $\omega_{\mu ij}=-\omega_{\mu ji}$ is then fixed from the conditions
\begin{align}
de_{i}+\omega_{i}{}^{j}\wedge e_{j} = 0 \ecq e_i\equiv e_{i\mu}dx^{\mu} \ed \label{torsionfree}
\end{align}
The Riemann tensor is
\begin{align}
\cR_{\mu\nu i j} = \partial_{\mu}\omega_{\nu i j} + \omega_{\mu i}{}^{k}\omega_{\nu kj} - (\mu\leftrightarrow\nu) \ec
\end{align}
while the Ricci tensor and scalar are defined by $\cR_{\mu\nu} = \cR^{\rho}{}_{\mu\rho\nu}$ and $\cR=\cR^{\mu}{}_{\mu}$, respectively. With this definition $\cR=6$ for a round unit 3-sphere.
	
The space covariant derivative of spinors is defined as
\begin{align}
\nabla_{\mu} \psi = (\partial_{\mu} +\frac{i}{4}\omega_{\mu ij}\epsilon^{ijk}\gamma_{k})\psi \ec
\end{align}
while the Lie derivative $\hat{\cL}_v$ along $v^{\mu}$ acting on scalars $\phi$, spinors $\psi$, and vector fields $A_{\mu}$, is given by
\begin{align}
\hat{\cL}_v\phi &= v^{\mu}\partial_{\mu}\phi \ec\\
\hat{\cL}_v \psi &= \left(v^{\mu}\nabla_{\mu} + \frac{i}{4}\epsilon^{\mu\nu\rho}\nabla_{\mu}v_{\nu}\gamma_{\rho}\right)\psi \ec \\
\hat{\cL}_v A_{\mu} &= v^{\nu}\partial_{\nu} A_{\mu} + \partial_{\mu}v^{\nu}A_{\nu} \ed
\end{align} 
	
\subsection{Differential Geometry on $S^3$}
	
We will hereby summarize various details on differential geometry on $S^3$ that are used in the main text. Let $S^3$ be the radius $r$ 3-sphere embedded into $\bC^2$ as
\begin{align}
	|z_1|^2+|z_2|^2 = 1\ecq r\vec{z}\in\bC^2 \ed
\end{align}
Each point on $S^3$ can be represented by an $SU(2)$ element
\begin{align}
g = \begin{pmatrix}
z_2 & i z_1 \\ i \bar{z}_1 & \bar{z}_2
\end{pmatrix} \ed
\end{align}
The $\mathfrak{su}(2)$-valued left/right invariant 1-forms $\omega^{(\ell / r)}$, and the frame 1-forms $e^{(\ell /r )}$ associated with them are defined as
\begin{align}
\omega^{(\ell)} \equiv g^{-1}dg = \frac{i}{r}e^{(\ell)}_i\gamma^i \ecq \omega^{(r)} \equiv dgg^{-1} = \frac{i}{r} e^{(r)}_i\gamma^i \ed
\end{align} 
They satisfy the Maurer-Cartan equations
\begin{align}
de^{(\ell)}_i  + \frac{1}{r}\epsilon_i{}^{jk}e^{(\ell)}_k \wedge e^{(\ell)}_ j = 0\ec \\
de^{(r)}_i  - \frac{1}{r}\epsilon_i{}^{jk}e^{(r)}_k \wedge e^{(r)}_j = 0\ec
\end{align}
from which the spin-connections can be directly read-off by using \eqref{torsionfree}. 
	
The $\mathfrak{su}(2)_{\ell}\oplus\mathfrak{su}(2)_r$ isometries of $S^3$ are generated, respectively, by the vector fields $\cL^i$ and $\cR^j$, which are dual to the 1-forms $e^{(r)}_i$ and $e^{(\ell)}_i$ up to proportionality constants that we define as\footnote{$\cL^i$ generates the left $SU(2)$ action $\cL^i g = -\frac{1}{2}\gamma^i g$, while $\cR^i$ generates the right action $\cR^ig = \frac{1}{2}g\gamma^i$.}
\begin{align}
e^{(\ell)}_i( \cR^j ) = -\frac{i r}{2}\delta_i{}^j \ecq e^{(r)}_i(\cL^j) = \frac{i r}{2}\delta_i{}^j \ed
\end{align}
They satisfy the $\mathfrak{su}(2)$ algebra
\begin{align}
[\cL^i,\cL^j] = i\varepsilon^{ijk}\cL^k \ecq [\cR^i,\cR^j] = i\varepsilon^{ijk}\cR^k\ed
\end{align}
	
In the round coordinates
\begin{align}
z_1 =  \cos(\theta)e^{i\tau} \ecq z_2 = \sin(\theta) e^{i\varphi} \ec
\end{align}
the metric on $S^3$ is given by
\begin{align}
ds^2 = e^{(\ell) i} e^{(\ell)}_{i} = e^{(r) i} e^{(r)}_{i} = r^2(d\theta^2 + \cos^2(\theta)d\tau^2+\sin^2(\theta)d\varphi^2) \ed
\end{align}
and the vectors $\cL^i = \cL^{i\mu}\partial_{\mu}$ $\cR^i=\cR^{i\mu}\partial_{\mu}$ are given by
\begin{align}
\cL^1 &= \frac{i}{2}\left(-\cos(\tau+\varphi)\partial_{\theta} - \tan(\theta)\sin(\tau+\varphi)\partial_{\tau} + \cot(\theta)\sin(\tau+\varphi)\partial_{\varphi}\right) \ec \label{L1} \\
\cL^2 &= \frac{i}{2}\left(\sin(\tau+\varphi)\partial_{\theta} - \tan(\theta)\cos(\tau+\varphi)\partial_{\tau} + \cot(\theta)\cos(\tau+\varphi)\partial_{\varphi}\right) \ec \label{L2} \\
\cL^3 &= \frac{i}{2}\left(\partial_{\tau}+\partial_{\varphi}\right) \ec \label{L3} \\
\cR^1 &= \frac{i}{2}\left(\cos(\tau-\varphi)\partial_{\theta} + \tan(\theta)\sin(\tau-\varphi)\partial_{\tau} + \cot(\theta)\sin(\tau-\varphi)\partial_{\varphi}\right) \ec \label{R1} \\
\cR^2 &= \frac{i}{2}\left(-\sin(\tau-\varphi)\partial_{\theta} + \tan(\theta)\cos(\tau-\varphi)\partial_{\tau} + \cot(\theta)\cos(\tau-\varphi)\partial_{\varphi}\right) \ec \label{R2} \\
\cR^3 &= \frac{i}{2}\left(\partial_{\tau}-\partial_{\varphi}\right) \ed \label{R3} 
\end{align}

It will also be useful to introduce stereographic coordinates. Let 
\begin{align}
r z_1 = X_1 + i X_2 \ecq r z_2 = X_3 +iX_4 \ed
\end{align} 
The stereographic coordinates $x_i$ ($i=1,2,3$) are defined as
\begin{alignat}{3}
X_{1,2} &= \frac{x_{1,2}}{1+\frac{x^2}{4r^2}} \ecq &X_4 &= \frac{x_3}{1+\frac{x^2}{4r^2}}\ecq &X_3 &= r\frac{1-\frac{x^2}{4r^2}}{1+\frac{x^2}{4r^2}} \ec \\
x_{1,2} &= \frac{2X_{1,2}}{1+X_3/r} \ecq &x_3 &= \frac{2X_{4}}{1+X_3/r} \ecq &x^2&\equiv x_1^2+x_2^2+x_3^2 \ed
\end{alignat}
In our definition the origin $\vec{x} = (0,0,0)$ is mapped to  $\vec{X}=(0,0,r,0)$.
The induced metric on $S^3$ is conformally flat
\begin{align}
g_{\mu\nu} = e^{2\Omega}\delta_{\mu\nu}\ecq e^{\Omega} = \frac{1}{1+\frac{x^2}{4r^2}} \ec
\end{align}
and we define the stereographic frame as
\begin{align}
e^i{}_{\mu} = e^{\Omega}\delta^i{}_{\mu} \ed
\end{align}
	
Let us summarize how Killing spinors on $S^3$ look in the different frames that we introduced. The spinor covariant derivatives in the left and right invariant frames are given, respectively, by
\begin{align}
\nabla_{\mu}\biggr|_{\text{left inv.}} = \partial_{\mu} + \frac{i}{2r}\gamma_{\mu} \ecq \nabla_{\mu}\biggr|_{\text{right inv.}} = \partial_{\mu} - \frac{i}{2r}\gamma_{\mu} \ed
\end{align}
Let $\xi^{(\ell)}$ and $\xi^{(r)}$ be spinors satisfying
\begin{align}
\nabla_{\mu}\xi^{(\ell)} = \frac{i}{2r}\gamma_{\mu}\xi^{(\ell)} \ecq \nabla_{\mu}\xi^{(r)} = -\frac{i}{2r}\gamma_{\mu}\xi^{(r)} \ed
\end{align}
Then in the left invariant frame $\xi^{(\ell)}$ is some constant spinor $\chi^{(\ell)}$, while $\xi^{(r)}$ is some constant spinor $\chi^{(r)}$ in the right invariant frame. In the stereographic frame one can check that
\begin{align}
\xi^{(\ell)} = e^{\Omega/2}\left(1 - \frac{i}{2r}x^i\gamma_i\right)\chi^{(\ell)} \ecq \xi^{(r)} = e^{\Omega/2}\left(1 + \frac{i}{2r}x^i\gamma_i\right)\chi^{(r)} \ed
\end{align}
	
\section{Closure of Superconformal Algebra}
\label{closuredetails}
	
For any two spinors $\xi_{a\dot a}$ and $\tilde{\xi}_{a\dot a}$ satisfying \eqref{CKSeq}, the anti-commutator of supreconformal transformations \eqref{Avar}--\eqref{qtpsitvar} acting on any field $\Phi$ closes up to equations of motion into
\begin{align}
\{\delta_{\xi}, \delta_{\tilde{\xi}}\} \Phi = \left(\hat{\cK}_{\xi,\tilde{\xi}}+\hat{\cG}_{\Lambda}\right)\cdot\Phi + \text{e.o.m.} \ed
\end{align}
The operator $\hat{\cK}_{\xi,\tilde{\xi}}$ is defined as
\begin{align}
\hat{\cK}_{\xi,\tilde{\xi}}  \equiv \hat{\cL}_{v} + \hat{R}_C + \hat{R}_H + \rho \hat{\Delta} \ec
\end{align}
where
\begin{itemize}
	\item $\cL_v$ is the Lie derivative along $v^{\mu}\equiv i\tilde{\xi}^{a\dot a}\gamma^{\mu}\xi_{a\dot a}$.
	\item $\hat{R}_{C/H}$ is an $\mathfrak{su}(2)_{C/H}$ transformation, acting on doublets with the matrices 
	\begin{align}
		\bar{R}_{\dot{a}\dot b} &\equiv i(\tilde{\xi}^c{}_{(\dot a}\xi'_{|c|\dot b)} + \xi^c{}_{(\dot a}\tilde{\xi}'_{|c|\dot b)}) \ec \label{Rb}\\
		R_{ab} &\equiv i(\tilde{\xi}_{(a}{}^{\dot c}\xi'_{b)\dot c} + \xi_{(a}{}^{\dot c}\tilde{\xi}'_{b)\dot c}) \ec \label{R}
	\end{align}
	such that, e.g., $\hat{R}_C\psi_{\dot a} = \bar{R}_{\dot a}{}^{\dot b}\psi_{\dot b}$, $\hat{R}_Hq_a = R_a{}^bq_b$, and with the obvious generalization for triplets: $\hat{R}_H D_{ab} = R_a{}^cD_{cb} + R_b{}^c D_{ac}$, etc.
	\item $\rho$ is the dilation transformation parameter
	\begin{align}
		\rho = i(\tilde{\xi}^{a \dot b}\xi'_{a\dot b} + \xi^{a \dot b}\tilde{\xi}'_{a\dot b}) \ed\label{rho}
	\end{align}
	The components of the vectormultiplet \eqref{Vmul} appear with dimensions $\hat{\Delta}[\cV] = (0,\frac{3}{2},1,2)$, and those of the hypermultiplet \eqref{Hmul} have $\hat{\Delta}[\cH] = (\frac{1}{2},1)$.
	\item $\hat{\cG}_{\Lambda}$ is a gauge transformation with parameter
	\begin{align}
		\Lambda = (\tilde{\xi}^c{}_{\dot a}\xi_{c\dot b})\Phi^{\dot{a}\dot b} - v^{\mu}A_{\mu} \ec
	\end{align}
	such that, e.g., $\hat{\cG}_{\Lambda}A_{\mu} = \cD_{\mu}\Lambda$, $\hat{\cG}_{\Lambda} \Phi_{\dot{a}\dot b} = i[\Lambda,\Phi_{\dot{a}\dot b}]$, $\hat{\cG}_{\Lambda} q_a = i \Lambda q_a$, $\hat{\cG}_{\Lambda}\tq_a = -i\Lambda \tq_a$, etc. 
\end{itemize}
	
\section{$\cN=4$ Algebras}
\label{algebras}
	
\subsection{Superconformal Algebra}
	
The 3d $\cN=4$ superconformal algebra is $\mathfrak{osp}(4|4)$, and its  bosonic sub-algebra $\mathfrak{so}(3,2)\oplus\mathfrak{su}(2)_C\oplus\mathfrak{su}(2)_H$ consists of conformal and R-symmetry transformations. In flat space, the conformal symmetry generators can be divided into translations $P_{\mu}$, rotations $\cM_{\mu\nu}$, dilatations $D$ and special conformal transformations $K_{\mu}$. The $\mathfrak{su}(2)_C$ and $\mathfrak{su}(2)_H$ R-symmetry generators will be denoted by $R_a{}^b$ and $\bar{R}_{\dot a}{}^{\dot b}$, respectively. The corresponding sub-algebra is
\begin{align}
	[M_{\alpha}{}^{\beta}, P_{\gamma\delta}] &= \delta_{\gamma}{}^{\beta}P_{\alpha\delta} + \delta_{\delta}{}^{\beta}P_{\alpha\gamma} - \delta_{\alpha}{}^{\beta}P_{\gamma\delta} \ec \label{MPcomm}\\
	[M_{\alpha}{}^{\beta}, K^{\gamma\delta}] &= - \delta_{\alpha}{}^{\gamma}K^{\beta\delta} - \delta_{\alpha}{}^{\delta} K^{\beta\gamma} + \delta_{\alpha}{}^{\beta} K^{\gamma\delta} \ec \\
	[M_{\alpha}{}^{\beta}, M_{\gamma}{}^{\delta}] &= -\delta_{\alpha}{}^{\delta}M_{\gamma}{}^{\beta} + \delta_{\gamma}{}^{\beta} M_{\alpha}{}^{\delta} \ecq [D, P_{\alpha\beta}] = P_{\alpha\beta} \ecq [D,K^{\alpha\beta}] = -K^{\alpha\beta} \ec \\
	[P_{\alpha\beta},K^{\gamma\delta}] &= 4\delta_{(\alpha}{}^{(\gamma}M_{\beta)}{}^{\delta)} + 4\delta_{(\alpha}{}^{\gamma}\delta_{\beta)}{}^{\delta}D \,,\label{KPcomm}\\
	[R_a{}^{b}, R_c{}^{ d}] &= -\delta_a{}^{d} R_c{}^{b} + \delta_c{}^{b} R_a{}^{d} \ecq   
	[\bar{R}_{\dot a}{}^{\dot b}, \bar{R}_{\dot c}{}^{\dot d}] = -\delta_{\dot a}{}^{\dot d} \bar{R}_{\dot c}{}^{\dot b} + \delta_{\dot c}{}^{\dot b} \bar{R}_{\dot a}{}^{\dot d} \ec \label{RRcomm}
\end{align}
where we defined
\begin{align}
	P_{\alpha\beta} \equiv (\gamma^{\mu})_{\alpha\beta}P_{\mu} \ecq K^{\alpha\beta} \equiv (\gamma^{\mu})^{\alpha\beta}K_{\mu} \ecq M_{\alpha}{}^{\beta} \equiv \frac{i}{2}(\gamma^{\mu}\gamma^{\nu})_{\alpha}{}^{\beta}\cM_{\mu\nu} \ed
\end{align}
The algebra \eqref{MPcomm}--\eqref{RRcomm} is represented on a dimension $\Delta$ scalar primary operator $\cO_{a\dot a}(x)$ in the $(\mathbf{2},\mathbf{2})$ irrep of $\mathfrak{su}(2)_C\oplus\mathfrak{su}(2)_H$ as
\begin{align}
	[P_{\mu}, \cO_{a\dot a}(x)] &= i\partial_{\mu} \cO_{a\dot a}(x) \ecq
	[K_{\mu}, \cO_{a\dot a}(x)] = i ( x^2\partial_{\mu} - 2 x_{\mu}(x\cdot\partial) - 2\Delta x_{\mu} )\cO_{a\dot a}(x) \ec \notag\\
	[\cM^{\mu\nu}, \cO_{a\dot a}(x)] &= i(x^{\mu}\partial^{\nu} - x^{\nu}\partial^{\mu})\cO(x) \ecq [D, \cO_{a\dot a}(x)] = (x\cdot\partial + \Delta) \cO_{a\dot a}(x) \ec \label{conformaldiff}\\
	[R_a{}^b, \cO_{c\dot c}(x)] &= \delta_c{}^b\cO_{a \dot c} - \frac{1}{2}\delta_a{}^b\cO_{c\dot c} \ecq [\bar{R}_{\dot a}{}^{\dot b}, \cO_{c\dot c}(x)] = \delta_{\dot c}{}^{\dot b}\cO_{c \dot a} - \frac{1}{2}\delta_{\dot a}{}^{\dot b}\cO_{c\dot c} \ec\notag
\end{align}
The transformations of the odd generators in $\mathfrak{osp}(4|4)$ can be read from the variations \eqref{Avar}--\eqref{qtpsitvar} as follows. The solution of the conformal Killing spinor equation \eqref{CKSeq} on $\mathbb{R}^3$ is
\begin{align}
	\xi_{a\dot a} = \epsilon_{a\dot a} + x^{i}\gamma_{i}\eta_{a\dot a} \ecq \xi'_{a\dot a} = \eta_{a\dot a} \ed
\end{align}
We then define the action of the Poincar\'e supercharges $Q_{\alpha a \dot a}$ and conformal supercharges $S_{\alpha a \dot a}$ by
\begin{align}
	\delta_{\xi}\cO \equiv \frac{i}{2} [\epsilon^{\alpha a \dot a}Q_{\alpha a \dot a} + \eta^{\alpha a \dot a}S_{\alpha a \dot a}, \cO ] \ed \label{delta2QS}
\end{align}
The commutators of the odd generators $Q_{\alpha a \dot a}$ and $S_{\alpha a \dot a}$ then follow by matching the action of $\hat{\cK}_{\xi, \tilde{\xi}}$ defined in Appendix \ref{closuredetails} with \eqref{conformaldiff}. The resulting odd-odd and even-odd part of the algebra is
\begin{alignat}{3}
	\{Q_{\alpha a\dot{a}} , Q_{\beta b\dot{b}}\} &= 4\varepsilon_{ab}\varepsilon_{\dot{a}\dot{b}}P_{\alpha\beta} \ecq & \{S^{\alpha}{}_{a\dot{a}}, S^{\beta}{}_{b\dot{b}}\} &= 4\varepsilon_{ab}\varepsilon_{\dot{a}\dot{b}} K^{\alpha\beta} \ec \label{QQSScomm}\\
	[K^{\alpha\beta},Q_{\gamma a\dot{a}}] &= i\left(\delta_{\gamma}{}^{\alpha} S^{\beta}{}_{a\dot{a}} + \delta_{\gamma}{}^{\beta} S^{\alpha}{}_{a\dot{a}} \right) \ecq & [P_{\alpha\beta} , S^{\gamma}{}_{a\dot{a}}] &= -i \left( \delta_{\alpha}{}^{\gamma} Q_{\beta a\dot{a}} + \delta_{\beta}{}^{\gamma} Q_{\alpha a\dot{a}} \right) \ec \\
	[M_{\alpha}{}^{\beta}, Q_{\gamma a\dot{a}} ] &= \delta_{\gamma}{}^{\beta} Q_{\alpha a\dot{a}} - \frac{1}{2} \delta_{\alpha}{}^{\beta} Q_{\gamma a\dot{a}} \ecq & [M_{\alpha}{}^{\beta}, S^{\gamma}{}_{a\dot{a}}] &= - \delta_{\alpha}{}^{\gamma} S^{\beta}_{}{a\dot{a}} + \frac{1}{2}\delta_{\alpha}{}^{\beta} S^{\gamma}{}_{ a\dot{a}} \ec \\
	[D,Q_{\alpha a\dot{a}}] &= \frac{1}{2} Q_{\alpha a\dot{a}} \ecq & [D, S^{\alpha}{}_{a\dot{a}}] &= -\frac{1}{2} S^{\alpha}{}_{a\dot{a}} \ec \\
	[R_a{}^b, Q_{\alpha c\dot{c}}] &= \delta_c{}^b Q_{\alpha a \dot{c}} - \frac{1}{2}\delta_a{}^b Q_{\alpha c \dot{c}} \ecq & [R_a{}^b, S^{\alpha}{}_{c\dot{c}}] &= \delta_c{}^b S^{\alpha}{}_{a\dot{c}} - \frac{1}{2}\delta_a{}^b S^{\alpha}{}_{c\dot{c}}  \ec \label{RQRS} \\
	[\bar{R}_{\dot a}{}^{\dot b}, Q_{\alpha c\dot{c}}] &= \delta_{\dot c}{}^{\dot b} Q_{\alpha c \dot{a}} - \frac{1}{2}\delta_{\dot a}{}^{\dot b} Q_{\alpha c \dot{c}} \ecq & [\bar{R}_{\dot a}{}^{\dot b}, S^{\alpha}{}_{c\dot{c}}] &= \delta_{\dot c}{}^{\dot b} S^{\alpha}{}_{c\dot{a}} - \frac{1}{2}\delta_{\dot a}{}^{\dot b} S^{\alpha}{}_{c\dot{c}}  \ec
	\label{RbQRbS}
\end{alignat}
and also 
\begin{align}
	\{Q_{\alpha a\dot{a}}, S^{\beta}{}_{b\dot{b}}\} = 4i\left[ \varepsilon_{ab}\varepsilon_{\dot{a}\dot{b}}\left(M_{\alpha}{}^{\beta} + \delta_{\alpha}{}^{\beta} D \right) + \delta_{\alpha}{}^{\beta} \left( \varepsilon_{\dot{a}\dot{b}}R_{ab} + \varepsilon_{ab}\bar{R}_{\dot{a}\dot{b}}\right) \right] \ed \label{QS}
\end{align} 
	
\subsection{Non-conformal $\cN=4$ Algebra on $S^3$}
	
We will now construct the $S^3$ $\cN=4$ algebra  $\mathfrak{su}(2|1)_{\ell}\oplus\mathfrak{su}(2|1)_r$  explicitly, as a sub-algebra of the  $\mathfrak{osp}(4|4)$ superconformal algebra defined in \eqref{MPcomm}--\eqref{RRcomm} and \eqref{QQSScomm}--\eqref{QS}. The matrices $h_a{}^b$ and $\bar{h}^{\dot a}{}_{\dot b}$ in \eqref{massiveSpinors} are traceless and square to $1$. Therefore, they can always be decomposed into commuting spinors (``twistors'') $u_{\pm}^a$ and $\bar{u}_{\pm}^{\dot a}$ as
\begin{align}
	h_a{}^b = u_{+a}u_-^b + u_{-a}u_+^b \ecq \bar{h}^{\dot a}{}_{\dot b} =\bar{u}^{\dot a}_+\bar{u}_{-\dot b} + \bar{u}^{\dot a}_-\bar{u}_{+\dot b} \ec \label{twistors}
\end{align}
where $(u_+u_-)=(\bar{u}_+\bar{u}_-)=1$. The decomposition to twistors \eqref{twistors} simplifies the construction of the non-conformal sub-algebra, because it eliminates the need to carry the $\mathfrak{su}(2)_C\oplus\mathfrak{su}(2)_H$ indices. The twistors $u_{\pm}$ and $\bar{u}_{\pm}$ are simply the eigenvectors of $h_a{}^b$ and $\bar{h}^{\dot a}{}_{\dot b}$ :
\begin{align}
	u_{\pm}^ah_a{}^b = \pm u_{\pm}^b \ecq \bar{h}^{\dot a}{}_{\dot b}\bar{u}_{\pm}^{\dot b} = \pm \bar{u}_{\pm}^{\dot a} \ed
\end{align}

Let us parameterize the Cartan of $\mathfrak{su}(2)_C\oplus\mathfrak{su}(2)_H$ as\footnote{Recall that $(u_+u_-)\equiv u_+^au_{-a}$ and $(u_+Ru_-)\equiv u_+^a R_a{}^bu_{-b}$, etc.}
\begin{align}
	R_H^3 \equiv \frac{1}{2}h_a{}^bR_b{}^a = (u_+Ru_-) \ecq R_C^3 \equiv \frac{1}{2}\bar{h}^{\dot a}{}_{\dot b}\bar{R}^{\dot b}{}_{\dot a} = (\bar{u}_+\bar{R}\bar{u}_-) \ed \label{cartan}
\end{align}
The generators of the $\mathfrak{u}(1)_{\ell}\oplus\mathfrak{u}(1)_r\subset\mathfrak{su}(2)_C\oplus\mathfrak{su}(2)_H$ R-symmetry of the $S^3$ $\cN=4$ algebra are then defined in terms of \eqref{cartan} to be
\begin{align}
	R_\ell = R_H^3 + R_C^3 \ecq R_r = R_H^3 - R_C^3 \ed
\end{align}
Furthermote, the odd generators of $\mathfrak{su}(2|1)_\ell\oplus\mathfrak{su}(2|1)_r$ are given by
\begin{align}
	\cQ^{(\ell_{\pm})}_{\alpha} \equiv \frac{1+i}{2}u_{\pm}^a\bar{u}_{\pm}^{\dot a} \left(Q_{\alpha a \dot a} + \frac{i}{2r} S_{\alpha a \dot a}\right) \ecq \cQ^{(r_{\pm})}_{\alpha} \equiv \frac{1+i}{2}u_{\pm}^a\bar{u}_{\mp}^{\dot a} \left(Q_{\alpha a \dot a} -\frac{i}{2r} S_{\alpha a \dot a}\right) \ed \label{Qlr}
\end{align}
The relative coefficients between $Q_{\alpha a\dot a}$ and $S_{\alpha a \dot a}$ in \eqref{Qlr} can be fixed up to one constant by demanding that $\cQ^{(\ell_\pm)}$ anti-commute with $\cQ^{(r_\pm)}$. The only non-trivial odd-odd commutators are
\begin{align}
	\{\cQ^{(\ell_+)}_{\alpha}, \cQ^{(\ell_-)}_{\beta}\} &= -\frac{4i}{r}\left(J_{\alpha\beta}^{(\ell)} + \frac{1}{2} \varepsilon_{\alpha\beta}R_\ell\right)\ec \\
	\{\cQ^{(r_+)}_{\alpha}, \cQ^{(r_-)}_{\beta}\} &= -\frac{4i}{r}\left(J_{\alpha\beta}^{(r)} + \frac{1}{2} \varepsilon_{\alpha\beta}R_r\right)\ec
\end{align}
where $J^{(\ell)}_{\alpha\beta}$ and $J^{(r)}_{\alpha\beta}$ are the $\mathfrak{su}(2)_\ell\oplus\mathfrak{su}(2)_r$ isometry generators of $S^3$ defined by
\begin{align}
	J^{(\ell)}_{\alpha\beta} &= -\frac{r}{2}\left(P_{\alpha\beta}-\frac{1}{4r^2}K_{\alpha\beta}-\frac{1}{r}M_{\alpha\beta}\right) \ec \\
	J^{(r)}_{\alpha\beta} &= \frac{r}{2}\left(P_{\alpha\beta}-\frac{1}{4r^2}K_{\alpha\beta}+\frac{1}{r}M_{\alpha\beta}\right) \ed
\end{align}
In particular, if we denote their components as
\begin{align}
	(J^{(\ell)})_{\alpha}{}^{\beta} = \begin{pmatrix}
	J_3^{(\ell)} & J_+^{(\ell)} \\ J_-^{(\ell)} & - J_3^{(\ell)}
	\end{pmatrix} \ecq (J^{(r)})_{\alpha}{}^{\beta} = \begin{pmatrix}
	J_3^{(r)} & J_+^{(r)} \\ J_-^{(r)} & - J_3^{(r)}
	\end{pmatrix} \ec
\end{align}
then the only non-trivial even-even commutators are
\begin{alignat}{3}
	[J_3^{(\ell)}, J^{(\ell)}_{\pm}] &=\pm J^{(\ell)}_{\pm} \ecq &[J^{(\ell)}_+,J^{(\ell)}_-]&= 2J^{(\ell)}_3 \ec\\
	[J_3^{(r)}, J^{(r)}_{\pm}] &=\pm J^{(r)}_{\pm} \ecq &[J^{(r)}_+,J^{(r)}_-]&= 2J^{(r)}_3 \ed
\end{alignat}
The action of the generators $J^{(\ell/r)}$ on a scalar operator $\cO(x)$ on $S^3$ is given by
\begin{alignat}{3}
	[J_3^{(\ell)}, \cO(x)] &= - \cL^3 \cO(x) &\ecq [J_3^{(r)}, \cO(x)] &= - \cR^3 \cO(x) \ec\\
	[J_{\pm}^{(\ell)}, \cO(x)] &= - (\cL^1\pm i\cL^2) \cO(x) &\ecq [J_{\pm}^{(r)}, \cO(x)] &= - (\cR^1\pm i\cR^2)\cO(x) \ec
\end{alignat}
where $\cL^i$ and $\cR^i$ were defined in \eqref{L1}--\eqref{R3}.
	
Finally, the non-trivial even-odd commutators are
\begin{alignat}{3}
	[R_{\ell}, \cQ_{\alpha}^{(\ell_{\pm})}] &= \pm \cQ_{\alpha}^{(\ell_{\pm})}\ecq & [R_r, \cQ_{\alpha}^{(r_{\pm})}] &= \pm \cQ_{\alpha}^{(r_{\pm})}\ec \\
	[(J^{(\ell)})_{\alpha}{}^{\beta}, \cQ^{(\ell_{\pm})}_{\gamma}] &= \delta_{\gamma}{}^{\delta}\cQ^{(\ell_{\pm})}_{\alpha} - \frac{1}{2}\delta_{\alpha}{}^{\beta}\cQ^{(\ell_{\pm})}_{\gamma} \ecq & [(J^{(r)})_{\alpha}{}^{\beta}, \cQ^{(r_{\pm})}_{\gamma}] &= \delta_{\gamma}{}^{\delta}\cQ^{(r_{\pm})}_{\alpha} - \frac{1}{2}\delta_{\alpha}{}^{\beta}\cQ^{(r_{\pm})}_{\gamma} \ed
\end{alignat}
	
For the choice of $h_a{}^b = -\sigma^2$ and $\bar{h}^{\dot a}{}_{\dot b}=-\sigma^3$, we can take
\begin{align}
	u_{+}^a = \frac{1}{2}\begin{pmatrix}
	1-i \\1+i
	\end{pmatrix} \ecq u_{-}^a = \frac{1}{2}\begin{pmatrix}
	1-i \\-1-i
	\end{pmatrix} \ecq \bar{u}_+^{\dot a} = \begin{pmatrix}
	0\\1
	\end{pmatrix} \ecq \bar{u}_-^{\dot a} = \begin{pmatrix}
	1\\0
	\end{pmatrix} \ed
\end{align}
One then finds that
\begin{align}
	\cQ^{(r_-)}_1+\cQ^{(\ell_+)}_1 &= Q_{11\dot 2}+\frac{1}{2r}S^2{}_{2\dot 2} \ec\\
	\cQ^{(\ell_-)}_2+\cQ^{(r_+)}_2 &= Q_{21\dot 1}+\frac{1}{2r}S^1{}_{2\dot 1} \ec
\end{align}
are the nilpotent supercharges used to define the Higgs branch cohomology in \cite{Chester:2014mea,Beem:2016cbd}.
	
\section{1d Green's function from 3d theory}
\label{propagator}
The Green's function \eqref{GotG} of the fundamental twisted Higgs branch operators \eqref{QQt} inserted on the $\theta=\frac{\pi}{2}$ circle in $S^3$, can be calculated directly from the 3d Gaussian action \eqref{ShyperLoc3d}. Without loss of generality let us consider a $U(1)$ gauge theory with one hypermultiplet. The bosonic part of the action is
\es{ScalarPart}{
	S_\text{free hyper} = \int d^3 x \, \sqrt{g} 
	\, \tq^a(x) {\cal D}_a{}^b(x) q_b(x) \,,
}
where the operator ${\cal D}_a{}^b(x)$ is defined by
\es{DOp}{
	{\cal D}_a{}^b(x) \equiv \begin{pmatrix}
	- \nabla^2 + \frac 3{4r^2} + \frac{\sigma^2}{r^2} & -\frac{\sigma}{r^2} \\
	\frac{\sigma}{r^2} & -\nabla^2 + \frac 3{4r^2} + \frac{\sigma^2}{r^2} 
	\end{pmatrix}  \,.
}
	
It is a straightforward exercise to determine the two-point function ${\cal G}_a{}^b(x, x') = \langle q_a(x) \tq^b(x') \rangle$ by solving the differential equation
\es{DiffEq}{
	{\cal D}_a{}^c(x) {\cal G}_c{}^b(x, x') = \frac{\delta_a^b}{\sqrt{g(x')}} \delta^3(x - x') \,.
}
The solution is
\es{calG}{
	{\cal G}_a{}^b(x, x') = \langle q_a(x) \tq^b(x') \rangle 
	= \frac{1}{8 \pi r \cosh(\sigma \pi)} \begin{pmatrix}
	\frac{\cosh (\sigma \pi - \sigma \gamma) }{\sin (\gamma/2)}  & 
	\frac{\sinh( \sigma \pi - \sigma \gamma)}{\cos(\gamma/2) } \\
	-\frac{\sinh( \sigma \pi - \sigma \gamma)}{\cos(\gamma/2) } & 
	\frac{\cosh (\sigma \pi - \sigma \gamma) }{\sin (\gamma/2)} 
	\end{pmatrix}
}
where $\gamma$ is the relative angle between the points $x$ and $x'$.  In the coordinates $(\theta, \tau, \varphi)$ used previously, it is given by
\es{GotGamma}{
	\cos \gamma = \cos \theta \cos \theta' \cos (\tau - \tau') + \sin \theta \sin \theta' \cos (\varphi - \varphi') \,.
}
In particular, when both $x$ and $x'$ belong to the circle at $\theta = \pi/2$, we have $\gamma = \abs{\varphi - \varphi'}$.  Using the definition \eqref{QQt} of $Q(\varphi)$ and $\tQ(\varphi)$ in terms of the fields $q_a(x)$ and $\tq^a(x)$ evaluated on this circle, we have
\es{QQFromG}{
	\langle Q(\varphi) \tQ(0) \rangle = - \cos \frac{\varphi}{2} {\cal G}_1{}^2(\varphi, 0) -
	\sin \frac{\varphi}{2} {\cal G}_2{}^2(\varphi, 0) \,,
}
which, when using \eqref{calG}, can be seen to agree precisely with \eqref{GotG}.
	
\section{$\cQ^H_{\beta}$ BPS equations}
\label{BPSEQUATIONS}
In this section we will study the full set of BPS equations
\begin{align}
\delta_{\xi}\lambda_{a\dot a}=\delta_{\xi, \nu} \psi_{\dot a} = \delta_{\xi, \nu} \tpsi_{\dot a} = 0 \ec \label{BPSequations}
\end{align}
where the transformations were defined in \eqref{lamvar}, \eqref{xinuTrans1} and \eqref{xinuTrans2}, the Killing spinor $\xi = \xi^H_{\beta}$ is defined in \eqref{xiQ1Q2}, and $\nu_{a\dot a}$ satisfies \eqref{nucond}.\footnote{ In this section, we will always write $\xi$ for the particular spinor $\xi^H_{\beta}$ defined in \eqref{xiQ1Q2} to avoid clutter.} Here, we study the consequences of \eqref{BPSequations} before the reality conditions are imposed on the fields.

Let us unpack the contents of these equations. The gaugino BPS equations, can be used to solve for the auxiliary fields $D_{ab}$. This solution can be written as
\begin{align}
i D_{ab} = \frac{1}{\mu}\left( - \frac{i}{4} \varepsilon^{\mu\nu\rho}v_{\rho}\mu_{ab} F_{\mu\nu} + \mu_a{}^c(\xi_a{}^{\dot a}\gamma^{\mu}\xi_b{}^{\dot b}) \cD_{\mu}\Phi_{\dot{a}\dot b} + 2\mu_a{}^c(\xi_c{}^{\dot a}\xi'_b{}^{\dot b})\Phi_{\dot{a}\dot b} - \frac{1}{2}\mu_{ab}\mu^{\dot{a}\dot b}\Phi_{\dot a}{}^{\dot c}\Phi_{\dot{c}\dot b}\right) \ec \label{Dsol}
\end{align}
where the symmetric matrices $\mu_{ab}$ and $\mu_{\dot{a}\dot b}$ are given by
\begin{align}
\mu_{ab} = (\xi_a{}^{\dot c}\xi_{b\dot c}) \ecq \mu_{\dot{a}\dot b} = (\xi^c{}_{\dot a}\xi_{c\dot b}) \ecq \mu \equiv \det(\mu_{ab}) = \det(\mu_{\dot{a}\dot b}) = \beta^2 \cos^2(\theta) \ec
\end{align}
and $v^{\mu}$ is the Killing vector generating translations along $\tau$:
\begin{align}
v^{\mu} \equiv i \xi^{a\dot a}\gamma^{\mu}\xi_{a\dot a} \ed
\end{align}
	
The remaining gaugino BPS equations imply that the fields are independent of $\tau$ up to a field dependent gauge transformation. The result is more conveniently expressed in terms of the twisted fields
\begin{align}
\widetilde{\Phi}_{\dot{1} \dot 1} \equiv e^{i\tau}\Phi_{\dot{1}\dot 1} \ecq \widetilde{\Phi}_{\dot{2} \dot 2} \equiv e^{-i\tau}\Phi_{\dot{2}\dot 2} \ec \label{tPhi}
\end{align}
which satisfy $[\cZ,\widetilde{\Phi}_{\dot{1} \dot 1}]=[\cZ,\widetilde{\Phi}_{\dot{2} \dot 2}]=0$, up to a gauge transformation, where $\cZ$ was defined in \eqref{cZS3def}. The BPS configurations are naturaly expressed in terms of \eqref{tPhi} since $\cQ_{\beta}^H$ squares to $\cZ$. Let us also define a modified connection $\cD^{\star}_{\mu}$ as\footnote{Note that in our conventions $(\Phi_{\dot{1}\dot 1})^{\dagger} = -\Phi_{\dot{2}\dot 2}$, so the connection \eqref{Dstar} is complex unless $\beta$ is pure imaginary.}
\begin{align}
\cD^{\star}_{\tau} = \cD_{\tau} + \frac{ir}{2}\cos(\theta)\left(\beta\widetilde{\Phi}_{\dot{1}\dot 1} + \frac{1}{\beta}\widetilde{\Phi}_{\dot{2}\dot 2}\right) \ecq \cD_{\theta,\varphi}^{\star} = \cD_{\theta,\varphi} \ed \label{Dstar}
\end{align}
Using the definitions \eqref{tPhi} and \eqref{Dstar}, one can show that $\delta_{\xi}\lambda_{a\dot b} = 0$ implies that
\begin{align}
F^{\star}_{\theta\tau} = F^{\star}_{\varphi\tau} = \cD^{\star}_{\tau}D_{ab} = \cD^{\star}_{\tau}\widetilde{\Phi}_{\dot{1}\dot 1} = \cD^{\star}_{\tau}\widetilde{\Phi}_{\dot{2}\dot 2} = \cD^{\star}_{\tau}\Phi_{\dot{1}\dot 2} = 0 \ec \label{DtauVector}
\end{align}
where $F^{\star}_{\mu\nu} = i[\cD^{\star}_{\mu},\cD^{\star}_{\nu}]$. As implied by \eqref{DtauVector}, the modified connection \eqref{Dstar} actually satisfies $\cD_{\tau}\left(\beta\widetilde{\Phi}_{\dot{1}\dot 1} + \frac{1}{\beta}\widetilde{\Phi}_{\dot{2}\dot 2}\right)=0$, and so is literally independent of $\tau$ up to a gauge transformation. It then follows that all fields in $\cV$ are similarly $\tau$-independent.
	
The analysis of the $\cH'$ hypermultiplet BPS equations $\delta_{\xi, \nu} \psi_{\dot a} = \delta_{\xi, \nu} \tpsi_{\dot a} = 0$, is similar. One first solves for the auxiliary fields:
\begin{align}
G_a &= \frac{1}{\mu}\mu_{ad}^{(\nu)}\left[(\nu^{d\dot a}\gamma^{\mu}\xi^b{}_{\dot a})D_{\mu}q_b - (\nu^{d\dot a}\xi^b{}_{\dot c})\Phi^{\dot c}{}_{\dot a}q_b+ (\nu^{d\dot a}\xi'^b{}_{\dot a}) q_b\right] \ec \label{Gsol}\\
\widetilde{G}_a &= \frac{1}{\mu}\mu_{ad}^{(\nu)}\left[(\nu^{d\dot a}\gamma^{\mu}\xi^b{}_{\dot a})D_{\mu}\tq_b +  (\nu^{d\dot a}\xi^b{}_{\dot c})\tq_b\Phi^{\dot c}{}_{\dot a}+ (\nu^{d\dot a}\xi'^b{}_{\dot a}) \tq_b \right]\ec\label{Gtsol} 
\end{align} 
where we defined $\mu^{(\nu)}_{ab}\equiv (\nu_a{}^{\dot c}\nu_{b\dot c})$.
The remaining equations then imply that
\begin{align}
	\cD_{\tau}^{\star} G_a = \cD_{\tau}^{\star} \widetilde{G}_a= \cD_{\tau}^{\star} q_a = \cD_{\tau}^{\star}\tq_a = 0 \ed \label{DtauHyper}
	\end{align}
Note that the solutions \eqref{Gsol} and \eqref{Gtsol} for the auxiliary fields depend on the spinors $\nu_{a\dot a}$. Nevertheless, the conditions these spinors satisfy \eqref{nucond} can be shown to imply that \eqref{DtauHyper} holds for any choice of $\nu_{a\dot a}$.
	
The solutions \eqref{Dsol}, \eqref{Gsol}, \eqref{Gtsol} for the auxiliary fields in terms of the dynamical ones, together with the $\tau$-independence conditions \eqref{DtauVector} and \eqref{DtauHyper}, comprise the full set of restrictions that follow from the BPS equations \eqref{BPSequations} without imposing additional reality conditions on the fields. These conditions are sufficient in order to show that the action $S_{\text{hyper}}'[\cH']$ defined in \eqref{Sphyper} localizes to the 1d action \eqref{S1d}. Indeed, after dimensional reduction on $\tau$, plugging \eqref{Dsol}, \eqref{Gsol} and \eqref{Gtsol} in $S_{\text{hyper}}'[\cH']$, one can show that
\begin{align}
S_{\text{hyper}}'[\cH']\biggr|_{\cQ_{\beta}^H-\text{BPS}} = \int_{D^2} d^2x\sqrt{g_{D^2}} \nabla_{\bar{\mu}} K^{\bar{\mu}} \ec \label{DK}
\end{align}
where 
\begin{align}
K^{\bar{\mu}} = -\frac{r\cos(\theta)}{\mu}\left( i\varepsilon^{\bar{\mu}\bar{\nu}\tau}\mu^{\dot{a}\dot{b}}(\xi^b{}_{\dot a}\gamma_{\tau}\xi^c{}_{\dot b})\tq_b D_{\bar{\nu}}q_c + \mu^{\dot{a}\dot{b}}(\xi^b{}_{\dot a}\gamma^{\bar{\mu}}\xi'^c{}_{\dot b})\tq_b q_c +\mu^{ac}(\xi_c{}^{\dot a}\gamma^{\bar{\mu}}\xi^{b\dot b}) \tq_a\Phi_{\dot{a}\dot{b}}q_b \right) \ec \label{K}
\end{align}
where $\bar{\mu}$ runs over the coordinates $\theta$ and $\varphi$ of $D^2$. Using the explicit form of \eqref{K}, one can check that the boundary term left from \eqref{DK} is precisely the 1d action \eqref{S1d}. This completes the derivation.
\bibliographystyle{ssg}
\bibliography{3dLocPaper}

\begingroup\raggedright\begin{thebibliography}{10}

\bibitem{Lee:1998bxa}
S.~Lee, S.~Minwalla, M.~Rangamani, and N.~Seiberg, ``{Three point functions of
  chiral operators in D = 4, N=4 SYM at large N},'' {\em Adv. Theor. Math.
  Phys.} {\bf 2} (1998) 697--718,
  \href{http://xxx.lanl.gov/abs/hep-th/9806074}{{\tt hep-th/9806074}}.

\bibitem{DHoker:1999ea}
E.~D'Hoker, D.~Z. Freedman, S.~D. Mathur, A.~Matusis, and L.~Rastelli,
  ``{Extremal correlators in the AdS / CFT correspondence},''
  \href{http://xxx.lanl.gov/abs/hep-th/9908160}{{\tt hep-th/9908160}}.

\bibitem{Baggio:2012rr}
M.~Baggio, J.~de~Boer, and K.~Papadodimas, ``{A non-renormalization theorem for
  chiral primary 3-point functions},'' {\em JHEP} {\bf 07} (2012) 137,
  \href{http://xxx.lanl.gov/abs/1203.1036}{{\tt 1203.1036}}.

\bibitem{Beem:2013sza}
C.~Beem, M.~Lemos, P.~Liendo, W.~Peelaers, L.~Rastelli, and B.~C. van Rees,
  ``{Infinite Chiral Symmetry in Four Dimensions},'' {\em Commun. Math. Phys.}
  {\bf 336} (2015), no.~3 1359--1433,
  \href{http://xxx.lanl.gov/abs/1312.5344}{{\tt 1312.5344}}.

\bibitem{Beem:2014kka}
C.~Beem, L.~Rastelli, and B.~C. van Rees, ``{$ \mathcal{W} $ symmetry in six
  dimensions},'' {\em JHEP} {\bf 05} (2015) 017,
  \href{http://xxx.lanl.gov/abs/1404.1079}{{\tt 1404.1079}}.

\bibitem{Beem:2014rza}
C.~Beem, W.~Peelaers, L.~Rastelli, and B.~C. van Rees, ``{Chiral algebras of
  class S},'' {\em JHEP} {\bf 05} (2015) 020,
  \href{http://xxx.lanl.gov/abs/1408.6522}{{\tt 1408.6522}}.

\bibitem{Chester:2014mea}
S.~M. Chester, J.~Lee, S.~S. Pufu, and R.~Yacoby, ``{Exact Correlators of BPS
  Operators from the 3d Superconformal Bootstrap},'' {\em JHEP} {\bf 03} (2015)
  130, \href{http://xxx.lanl.gov/abs/1412.0334}{{\tt 1412.0334}}.

\bibitem{Beem:2016cbd}
C.~Beem, W.~Peelaers, and L.~Rastelli, ``{Deformation quantization and
  superconformal symmetry in three dimensions},''
  \href{http://xxx.lanl.gov/abs/1601.05378}{{\tt 1601.05378}}.

\bibitem{Tachikawa:2016kfc}
Y.~Tachikawa, ``{A brief review of the 2d/4d correspondences},'' 2016.
\newblock \href{http://xxx.lanl.gov/abs/1608.02964}{{\tt 1608.02964}}.

\bibitem{Dumitrescu:2016ltq}
T.~T. Dumitrescu, ``{An introduction to supersymmetric field theories in curved
  space},'' 2016.
\newblock \href{http://xxx.lanl.gov/abs/1608.02957}{{\tt 1608.02957}}.

\bibitem{Morrison:2016bps}
D.~R. Morrison, ``{Gromov-Witten invariants and localization},'' 2016.
\newblock \href{http://xxx.lanl.gov/abs/1608.02956}{{\tt 1608.02956}}.

\bibitem{Pasquetti:2016dyl}
S.~Pasquetti, ``{Holomorphic blocks and the 5d AGT correspondence},'' 2016.
\newblock \href{http://xxx.lanl.gov/abs/1608.02968}{{\tt 1608.02968}}.

\bibitem{Kim:2016usy}
S.~Kim and K.~Lee, ``{Indices for 6 dimensional superconformal field
  theories},'' 2016.
\newblock \href{http://xxx.lanl.gov/abs/1608.02969}{{\tt 1608.02969}}.

\bibitem{Pestun:2016jze}
V.~Pestun and M.~Zabzine, ``{Introduction to localization in quantum field
  theory},'' \href{http://xxx.lanl.gov/abs/1608.02953}{{\tt 1608.02953}}.

\bibitem{Zarembo:2016bbk}
K.~Zarembo, ``{Localization and AdS/CFT Correspondence},'' 2016.
\newblock \href{http://xxx.lanl.gov/abs/1608.02963}{{\tt 1608.02963}}.

\bibitem{Marino:2016new}
M.~Marino, ``{Localization at large N in Chern-Simons-matter theories},'' 2016.
\newblock \href{http://xxx.lanl.gov/abs/1608.02959}{{\tt 1608.02959}}.

\bibitem{Willett:2016adv}
B.~Willett, ``{Localization on three-dimensional manifolds},'' 2016.
\newblock \href{http://xxx.lanl.gov/abs/1608.02958}{{\tt 1608.02958}}.

\bibitem{Minahan:2016xwk}
J.~A. Minahan, ``{Matrix models for 5d super Yang-Mills},'' 2016.
\newblock \href{http://xxx.lanl.gov/abs/1608.02967}{{\tt 1608.02967}}.

\bibitem{Hosomichi:2016flq}
K.~Hosomichi, ``{${\cal N}=2$ SUSY gauge theories on $S^4$},'' 2016.
\newblock \href{http://xxx.lanl.gov/abs/1608.02962}{{\tt 1608.02962}}.

\bibitem{Dimofte:2016pua}
T.~Dimofte, ``{Perturbative and nonperturbative aspects of complex Chern-Simons
  Theory},'' 2016.
\newblock \href{http://xxx.lanl.gov/abs/1608.02961}{{\tt 1608.02961}}.

\bibitem{Qiu:2016dyj}
J.~Qiu and M.~Zabzine, ``{Review of localization for 5d supersymmetric gauge
  theories},'' 2016.
\newblock \href{http://xxx.lanl.gov/abs/1608.02966}{{\tt 1608.02966}}.

\bibitem{Pestun:2016qko}
V.~Pestun, ``{Review of localization in geometry},'' 2016.
\newblock \href{http://xxx.lanl.gov/abs/1608.02954}{{\tt 1608.02954}}.

\bibitem{Benini:2016qnm}
F.~Benini and B.~Le~Floch, ``{Supersymmetric localization in two dimensions},''
  2016.
\newblock \href{http://xxx.lanl.gov/abs/1608.02955}{{\tt 1608.02955}}.

\bibitem{Pufu:2016zxm}
S.~S. Pufu, ``{The F-Theorem and F-Maximization},'' 2016.
\newblock \href{http://xxx.lanl.gov/abs/1608.02960}{{\tt 1608.02960}}.

\bibitem{Rastelli:2016tbz}
L.~Rastelli and S.~S. Razamat, ``{The supersymmetric index in four
  dimensions},'' 2016.
\newblock \href{http://xxx.lanl.gov/abs/1608.02965}{{\tt 1608.02965}}.

\bibitem{Closset:2012vg}
C.~Closset, T.~T. Dumitrescu, G.~Festuccia, Z.~Komargodski, and N.~Seiberg,
  ``{Contact Terms, Unitarity, and F-Maximization in Three-Dimensional
  Superconformal Theories},'' {\em JHEP} {\bf 10} (2012) 053,
  \href{http://xxx.lanl.gov/abs/1205.4142}{{\tt 1205.4142}}.

\bibitem{Closset:2012ru}
C.~Closset, T.~T. Dumitrescu, G.~Festuccia, and Z.~Komargodski,
  ``{Supersymmetric Field Theories on Three-Manifolds},'' {\em JHEP} {\bf 05}
  (2013) 017, \href{http://xxx.lanl.gov/abs/1212.3388}{{\tt 1212.3388}}.

\bibitem{Gerchkovitz:2016gxx}
E.~Gerchkovitz, J.~Gomis, N.~Ishtiaque, A.~Karasik, Z.~Komargodski, and S.~S.
  Pufu, ``{Correlation Functions of Coulomb Branch Operators},''
  \href{http://xxx.lanl.gov/abs/1602.05971}{{\tt 1602.05971}}.

\bibitem{Papadodimas:2009eu}
K.~Papadodimas, ``{Topological Anti-Topological Fusion in Four-Dimensional
  Superconformal Field Theories},'' {\em JHEP} {\bf 08} (2010) 118,
  \href{http://xxx.lanl.gov/abs/0910.4963}{{\tt 0910.4963}}.

\bibitem{Baggio:2014sna}
M.~Baggio, V.~Niarchos, and K.~Papadodimas, ``{Exact correlation functions in
  $SU(2) \mathcal N=2$ superconformal QCD},'' {\em Phys. Rev. Lett.} {\bf 113}
  (2014), no.~25 251601, \href{http://xxx.lanl.gov/abs/1409.4217}{{\tt
  1409.4217}}.

\bibitem{Baggio:2014ioa}
M.~Baggio, V.~Niarchos, and K.~Papadodimas, ``{tt$^{*}$ equations, localization
  and exact chiral rings in 4d $ \mathcal{N} $ =2 SCFTs},'' {\em JHEP} {\bf 02}
  (2015) 122, \href{http://xxx.lanl.gov/abs/1409.4212}{{\tt 1409.4212}}.

\bibitem{Baggio:2015vxa}
M.~Baggio, V.~Niarchos, and K.~Papadodimas, ``{On exact correlation functions
  in SU(N) $ \mathcal{N}=2 $ superconformal QCD},'' {\em JHEP} {\bf 11} (2015)
  198, \href{http://xxx.lanl.gov/abs/1508.03077}{{\tt 1508.03077}}.

\bibitem{Seiberg:1996nz}
N.~Seiberg and E.~Witten, ``{Gauge dynamics and compactification to
  three-dimensions},'' in {\em {The mathematical beauty of physics: A memorial
  volume for Claude Itzykson. Proceedings, Conference, Saclay, France, June
  5-7, 1996}}, 1996.
\newblock \href{http://xxx.lanl.gov/abs/hep-th/9607163}{{\tt hep-th/9607163}}.

\bibitem{Intriligator:1996ex}
K.~A. Intriligator and N.~Seiberg, ``{Mirror symmetry in three-dimensional
  gauge theories},'' {\em Phys. Lett.} {\bf B387} (1996) 513--519,
  \href{http://xxx.lanl.gov/abs/hep-th/9607207}{{\tt hep-th/9607207}}.

\bibitem{deBoer:1996mp}
J.~de~Boer, K.~Hori, H.~Ooguri, and Y.~Oz, ``{Mirror symmetry in
  three-dimensional gauge theories, quivers and D-branes},'' {\em Nucl. Phys.}
  {\bf B493} (1997) 101--147,
  \href{http://xxx.lanl.gov/abs/hep-th/9611063}{{\tt hep-th/9611063}}.

\bibitem{Hanany:1996ie}
A.~Hanany and E.~Witten, ``{Type IIB superstrings, BPS monopoles, and
  three-dimensional gauge dynamics},'' {\em Nucl. Phys.} {\bf B492} (1997)
  152--190, \href{http://xxx.lanl.gov/abs/hep-th/9611230}{{\tt
  hep-th/9611230}}.

\bibitem{deBoer:1996ck}
J.~de~Boer, K.~Hori, H.~Ooguri, Y.~Oz, and Z.~Yin, ``{Mirror symmetry in
  three-dimensional theories, SL(2,Z) and D-brane moduli spaces},'' {\em Nucl.
  Phys.} {\bf B493} (1997) 148--176,
  \href{http://xxx.lanl.gov/abs/hep-th/9612131}{{\tt hep-th/9612131}}.

\bibitem{Gaiotto:2008ak}
D.~Gaiotto and E.~Witten, ``{S-Duality of Boundary Conditions In N=4 Super
  Yang-Mills Theory},'' {\em Adv. Theor. Math. Phys.} {\bf 13} (2009), no.~3
  721--896, \href{http://xxx.lanl.gov/abs/0807.3720}{{\tt 0807.3720}}.

\bibitem{Kapustin:2009kz}
A.~Kapustin, B.~Willett, and I.~Yaakov, ``{Exact Results for Wilson Loops in
  Superconformal Chern-Simons Theories with Matter},'' {\em JHEP} {\bf 03}
  (2010) 089, \href{http://xxx.lanl.gov/abs/0909.4559}{{\tt 0909.4559}}.

\bibitem{Bullimore:2015lsa}
M.~Bullimore, T.~Dimofte, and D.~Gaiotto, ``{The Coulomb Branch of 3d
  $\mathcal{N}=4$ Theories},'' \href{http://xxx.lanl.gov/abs/1503.04817}{{\tt
  1503.04817}}.

\bibitem{Jafferis:2010un}
D.~L. Jafferis, ``{The Exact Superconformal R-Symmetry Extremizes Z},'' {\em
  JHEP} {\bf 05} (2012) 159, \href{http://xxx.lanl.gov/abs/1012.3210}{{\tt
  1012.3210}}.

\bibitem{Hama:2010av}
N.~Hama, K.~Hosomichi, and S.~Lee, ``{Notes on SUSY Gauge Theories on
  Three-Sphere},'' {\em JHEP} {\bf 03} (2011) 127,
  \href{http://xxx.lanl.gov/abs/1012.3512}{{\tt 1012.3512}}.

\bibitem{Pestun:2007rz}
V.~Pestun, ``{Localization of gauge theory on a four-sphere and supersymmetric
  Wilson loops},'' {\em Commun. Math. Phys.} {\bf 313} (2012) 71--129,
  \href{http://xxx.lanl.gov/abs/0712.2824}{{\tt 0712.2824}}.

\bibitem{Pestun:2009nn}
V.~Pestun, ``{Localization of the four-dimensional N=4 SYM to a two-sphere and
  1/8 BPS Wilson loops},'' {\em JHEP} {\bf 12} (2012) 067,
  \href{http://xxx.lanl.gov/abs/0906.0638}{{\tt 0906.0638}}.

\bibitem{Berkovits:1993hx}
N.~Berkovits, ``{A Ten-dimensional superYang-Mills action with off-shell
  supersymmetry},'' {\em Phys. Lett.} {\bf B318} (1993) 104--106,
  \href{http://xxx.lanl.gov/abs/hep-th/9308128}{{\tt hep-th/9308128}}.

\bibitem{Assel:2015oxa}
B.~Assel and J.~Gomis, ``{Mirror Symmetry And Loop Operators},'' {\em JHEP}
  {\bf 11} (2015) 055, \href{http://xxx.lanl.gov/abs/1506.01718}{{\tt
  1506.01718}}.

\bibitem{Aharony:1997bx}
O.~Aharony, A.~Hanany, K.~A. Intriligator, N.~Seiberg, and M.~J. Strassler,
  ``{Aspects of N=2 supersymmetric gauge theories in three-dimensions},'' {\em
  Nucl. Phys.} {\bf B499} (1997) 67--99,
  \href{http://xxx.lanl.gov/abs/hep-th/9703110}{{\tt hep-th/9703110}}.

\bibitem{Giombi:2009ds}
S.~Giombi and V.~Pestun, ``{Correlators of local operators and 1/8 BPS Wilson
  loops on $S^2$ from 2d YM and matrix models},'' {\em JHEP} {\bf 10} (2010)
  033, \href{http://xxx.lanl.gov/abs/0906.1572}{{\tt 0906.1572}}.

\bibitem{Giombi:2012ep}
S.~Giombi and V.~Pestun, ``{Correlators of Wilson Loops and Local Operators
  from Multi-Matrix Models and Strings in AdS},'' {\em JHEP} {\bf 01} (2013)
  101, \href{http://xxx.lanl.gov/abs/1207.7083}{{\tt 1207.7083}}.

\bibitem{Drukker:2009sf}
N.~Drukker and J.~Plefka, ``{Superprotected n-point correlation functions of
  local operators in N=4 super Yang-Mills},'' {\em JHEP} {\bf 04} (2009) 052,
  \href{http://xxx.lanl.gov/abs/0901.3653}{{\tt 0901.3653}}.

\bibitem{WittenQM}
E.~Witten, ``{A New Look At The Path Integral Of Quantum Mechanics},''
  \href{http://xxx.lanl.gov/abs/1009.6032}{{\tt 1009.6032}}.

\bibitem{WittenCS}
E.~Witten, ``{Analytic Continuation Of Chern-Simons Theory},'' {\em AMS/IP
  Stud. Adv. Math.} {\bf 50} (2011) 347--446,
  \href{http://xxx.lanl.gov/abs/1001.2933}{{\tt 1001.2933}}.

\bibitem{Ded}
M.~Dedushenko, ``{Violation of the phase space general covariance as a
  diffeomorphism anomaly in quantum mechanics},'' {\em JHEP} {\bf 10} (2010)
  054, \href{http://xxx.lanl.gov/abs/1007.5292}{{\tt 1007.5292}}.

\bibitem{Catt}
A.~S. Cattaneo and G.~Felder, ``{A Path integral approach to the Kontsevich
  quantization formula},'' {\em Commun. Math. Phys.} {\bf 212} (2000) 591--611,
  \href{http://xxx.lanl.gov/abs/math/9902090}{{\tt math/9902090}}.

\bibitem{Joung:2014qya}
E.~Joung and K.~Mkrtchyan, ``{Notes on higher-spin algebras: minimal
  representations and structure constants},'' {\em JHEP} {\bf 05} (2014) 103,
  \href{http://xxx.lanl.gov/abs/1401.7977}{{\tt 1401.7977}}.

\bibitem{Kapustin:2010xq}
A.~Kapustin, B.~Willett, and I.~Yaakov, ``{Nonperturbative Tests of
  Three-Dimensional Dualities},'' {\em JHEP} {\bf 10} (2010) 013,
  \href{http://xxx.lanl.gov/abs/1003.5694}{{\tt 1003.5694}}.

\bibitem{Gulotta:2011si}
D.~R. Gulotta, C.~P. Herzog, and S.~S. Pufu, ``{From Necklace Quivers to the
  F-theorem, Operator Counting, and T(U(N))},'' {\em JHEP} {\bf 12} (2011) 077,
  \href{http://xxx.lanl.gov/abs/1105.2817}{{\tt 1105.2817}}.

\bibitem{Kapustin:1999ha}
A.~Kapustin and M.~J. Strassler, ``{On mirror symmetry in three-dimensional
  Abelian gauge theories},'' {\em JHEP} {\bf 04} (1999) 021,
  \href{http://xxx.lanl.gov/abs/hep-th/9902033}{{\tt hep-th/9902033}}.

\end{thebibliography}\endgroup

\end{document}